\documentclass[10pt,aps,showpacs,nofootinbib,prd,aps,epsf,floats,
               amsmath,amssymb,amsfonts,axodraw,floatfix,graphicx,twocolumn]{revtex4-2}
\usepackage{graphicx}
\usepackage{dcolumn}
\usepackage{microtype}
\usepackage{amsmath,amssymb}
\usepackage{multirow}
\usepackage{paralist}
\usepackage[utf8x]{inputenc}
\usepackage[T1]{fontenc}
\usepackage{ae}
\usepackage{aecompl}
\usepackage{titlesec}
\usepackage{bm}
\usepackage{dsfont}
\usepackage[export]{adjustbox}
\usepackage[title]{appendix}
\def\bfseries{\fontseries \bfdefault \selectfont \boldmath}
\usepackage{bm}
\usepackage{subfigure}
\usepackage[normalem]{ulem} 
\usepackage{xcolor,soul}
\renewcommand{\theequation}{\thesection.\arabic{equation}}
\newcounter{saveeqn}
\newcommand{\add}{\addtocounter{equation}{1}}
\newcommand{\alphaeqn}{\setcounter{saveeqn}{\value{equation}}%
\setcounter{equation}{0}%
\renewcommand{\theequation}{\mbox{\thesection.\arabic{saveeqn}{\alpha{equation}}}}}
\newcommand{\reseteqn}{\setcounter{equation}{\value{saveeqn}}%
\renewcommand{\theequation}{\thesection.\arabic{equation}}}

\definecolor{olive}{rgb}{0.3, 0.54, .1}

\titleformat*{\section}{\large\bfseries}

\newcommand{\bea}{\begin{eqnarray}}
\newcommand{\eea}{\end{eqnarray}}
\newcommand{\beaa}{\begin{align}}
\newcommand{\eeaa}{\end{align}}

\newcommand{\Dsf}{\slash\hspace{-0.28cm}\mathcal{D}^{(f)}}

\newcommand{\no}{\nonumber}
\begin{document}

\title{ 
Chiral phase transition of a dense,   magnetized and rotating quark matter
}
\author{S.~M.~A.~Tabatabaee Mehr}\email{tabatabaee@ipm.ir}
\author{F.~Taghinavaz}\email{ftaghinavaz@ipm.ir}
\affiliation{ IPM, School of Particles and Accelerators, P.O. Box 19395-5531, Tehran, Iran}
\begin{abstract}
   We investigate the  chiral symmetry restoration/breaking of a dense, magnetized and rotating  quark matter  within the Nambu Jona-Lasinio  model including $N_f=2$ and $N_c=3$ numbers of flavors and colors, respectively.  Imposing the spectral boundary conditions as well as the positiveness of energy levels lead to a  correlation between the magnetic and rotation fields such that strongly magnetized plasma can not rotate anymore. We solve the gap equation at zero and finite temperature. At finite temperature and baryon chemical potential $\mu_B$, we sketch the phase diagrams $T_c(\mu_B)$ and $T_c(R\Omega)$ in different cases. As a result, we always observe inverse-rotational catalysis mean to decrease $T_c$ by increasing $R\Omega$. But the magnetic field has a more complex structure in the phase diagram. For slowly  rotating plasma, we find that $T_c$ decreases by increasing $eB$, while in the fast rotating plasma we see that $T_c$ increases by increasing $eB$.  Also, we locate exactly the position of Critical End Point  by solving  the equations of first and second derivatives of effective action with respect to the order parameters, simultaneously.  
\end{abstract}

\maketitle
\section{Introduction}
Quark Matter (QM) experiences phase transitions under extreme external conditions \cite{Shuryak:2004cy}. These external conditions could be temperature, chemical potential, magnetic or rotation fields, and so on. Examining the phase transition of QM has revealed much valuable information including the existence of the Quark-Gluon Plasma (QGP) phase that contains deconfined quarks and gluons at very high temperatures \cite{ALICE:2012eyl, STAR:2005gfr}. At very dense systems, the quarks bind to each other to break the color symmetry and produce the so-called color-superconductivity phase  \cite{Alford:1997zt}. It may arise weird phases under special conditions which depend on the considered symmetry \cite{Fukushima:2010bq}. The phase transition properties can serve as Equation of State (EoS) data or inputs to the dynamical evolutions of QM at Large-Hadron Colliders (LHC) and that is why they are so important.    

Magnetic field $eB$ and rotational fields $\Omega$ have much relevance to the QM studies. In a non-central collisions of heavy ions the spectator particles create a strong magnetic field of order $\sqrt{eB} \sim 0.1 GeV$ for the RHIC and $\sqrt{eB} \sim 0.5 GeV$ for the LHC \cite{Skokov:2009qp}. Also, the non-central collision may produce a large amount of angular momenta which results in the most vortical fluid ever seen \cite{STAR:2017ckg}. The strength of the rotational field in LHC is about $10^{22} s^{-1}$ which is very larger than any observed vortical fluids in nature. Needless to say that these huge magnetic and rotational fields can potentially affect the QM properties.

There is much more known about the properties of QM under the external magnetic field. Most of the low-energy models predicted that chiral symmetry breaking is enhanced as a function of $eB$, see the review article \cite{Miransky:2015ava}. Magnetic fields have a potential to induce  variety of exciting effects in the thermodynamics of QCD. Some effective model calculations have shown that the transition temperature between the hadron matter and QM increases with $eB$ \cite{Mizher:2010zb, Fukushima:2012xw}.  This conclusion was proved in the Sakai-Sugimoto model \cite{Johnson:2008vna} and within the holographic approach \cite{Evans:2010xs}. Furthermore, splitting between the deconfinement and chiral transitions was predicted to take place for large $eB$s.  Also, the strength of the transition was  seen to increase and results in a first-order phase transition \cite{Fraga:2008um}.    However, it was observed the opposite effect of decreasing transition temperature  with increasing $eB$.  This observation was confirmed in the lattice QCD \cite{Bali:2011qj, Endrodi:2015oba, Bruckmann:2013oba}, in the holographic models \cite{Preis:2010cq} and in the quark-hadron phase transition \cite{Agasian:2008tb}. The Strong magnetic field can also reveal the non-trivial structures of gauge-field topology as a chiral magnetic effect \cite{Fukushima:2008xe, Kharzeev:2007jp}.

In recent years, studies of QM under the strong rotation have got lots of interests. Here, we refer to some of them. The uncharged rotating fermionic matter phase diagrams were studied within the Nambu-Jona Lasinio (NJL) model \cite{Chernodub:2016kxh, Chernodub:2017ref}. A two-flavor NJL model was used to study the phase diagrams of  charged quarks under a rotation \cite{Wang:2018sur}. Also, the chiral phase transition of a two-flavor NJL model in a rotating sphere was investigated by considering the finite-size effects \cite{Zhang:2020jux}. Lattice QCD techniques have been developing to study the phase diagrams of rotating QM  and they need more attention \cite{Braguta:2021jgn}. The influence of helicity imbalance on the phase diagrams of dense QM was already discussed \cite{Chernodub:2020yaf}. Another interesting topic is the possibility of charged pion condensation in external magnetic and rotation fields and it takes much attention \cite{Liu:2017spl, Liu:2017zhl, Chen:2019tcp}. There is a good review on the QCD phase diagram under the rotation and magnetic field that states the basics of this topic \cite{Chen:2021aiq}. In the paper \cite{Sadooghi:2021upd}, the authors study the uncharged fermions phase diagrams within the one-flavor NJL model in presence of constant magnetic and rotation fields and a global boundary condition.

A missing chain of the QM phase diagram studies in magnetic and rotation fields is to consider situations that could be so close to the real world experiments. Motivated by this in the current paper, we investigate the chiral phase transition of a bounded two-flavor and three-colors NJL model  at finite  chemical potential and temperature in presence of constant magnetic and rotation fields.   

A part of our work is devoted to developing a general framework  to obtain the effective action of chiral order parameter $\sigma$ starting from the NJL model in curved space. It has used the Ritus method as a general approach that can solve the Dirac equation in many cases \cite{Ritus:1972ky}.  The obtained master formula for the in-medium effective action is characterized by the Ritus eigenvalues and we benefit from it to study the phase diagrams. 

We also try  to solve the Dirac equation for quarks with different charges in a constantly rotating and magnetized plasma. This will be done by using the Ritus method, similar to the paper \cite{Sadooghi:2021upd}. Eigenfunctions are written in terms of the hypergeometric functions and eigenvalues (generalized Landau levels) are determined by applying appropriate  boundary condition. We select the spectral boundary condition on which QM is confined in a cylinder with radius \textrm{R} and $R\Omega \leq 1$ because of the causality. Furthermore, the net flux going through the boundary $R_c = \frac{1}{\Omega}$ is zero. Because of this boundary condition and the positiveness condition of energy eigenvalues, the magnetic field strength 
\footnote{To deal with the magnetic field, we use the dimensionless quantity $\alpha \equiv \frac{|eB| R^2}{2}$, everywhere.}
and the rotation field $R\Omega$ turn out to be correlated. We find that if $\alpha\leq \alpha_c = 7$ then  $0\leq R\Omega\leq $1, while for $\alpha \geq \alpha_c$ we obtain $0\leq R\Omega\leq R \Omega_\textrm{Max}$ where this maximum value  $R\Omega_\textrm{Max}$ decreases by increasing $\alpha$. It is such that for $\alpha \sim 30$ and beyond, the only allowed values of $R\Omega_\textrm{Max} \sim 0$. Therefore,  strongly magnetized plasma can not rotate anymore.    To the best of our knowledge, this is a non-trivial and novel effect that is rooted in the applied boundary condition. 

The gap equation is solved numerically at zero and finite temperature for the QM with $N_c=3$ and $N_f=2$.
 Some features of the phase diagram in our model are in common with the normal NJL model (no rotation or magnetic field) such as increasing of order parameter by increasing the coupling constant or momentum cut-off. A consistency occurs for the needed coupling constant to provide a non-zero dynamical mass at small magnetic fields with the normal NJL coupling threshold, i.e.  $G_c\Lambda^2 \sim \frac{2\pi^2}{N_c N_f}$. The strength of this threshold coupling decreases by increasing $\alpha$ which is a clear sign of magnetic-catalysis. Near the border $x \sim \alpha$, 
 \footnote{$x \equiv \frac{|eB|r^2}{2}$ and $r$ is distance from the rotation axis.}
 the dynamical mass starts to fall because of the violation of slowly varying field assumption, $\partial_r \sigma \ll \sigma^2$.  We observe an interesting effect near the boundary,  namely the surface magnetic catalysis that is due to the mode accumulation \cite{Chen:2017xrj}. Since vorticity does not play a role at zero temperature, we don not observe the rotational-magnetic inhibition \cite{Chen:2015hfc}. The profile $\sigma(\alpha)$ seems to be very oscillatory at  small magnetic fields and no assertive statement can be made on the magneto-catalysis or inverse magneto-catalysis.
 
 At finite temperature, the phase diagram can be viewed as $T_c(\mu)$ or $T_c(R\Omega)$ planes.\footnote{For the sake of simplicity, we use $\mu$ instead of $\mu_B$.} Below these curves  we get the chirally broken phase, while above them the chiral symmetry is restored. We fit the transition points of the diagrams to a polynomial function in order to compare them better with the Lattice results \cite{Ding:2015ona}. It will be found that the magnetic and rotation fields increase the curvature of the phase diagram compared to the normal model. The important feature is the interplay among  the magnetic and rotation fields that has enriched the physics of the phase diagram. As a consequence, by fixing $\alpha$ and $\mu$, $T_c$ always decreases with $R\Omega$  which is a sign of inverse-rotational catalysis. But fixing $R\Omega$ gives complicated diagrams that in slowly rotating systems we have inverse-magneto catalysis, while in fast rotating matters the dominant pattern of the phase diagram  is magneto-catalysis. To the best of our knowledge, this is a novel situation that has not been seen so far. This feature comes from the boundary condition and the relation between magnetic and rotation fields. We will present a new technique to find the location of Critical End Point(CEP) by solving the gap equation and its derivatives with respect to the $\sigma$ field, simultaneously. Another finding is that the more rotational the system is, the smaller the CEP is. This is a clear sign of inverse-rotational catalysis. Also, the location of CEP and its profile in the $(R \Omega, T_c)$ plane depend on $\alpha$.   Slope of the curve $T_{\textrm{CEP}}(R \Omega)$ or $\frac{d T_{\textrm{CEP}}}{d R\Omega}$ at each fixed $\mu$ for smaller $\alpha$ is bigger than larger $\alpha$. It is such that at very large $\alpha$, this slope closes to zero. As already mentioned, at large $\alpha$ the $R\Omega_\textrm{Max} \to 0$ and thus  ($R\Omega-T_c$) diagram for large $\alpha$ is a point on the vertical axis. 
 
 The Organization of the paper is as follows. First in section 2,  we construct our general framework to obtain the effective action of the $"\sigma"$ field in curved space by utilizing the Ritus method.  Then in section 3, we try to  solve the Dirac equation  for different quark flavors and this task is done   similarly to the procedures  of the paper \cite{Sadooghi:2021upd}. We derive the energy levels as well as the generalized Landau levels and impose the spectral boundary condition  as well as the positiveness condition of energy levels to set a proper quantization scheme. Next in section 4,  by using the master effective action formula we solve numerically the gap equation  at zero and finite temperature by keeping ($\alpha, R\Omega$) fixed. Many plots are sketched and their physical consequences are discussed separately. Finally in section 5, we close our paper with a brief conclusion and outlook to future works.

\section{Effective action}
\setcounter{equation}{0}
In this section, we aim to derive a concrete expression for the effective action of static auxiliary fields $(\sigma, \pi_i)$ in the NJL model with  $N_f=2$ and $N_c$ number of flavors and colors,  respectively. We can assume  the presence of external fields in this general framework  which could be either constant magnetic fields or rotating frames or even both of them.  Although this subject has been known in detail, we shall try to give a general formula for the effective action. After that, we describe its variants by including those fields.     

The quantum field theory of spin $\frac{1}{2}$ particles in curved spaces needs the use of vierbein tensors $e_\mu^a$. They link the global curved space with metric $g_{\mu \nu}$ to the local flat space with metric $\eta_{a b}$ via the identity $g_{\mu \nu} = e_\mu^a e_\nu^b \eta_{a b}$.  Therefore,  a quantity like $\mathcal{A}_\mu$   connects to  $\mathcal{A}_a$ or vice versa through the relation
\footnote{In this paper, we use the Greek indices for the global curved coordinates, while  small Latin indices and numbers refer to the local flat coordinates.}
\bea
\mathcal{A}_\mu = e_\mu^a \mathcal{A}_a, \qquad \mathcal{A}_a = e_a^\mu \mathcal{A}_\mu.
\eea
We use the orthogonality relations $e^\mu_b  e^a_\mu = \delta^a_b$ or $e^\mu_a  e^a_\nu = \delta^\mu_\nu$ for transformation between the local and global coordinates.  Consider the $N_f=2$  local NJL action in the massless limit in a curved space
\footnote{The term "local"  means that every variable of the Lagrangian has space-time dependence. That is why we omit the space-time dependence of each variable. Our concern in this paper is constant external fields and hence we take  this type of action in curved spaces.}
\bea\label{eqsec2f1}
&&S_\textrm{NJL} = \int_x \sqrt{-g} \mathcal{L}_\textrm{NJL},\\
&&\mathcal{L}_\textrm{NJL} = \sum_{f= u, d} \bigg(\Bar{\psi}_f \left(i \, \, \Dsf + \mu_B \gamma^0\right) \psi_f \no\\
&&\hspace{2cm}+\frac{G}{2} \left( (\Bar{\psi}_f \psi_f)^2 + (\Bar{\psi}_f i \gamma_5 \tau_i \psi_f)^2\right) \bigg).\no
\eea
Here, $"g"$ denotes the determinant of metric, $\psi_f = \bigg( \begin{array}{c}
    \psi_u  \\
     \psi_d
\end{array}\bigg)$ represents the fermion doublet, $G$ is the strength of interaction, $\mu_B$ is the baryonic chemical potential and $\tau_i (i= 1, 2, 3)$ are the Pauli matrices. We use the convention $\int_x \equiv \int d^4x$ and $\hspace{0.1cm} \Dsf = \gamma^\mu \mathcal{D}_\mu^{(f)} = \gamma^a \mathcal{D}_a^{(f)}$. Generally, the kernel  $\mathcal{D}_\mu^{(f)}$ is a matrix in Dirac space and includes external fields contribution for each flavor as well as the free partial term
\bea\label{eqsec2f2}
\mathcal{D}_\mu^{(f)} = \partial_\mu + D_\mu^{(f)}.
\eea

Introducing the auxiliary fields $(\sigma, \pi_i)$ would enable us to derive the effective action $\mathcal{V}_\textrm{eff}$ of those fields in the Eq. \eqref{eqsec2f1} via  the bosonization procedure
\bea\label{eqsec2f5}
Z&&= \int \prod_{i=1}^3 \mathcal{D} \sigma \, \mathcal{D} \pi_i \prod_{f=u,d} \mathcal{D}\psi_f\, \mathcal{D}\Bar{\psi}_f \,  \, e^{i \int_x \sqrt{-g} \mathcal{L}_B} \no\\
&&= \int  \prod_{i=1}^3 \mathcal{D} \sigma \, \mathcal{D} \pi_i \, e^{i \int_x \sqrt{-g} \mathcal{V}_\textrm{eff}},
\eea
 with the following action
\bea\label{eqsec2f3}
&&\mathcal{L}_\textrm{B}= - \frac{1}{2 G} \left( \sigma^2 + \pi_i \pi_i\right) \no\\
&&+ \sum_{f= u, d} \Bar{\psi}_f \left(i \,\, \Dsf  + \mu_B \gamma^0 - \sigma - i \gamma_5 \tau_i  \pi_i \right) \psi_f.
\eea
This action equals to its counterpart given by the Eq. \eqref{eqsec2f1} in the level of equations of motion
\bea\label{eqsec2f4}
\sigma = - G \sum_{f= u, d} \langle \Bar{\psi}_f \psi_f\rangle, \quad \pi_i = - G \sum_{f= u, d} \langle \Bar{\psi}_f i \gamma_5 \tau_i \psi_f \rangle.\no
\eea
The equivalence between the action \eqref{eqsec2f1} and \eqref{eqsec2f3} can also be seen from the partition function point of view, which by integrating over the $(\sigma, \pi_i)$  fields, we obtain the original action. Throughout this paper,  we  set $\pi_i=0$. Thus,  $\mathcal{V}_\textrm{eff}$ can be given as  follows
\bea\label{eqsec2f6}
\mathcal{V}_\textrm{eff} = - \frac{1}{2 G} \sigma^2 + \sum_{f= u, d}  \log \det \left( i \,\, \Dsf + \mu_B \gamma^0 - \sigma \right).\no\\
\eea
Our interesting systems let us to simplify the second term of the Eq. \eqref{eqsec2f6} due to the property $\left[\gamma_5, \mathcal{D}_a^{(f)}\right]=0$
\begin{align}\label{eqsec2f7}
&\log \det \left( i \,\, \Dsf + \mu_B \gamma^0 - \sigma \right) = Tr \log \left( i \,\, \Dsf + \mu_B \gamma^0 - \sigma \right) \no\\
&= \frac{1}{2} Tr \log  \left( i \,\, \Dsf + \mu_B \gamma^0 - \sigma \right) \no\\
&+ \frac{1}{2} Tr \log \left(\gamma_5 \left(i \,\, \Dsf + \mu_B \gamma^0 - \sigma \right)\gamma_5\right)\no\\
&= \frac{1}{2} Tr \log \bigg( (\,\,\,\Dsf)^2 -  i \mu_B \left\{ \hspace{0.1cm} \Dsf, \gamma^0 \right\} - i \left[ \hspace{0.1cm} \Dsf, \sigma\right]- \mu_B^2 + \sigma^2 \bigg).\no\\
\end{align}
The third term of the last line expression can be ignored because of the slowly varying field assumption, i.e. $\partial_r \sigma \ll \sigma^2$  and the second term goes to $\left\{ \hspace{0.1cm} \Dsf, \gamma^0 \right\} = 2 \mathcal{D}_0^{(f)}$. Therefore, the  effective potential results in
\begin{align}\label{eqsec2f9}
\mathcal{V}_\textrm{eff} = - \frac{1}{2 G} \sigma^2 + \frac{1}{2} \sum_{f= u, d}  Tr \log \left( (\,\,\,\Dsf)^2 -  2 i \mu_B \mathcal{D}_0^{(f)} - \mu_B^2 + \sigma^2\right).\no\\
\end{align}

We utilize  the Ritus method to obtain the contribution of the second term of Eq. \eqref{eqsec2f9}. It  can help us to solve the Dirac equation and obtain the eigenvalues and eigenstates \cite{Tabatabaee:2020efb, Sadooghi:2016jyf}
\bea\label{eqsec2f10}
&& i \hspace{0.15cm} \Dsf \psi_{\textrm{M}, \kappa}^{(f)}(\Bar{p}) = \kappa \psi_{\textrm{M}, \kappa}^{(f)}(\Bar{p}) \gamma^a \Tilde{p}^{(f)}_a,\no\\
&& i \hspace{0.05cm} \mathcal{D}_0^{(f)} \psi_{\textrm{M}, \kappa}^{(f)} (\Bar{p}) = \kappa \Tilde{p}_0^{(f)} \psi_{\textrm{M}, \kappa}^{(f)}(\Bar{p}).
\eea
Here, $\kappa = \pm 1$ refers to the positive and negative frequency eigenmodes and $\psi_{\textrm{M}, \kappa}^{(f)}(\Bar{p})$ stands for eigenstates of the Dirac equation. Also, the corresponding eigen-four-momentum $\Tilde{p}^{(f)}_a= \left( \Tilde{p}_0^{(f)}, \Tilde{p}_1^{(f)}, \Tilde{p}_2^{(f)}, \Tilde{p}_3^{(f)}\right)$ is a model-dependent quantity and the collection of quantum numbers which identifies the eigenstates is called by $\textrm{M}$. The center-four-momentum $\Bar{p}_a= \left( \Bar{p}_0, \Bar{p}_1, \Bar{p}_2, \Bar{p}_3\right)$ is the eigenvalues of center members of free kernels $ \Bar{\partial}_c^{(f)}$, those that commute with  other kernels $\mathcal{D}_k^{(f)}$. Therefore, this center-momentum appears in the Fourier expansion
\bea
\psi_{\textrm{M}, \kappa} ^{(f)}(\Bar{p}) = e^{i \kappa \Bar{p} \cdot \Bar{x}} \psi_{\textrm{M}, \kappa}^{(f)}(x_\perp),
\eea
where $x_\perp$ is the perpendicular subspace to the $\Bar{x}$ space. Usually, the center-momentum space includes energy and one of the original momenta of particles. Due to the independence of the trace from the chosen bases, we select the eigenstates of the Eq. \eqref{eqsec2f10} to compute the second term of the Eq. \eqref{eqsec2f9}.  We sandwich the operator of the Eq. \eqref{eqsec2f9} between the eigenstates with the same quantum numbers given by the Ritus ansatz
\begin{widetext}
 \begin{align}\label{eqsec2f12}
 \mathcal{V}_\textrm{eff}
 &= - \frac{1}{2 G} \sigma^2 + \frac{1}{2} \sum_{f, \textrm{M}}  \int \frac{d\Bar{p}}{(2\pi)^d} Tr \bigg( \psi_{\textrm{M}, +1}^{(f) \dagger}(\Bar{p}) \, \log \left( (\,\,\Dsf)^2 -  2 i \mu_B \mathcal{D}_0^{(f)} - \mu_B^2 + \sigma^2\right) \psi_{\textrm{M}, +1}^{(f)}(\Bar{p}) \bigg) \no\\
 & = - \frac{1}{2 G} \sigma^2 + \frac{N_c}{2} \sum_{f, \textrm{M}}   \int \frac{d\Bar{p}}{(2\pi)^d} Tr \bigg( \psi_{\textrm{M}, +1}^{(f) \dagger}(\Bar{p}) \, \psi_{\textrm{M}, +1}^{(f)}(\Bar{p}) \bigg) \, \log \left(-\left(\Tilde{p}_0^{(f)} + \mu_B\right)^2 +\mathcal{E}^{(f)^2}\right),
 \end{align}
 \end{widetext}
 where $\mathcal{E}^{(f)} \equiv \sqrt{\sigma^2 + \sum\limits_{i=1}^3 (\Tilde{p}_i^{(f)})^2}$ and $d$ is the number of non-zero elements in $\Bar{p}_a$ space. In the last line trace acts on the Ritus bases. The Eq. \eqref{eqsec2f12} is our general formula for the effective action  in the zero temperature  limit $(T=0)$ and in curved space irrespective of external fields. In some cases we could simplify further the effective action such as integrating over the $\Bar{p}$ space. Likewise, the  Ritus functions are chosen to satisfy the orthogonality condition as  follows  
 \bea\label{eqsec2f13}
 \int_x\, \psi_{\textrm{M}, \kappa}^{(f) \dagger}
 (\Bar{p}')  \, \psi_{\textrm{M}', \kappa'}^{(f)}(\Bar{p}) =  \, \delta_{\textrm{M}, \textrm{M}'}\, \delta_{\kappa, \kappa'} \delta(\Bar{p}' - \Bar{p}).\quad
 \eea
 
 Generalization of the Eq. \eqref{eqsec2f12} to  $T\neq 0$ is straightforward. To this purpose, we have to notice that
 \bea\label{eqsec2f14}
 \tilde{p}_0^{(f)} = \Bar{p}_0 + L_\textrm{M}^{(f)},
 \eea
 where $L_\textrm{M}^{(f)}$ is a generalized and model-dependent angular-momentum contribution. This new part arises due to the curvature of space via the spin-connection term. It means that in general $\Tilde{p}_0^{(f)} \neq \Bar{p}_0^{(f)}$, but in flat space $L_M^{(f)} = 0$.  To obtain the effective action at  finite temperature in the imaginary time formalism the following replacements should  be made
 \bea\label{eqsec2f15}
 \Bar{p}_0 \to i \omega_n = i (2 n +1) \pi T, \quad \int \frac{d\Bar{p}_0}{2\pi} \to i T \sum\limits_{n=-\infty}^\infty.\no
 \eea
 We follow the normal procedure to sum over the Matsubara frequencies of the Eq. \eqref{eqsec2f12} \cite{Kapusta:2006pm} and then we arrive at
 \begin{align}\label{eqsec2f16}
 &\mathcal{V}_\textrm{eff} = - \frac{1}{2 G} \sigma^2 + \frac{N_c}{2} \sum\limits_{f, M}  \int \frac{d\Bar{\mathbf{p}}}{(2\pi)^{d-1}} Tr \bigg( \psi_{\textrm{M},+1}^{(f) \dagger}(\Bar{\mathbf{p}})  \, \psi_{\textrm{M},+1}^{(f)}(\Bar{\mathbf{p}}) \bigg) \, \mathcal{V}_T,\no\\
 & \mathcal{V}_T \equiv   \mathcal{E}^{(f)} + T \log \left(1 + e^{-\beta \left( \mathcal{E}^{(f)} + \mu_B + L_M^{(f)}\right)}\right) \no\\
 &\hspace{1.6cm}+ T \log \left(1 + e^{-\beta \left( \mathcal{E}^{(f)} - \mu_B - L_M^{(f)}\right)}\right).
 \end{align}
 In this relation $\Bar{\mathbf{p}}$ is the non-vanishing spatial part of the center-momenta and $\beta = \frac{1}{T}$. The Eq. \eqref{eqsec2f16} is our general master formula for the in-medium  effective action of  $\sigma$ field in curved space. In what follows  we try to clarify our proposal with a few examples. 
  \subsection{$\mathcal{V}_\textrm{eff}$ in $eB = \Omega =0$}
 Effective action of $\sigma$ field in the medium of free quarks is a very well-known issue. We would like to compare our results with the existing results. In the case of $eB= \Omega= 0$, the eigenstates and eigenmomenta defined in the Eq. \eqref{eqsec2f10},  are given by
 \bea
 \psi_\textrm{M}^{(f)}(\Bar{p}) = e^{i \bar{p} \cdot x} \mathds{1}_{4\times 4}, \,\, \Bar{p}_a = \Tilde{p}_a^{(f)} = \left(p_0, p_1, p_2, p_3\right), \,\, L_M^{(f)}=0,\no
 \eea
where $\mathds{1}_{4 \times 4}$ is the rank 4 identity matrix. Inserting these values into the Eq. \eqref{eqsec2f16}, we obtain 
 \begin{align}\label{eqsec2f18}
 \mathcal{V}^0_\textrm{eff} &= - \frac{\sigma^2}{2G} + 2 N_c N_f  \int \frac{d^3p}{(2\pi)^3} \mathcal{V}^0_T\no\\
 \mathcal{V}^0_T & =  \mathcal{E}_0 + T \log \left(1 + e^{-\beta ( \mathcal{E}_0 + \mu_B)}\right) \no\\
 &\hspace{0.85cm}+ T \log \left(1 + e^{-\beta ( \mathcal{E}_0 - \mu_B )}\right),
 \end{align}
 where  $\mathcal{E}_0 \equiv \sqrt{\sigma^2 + \sum\limits_{i=1}^3 p_i^2}$ and $d^3p = dp_1\, dp_2\, dp_3$. 
 \subsection{$\mathcal{V}_\textrm{eff}$ in $eB  \neq 0, \Omega=0$}
 In a non-rotating  medium ($\Omega=0$) with a constant magnetic field aligned in the $z$ axis $\vec{B}= B \hat{z}$,  the eigensatets and eigenmomenta of  particles with charge $Q_f$ are given in terms of harmonic oscillator functions  \cite{Sadooghi:2016jyf}
\begin{align}\label{eqsec2f19}
&\psi_\textrm{M}^{(f)}(\Bar{p}) = e^{i \Bar{p} \cdot \Bar{x}} \bigg( \mathcal{P}_+^{s_f} f_n^{+ s_f}(\xi_x) + \Pi_n \mathcal{P}_-^{s_f} f_n^{- s_f}(\xi_x)\bigg), \no\\
& \mathcal{P}_\pm^{s_f} = \frac{1 \pm i s_f \gamma_1 \gamma_2}{2} , \,\, s_f = \mbox{sgn} (e Q_f B), \,\,\, \Pi_n = 1 - \delta_{n, 0}, \no\\
& f_+^n(\xi_x) = \frac{1}{\sqrt{\ell_B 2^n n! \sqrt{\pi}}} e^{- \frac{\xi_x^2}{2}} H_n(\xi_x),\,\, \xi_x = \frac{x_1 - s_f p_2 \ell_B^2}{\ell_B}, \no\\
&\ell_B^2 = \frac{1}{|e Q_f B|}, \,\,\,f_n^{- s_f}(\xi_x) = f_{n-1}^{+ s_f}(\xi_x),\,\,\,L_M^{(f)}=0,\no\\
& \Bar{p}_a^{(f)} = \left( p_0, 0, p_2, p_3\right), \,\, \Tilde{p}_a^{(f)} = \left( p_0, 0, - s_f \sqrt{2 n |e Q_f B|}, p_3\right).
\end{align}
The number "$n = j + (1 - s_f \sigma_3)/2$" labels the Landau levels and $\sigma_3=\pm 1$ represent the eigenvalues of spin. From this definition we see that  the lowest Landau level has one spin degrees of freedom, while two spin degrees of freedom occupy  higher Landau levels. Plugging the eigenstates and eigenmomenta of the Eq. \eqref{eqsec2f19} into the Eq. \eqref{eqsec2f16}, we obtain the following result
\begin{align}
    &\mathcal{V}^B_\textrm{eff} = - \frac{\sigma^2}{2G} +  N_c \sum\limits_{f= u, d}  \sum\limits_{n=0}^\infty \int \frac{dp_2 \, dp_3}{(2\pi)^2} \mathcal{B}_n^{(f)}(x)  \mathcal{V}_T^B,\no\\
    & \mathcal{B}_n^{(f)}(x) \equiv (f_n^{+s_f}(\xi_x))^2 + \Pi_n (f_n^{-s_f}(\xi_x))^2, \no\\
    & \mathcal{V}_T^B = \mathcal{E}_B^{(f)} + T \log \left(1 + e^{-\beta ( \mathcal{E}_B^{(f)} + \mu_B)}\right)\no\\
    &\hspace{1.63cm}+ T \log \left(1 + e^{-\beta ( \mathcal{E}_B^{(f)} - \mu_B )}\right),\no\\
    & \mathcal{E}_B^{(f)} = \sqrt{\sigma^2 + 2 n |e Q_f B| + p_3^2}.
\end{align}
The $p_2$ integration is doable by a simple shift of  $f_n^{\pm s_f}(\xi_x)$ argument
\bea
\int dp_2 (f_n^{\pm s_f}(\xi_x))^2 = \ell_B^{-1} \int d\xi_x (f_n^{\pm s_f}(\xi_x))^2 =  \ell_B^{-2}.\no
\eea
Therefore, the final result of the effective action simplifies as 
\bea
\mathcal{V}^B_\textrm{eff} =  - \frac{\sigma^2}{2G} +  N_c  \sum\limits_{f= u, d}   \sum\limits_{n=0}^\infty \frac{|e Q_f B|}{2\pi} \alpha_n \int \frac{dp_3}{2\pi}  \mathcal{V}_T^B,\no\\
\eea
in which $\alpha_n = 2 -\delta_{n, 0}$ counts the degeneracy of Landau levels.
 \subsection{$\mathcal{V}_\textrm{eff}$ in $eB =0, \Omega \neq 0$ with boundary condition}
 This problem is less known in community compared to the former cases. However, we can derive it by using the master Eq. \eqref{eqsec2f16}. In a non-magnetized  system ($eB=0$) which rotates uniformly along the $\hat{z}$ axis $\vec{\Omega} = \Omega \hat{z}$, the radial size constrains to $0 \leq r \leq R = \frac{1}{\Omega}$ because of the causality requirement. Eigenstates and eigenmomenta of particles in this system are given  below
\begin{align}\label{eqsec2f23}
&\psi_\textrm{M}^{(f)}(\Bar{p}) = e^{i \Bar{p} \cdot \Bar{x}} \bigg( \mathcal{P}_+ f_\ell^{+}(x) + \mathcal{P}_- f_\ell^{-}(x)\bigg), \,\,\, \mathcal{P}_\pm = \frac{1 \pm i  \gamma_1 \gamma_2}{2} ,\no\\
& f_{\ell_\pm}(x) = \sqrt{2} \Omega e^{i \ell_\pm \phi} \frac{J_{\ell_\pm}(q_{\ell_\pm, j} r \Omega)}{J_{\ell_\mp}(q_{\ell,j})},  \,\,\, \ell_{\pm} = j \mp \frac{1}{2},\no\\
& \Bar{p}_a^{(f)} = \left( p_0, 0, 0, p_3\right), \quad \Tilde{p}_a^{(f)} = \left( \tilde{p}_0, 0,   q_{\ell, j} \Omega, p_3\right), \no\\
&L_M^{(f)}= L_\ell^{\Omega}=  \Omega \, j,
\end{align}
where $"\ell"$ is the eigenvalue of orbital angular momentum $L_z = -i \partial_\phi$ and $j=\ell+\frac{1}{2}$ is the eigenvalue of total angular momentum. $J_{\ell}(x)$ is the first kind of Bessel functions with order $\ell$ and $q_{\ell, j}$ is the j'th zero of $\ell$'th Bessel function. We use the  global (spectral) boundary condition on which the total flux of fermion's current on the cylinder surface $R= \frac{1}{\Omega}$ becomes zero. Final result of the effective action in this medium is derived after plugging the eigenstates and eigenmomenta of the Eq. \eqref{eqsec2f23} into the Eq. \eqref{eqsec2f16}
\begin{align}
&\mathcal{V}^\Omega_\textrm{eff} =  - \frac{\sigma^2}{2G} +  \frac{2 N_c N_f \Omega^2}{\pi}   \sum\limits_{\ell=-\infty}^\infty  \sum\limits_{j=1}^\infty \int \frac{dp_3}{2\pi}  \mathcal{J}_\ell(r)\mathcal{V}_T^\Omega,\no\\
&\mathcal{J}_\ell(r) \equiv \frac{J_{\ell}^2( q_{\ell, j} r \Omega) + J_{\ell+1}^2( q_{\ell, j} r \Omega)}{J_{\ell+1}^2(q_{\ell, j})}, \no\\
&\mathcal{V}_T^\Omega = \mathcal{E}^\Omega + T \log \left( 1 + e^{-\beta ( \mathcal{E}^\Omega + \mu_B + L_\ell^{\Omega})}\right) \no\\
&\hspace{1.43cm}+ T \log \left( 1 + e^{-\beta ( \mathcal{E}^\Omega - \mu_B - L_\ell^{\Omega})}\right),\no\\
& \mathcal{E}^{\Omega} = \sqrt{\sigma^2 + q_{\ell, j}^2 \Omega^2 + p_3^2}.
\end{align}
 \subsection{$\mathcal{V}_\textrm{eff}$ in $eB  \neq 0, \Omega \neq 0$ with  boundary condition}
 To the best of our knowledge, this problem has not been solved so far and we intend to obtain the effective action of $\sigma$ field in a magnetized and rotating medium from the master formula, the Eq. \eqref{eqsec2f16}. Indeed,  the main purpose of this section is to derive this relation. We  consider a bounded system  which  $\vec{\Omega} = \Omega \hat{z}$ and  $\vec{B} = B \hat{z}$. Eigenstates and eigenmomenta for particles with charge $Q_f$ are given  below \cite{Sadooghi:2021upd}
 \begin{align}
 &\psi_\textrm{M}^{(f)}(\Bar{p}) = e^{i \Bar{p} \cdot \bar{X}} \left( \mathcal{P}_+^{(s_f)} f^+_{\lambda, \ell, s_f}(X) + \mathcal{P}_-^{(s_f)} f^-_{\lambda, \ell, s_f}(X)\right), \no\\
 & f^\pm_{\lambda, \ell, s_f}(X) = \mathcal{C}^\pm_{\lambda, \ell, s_f} e^{i \ell_\pm \phi} \Psi^\pm_{\lambda, \ell, s_f}(X), \no\\
 & \Psi^\pm_{\lambda, \ell, s_f}(X) = \mathfrak{a}_\pm  e^{- \frac{X}{2}} X^{\frac{|\ell_\pm|}{2}} {}_1 \mathcal{F}_1\left(- \mathcal{N}^\pm_{\lambda, \ell, s_f}; |\ell_\pm|+1; X \right),\no\\
 & \mathfrak{a}_\pm \equiv \frac{1}{|\ell_\pm|!} \bigg(\frac{|e Q_f B|}{2\pi} \frac{ (\mathcal{N}^\pm_{\lambda, \ell, s_f} + |\ell_\pm|)!}{ \mathcal{N}^\pm_{\lambda, \ell,  s_f}!}\bigg)^{\frac{1}{2}}, \\
 & \mathcal{C}^\pm_{\lambda, \ell, s_f} =  \bigg(\frac{|e Q_f B|}{2\pi \int_0^{\alpha_Q} dX (\Psi^\pm_{\lambda, \ell, s_f}(X))^2}\bigg)^{\frac{1}{2}},\no\\
 & \bar{p}_a^{(f)} = \left( \bar{p}_0, 0, 0, p_3\right), \quad \tilde{p}_a^{(f)} = \left( \tilde{p}_0, 0, \kappa s_\ell \sqrt{2 \lambda |e Q_f B|}, p_3\right),\no
 \end{align}
 where
 \begin{align}
     &\lambda = \frac{ (\tilde{p}^f_0)^2- p_3^2 - \sigma^2}{2 |e Q_f B|}, \,\, \alpha_Q \equiv \frac{|e Q_f B|R^2}{2},\no\\
     & \mathcal{N}^\pm_{\lambda, \ell, s_f} \equiv \lambda + \frac{s_f \ell_\mp - |\ell_\pm|-1}{2} , \\
    & X \equiv  \frac{|e Q_f B|r^2}{2},\,\, L_M^{(f)} = L_\ell^\Omega = \Omega \, j.\no
 \end{align}
The ${}_1 \mathcal{F}_1\left(a; b; X \right)$ is the first kind of confluent hypergeometric functions and $\ell_\pm = j \mp \frac{1}{2}$. We elaborate on these functions in the next section. Inserting the appropriate quantities into the Eq. \eqref{eqsec2f16} would enable us to obtain the effective action of the $\sigma$ field 
 \begin{align}\label{eqsec2f28}
     &\mathcal{V}^{B, \Omega}_\textrm{eff} = - \frac{\sigma^2}{2 G} + N_c \sum\limits_{f= u, d} \sum\limits_{\ell=-\infty}^\infty \int \frac{dp_3}{2\pi}  \mathcal{G}_{\lambda, \ell, s_f}(X) \mathcal{V}_T^{B, \Omega},\no\\
     & \mathcal{G}_{\lambda, \ell, s_f}(X) \equiv  (\mathcal{C}_{\lambda, \ell, s_f}^{+})^2 \left((\Psi^+_{\lambda, \ell, s_f}(X))^2 + (\Psi^-_{\lambda, \ell, s_f}(X))^2 \right),\no\\
     &\mathcal{V}_T^{B, \Omega} \equiv \mathcal{E}^{B, \Omega} + T \log \left( 1 + e^{-\beta ( \mathcal{E}^{B, \Omega} + \mu_B + L_\ell^{\Omega})}\right) \no\\
     &\hspace{2cm}+ T \log \left( 1 + e^{-\beta ( \mathcal{E}^{B, \Omega} - \mu_B - L_\ell^{\Omega})}\right),\no\\
     &\mathcal{E}^{B, \Omega} \equiv \sqrt{2 \lambda |e Q_f B| + p_3^2 +\sigma^2}. 
 \end{align}
 Throughout this paper, we take this effective action as our starting point for the numerical computations and derive the chiral phase diagram of $N_f=2$ and $N_c=3$ QM.
 \section{Analytical method}
We put the QM system  in a cylinder with radius \textrm{R} that rotates uniformly along the $\hat{z}$ axis, $\boldsymbol{\Omega} = \Omega \hat{z}$. Hence, the line element associated with the particle trajectories in 3+1 dimensions is 
 \begin{align}\label{eqsec3f1}
     ds^2 &= (1-r^2 \Omega^2) dt^2 - dx^2 -dy^2 - dz^2 \no\\
     &\quad+ 2 \Omega dt(y dx - x dy).
 \end{align}
 The coordinate system is $x^\mu = \left( t, x, y, z\right) = \left( t, r \cos \phi, r \sin \phi, z\right)$ and  $\sqrt{-g}=1$. The  metric of Eq. \eqref{eqsec3f2} is not curved because all the Riemann tensor components are zero, $R^{a}_{bcd}=0$, but it is curvilinear which means $\Gamma^{a}_{bc}\neq 0$.  To connect the global curved space to the local flat space, we use the vierbeins. Due to the local Lorentz symmetry, there are more choices to the $e^\mu_a$. We choose the following bases\cite{Chernodub:2016kxh, Sadooghi:2021upd}
 \begin{align}\label{eqsec3f2}
    e^t_0=e^x_1=e^y_2=e^z_3=1, \qquad e^x_0 = y \Omega, \,\,\, e^y_0= - x \Omega. 
 \end{align}
 Likewise, the orthogonality condition $e^\mu_a e^b_\mu = \delta^b_a$ rules
 \begin{align}\label{eqsec3f3}
     e^0_t = e^1_x = e^2_y = e^3_z =1, \qquad e^1_t = - y \Omega, \,\,\, e^2_t= x \Omega.
 \end{align}
 Our model Lagrangian is given by Eq. \eqref{eqsec2f1} with turning off the Pauli term. It results in the following Dirac equation for massive charged fermion with flavor $f$
 \begin{align}\label{eqsec3f4}
    \left( i \gamma^\mu \mathcal{D}_\mu^{(f)} - \sigma\right) \psi_f(x)=0. 
 \end{align}
 The corresponding $\mathcal{D}_\mu^{(f)}$ kernel contains free partial term as well as the interacting  $D_\mu^{(f)}$  contribution which comes from the magnetic and rotational fields
 \begin{align}\label{eqsec3f5}
     D_\mu^{(f)} = - i e Q_f A_\mu - \frac{i}{4} \omega_{\mu a b} \sigma^{a b},
 \end{align}
 where $\sigma^{a b} \equiv \frac{i}{2} \left[\gamma^a, \gamma^b \right]$ is the spin matrix, $\omega_{\mu a b} \equiv g_{\alpha \beta} e^\alpha_a \left( \partial_\mu e^\beta_b + \Gamma^\beta_{\mu \nu} e^\nu_b\right)$ is the spin connection and the Christoffel symbol is $\Gamma^{\beta}_{\mu \nu} = \frac{1}{2}g^{\beta \alpha} \left( \partial_\mu g_{\alpha \nu} + \partial_\nu g_{\alpha \mu} - \partial_\alpha g_{\mu \nu}\right)$.  We set the local gauge field to be $A_a = (0, \boldsymbol{A}) = (0, -\frac{B y}{2}, \frac{B x}{2}, 0)$ and therefore the magnetic field is $\boldsymbol{B}_a = B \hat{z}$. The vierbein choices in Eqs. \eqref{eqsec3f2} and \eqref{eqsec3f3} leads to the following components for $D_\mu^{(f)}$
 \begin{align}\label{eqsec3f6}
     &D_t^{(f)} = -i e Q_f \frac{B \Omega r^2}{2} - \frac{i}{2} \Omega \sigma^{12}, \quad D_x^{(f)} = i e Q_f \frac{B y}{2}, \no\\
     & D_y^{(f)} = - i e Q_f \frac{B x}{2}, \quad D_z^{(f)}=0.
 \end{align}
 Also, the Dirac matrices in the global curved space are
 \begin{align}\label{eqsec3f7}
     \gamma^t = \gamma^0, \quad \gamma^x = \gamma^1 + y \Omega \gamma^0, \quad \gamma^y = \gamma^2 - x \Omega \gamma^0, \quad \gamma^z = \gamma^3.
 \end{align}
 Therefore, the Dirac equation \eqref{eqsec3f4} reads to
 \begin{align}\label{eqsec3f8}
     &\left(\gamma \cdot \mathcal{D}^{(f)} - \sigma \right)\psi_f =0, \\
     & \gamma \cdot \mathcal{D}^{(f)} \equiv i \gamma^0 \left( \partial_t - i \Omega \hat{J}_z\right) + i \gamma^1 \left( \partial_x + \frac{i}{2} e Q_f B y\right) \no\\
     &\hspace{1.35cm}+ i \gamma^2 \left( \partial_y - \frac{i}{2} e Q_f B x\right) + i \gamma^3 \partial_z,\nonumber
 \end{align}
 which $\hat{J}_z = \hat{L}_z + \frac{\Sigma_z}{2}$ is total angular momentum operator with $\hat{L}_z = - i (x \partial_y - y \partial_x) = - i \partial_\phi$ is the azimuthal angular momentum  and $\Sigma_z = i \gamma^1 \gamma^2$ is the spin matrix along the $\hat{z}$ direction. 
 
 According to the Ritus ansatz, we choose  $\psi^{(f)}_\kappa = \mathbb{E}^{(f)}_{\lambda, \ell, \kappa} u_\kappa(\tilde{p}_{\lambda, \ell, \kappa})$ in such a way that $\gamma \cdot \mathcal{D}^{(f)} \mathbb{E}^{(f)}_{\lambda, \ell, \kappa} = \kappa \mathbb{E}^{(f)}_{\lambda, \ell, \kappa} \gamma \cdot \tilde{p}_{\lambda, \ell, \kappa}$. The main job of Ritus method is to find the eigenbasis $\mathbb{E}^{(f)}_{\lambda, \ell, \kappa}$ and the eigenmomenta $\tilde{p}^\mu_{\lambda, \ell, \kappa}$.  Due to the Hamiltonian structure of the Eq. \eqref{eqsec3f8}, the eigenbasis $\mathbb{E}^{(f)}_{\lambda, \ell, \kappa}$ has well-defined quanta under the $\left( \partial_t, \hat{J}_z, \partial_z\right)$  operators
 \begin{align}\label{eqsec3f9}
     &i \partial_t \mathbb{E}^{(f)}_{\lambda, \ell, \kappa} = \kappa \,E^{(f)}_{\lambda, \kappa, \ell}\mathbb{E}^{(f)}_{\lambda, \ell, \kappa}, \quad  i \partial_z \mathbb{E}^{(f)}_{\lambda, \ell, \kappa} = \kappa \, p_z \mathbb{E}^{(f)}_{\lambda, \ell, \kappa},\no\\
     &\hat{J}_z \mathbb{E}^{(f)}_{\lambda, \ell, \kappa} = j \, \mathbb{E}^{(f)}_{\lambda, \ell, \kappa},
 \end{align}
 with $j \equiv \ell+\frac{1}{2}$. Regarding the latter equation, the eigenbasis can be written  as 
 \begin{align}\label{eqsec3f10}
    \mathbb{E}^{(f)}_{\lambda, \ell, \kappa} = e^{- i \kappa (E^{(f)}_{\lambda, \kappa, \ell} t - p_z z)} \sum_{p=\pm 1} \mathcal{P}_p^{(f)} f^{p \, s_f}_{\lambda, \ell, s_f}(r, \phi),
 \end{align}
 which $\mathcal{P}_\pm^{(f)} = \frac{1 \pm i s_f \gamma^1 \gamma^2}{2}$ is the spin projector operator,  $s_f \equiv \mbox{sgn}(e Q_f B)$ and $\lambda$ is the  generalized Landau levels. They are integers for systems without boundaries, while in systems with boundaries they are rational numbers. If we write $f^\pm_{\lambda, \ell, \kappa}(r, \phi) = e^{i\ell_\pm \phi} F^\pm_{\lambda, \ell, \kappa}(r)$, then the governing equation for $F^\pm_{\lambda, \ell, \kappa}(r)$ will be \footnote{$\pm$ signs over the $F^\pm_{\lambda, \ell}(r)$ show the spin direction, same as in $\ell_\pm$.}
 \begin{align}\label{eqsec3f11}
     \left( X \partial_X^2 + \partial_X + \lambda - \frac{\ell_{\pm}^2}{4X} + \frac{s_f \ell_\mp}{2} - \frac{X}{4}\right) F^{(f) \,\pm}_{\lambda, \ell}(X)=0,
 \end{align}
 where $X \equiv x |Q_f| = \frac{|eQ_f B|r^2}{2}$ and
 \begin{align}\label{eqsec3f12}
  \lambda \equiv \frac{( E^{(f)}_{\lambda, \ell, \kappa}+ \kappa \Omega j)^2-p_z^2 - \sigma^2}{2 |e Q_f B|}.   
 \end{align}
 The quanta $\lambda$ can be seen as the eigenvalues of $\mathbb{E}^{(f)}_{\lambda, \ell, \kappa}$ in the perpendicular direction to the $eB$ and $\Omega$ axis
 \begin{align}\label{eqsec3f13}
     \gamma_\perp \cdot \mathcal{D}^{(f)}_\perp \mathbb{E}^{(f)}_{\lambda, \ell, \kappa} = \kappa s_\ell \sqrt{2 \lambda |eQ_f B|}\, \gamma^2\, \mathbb{E}^{(f)}_{\lambda, \ell, \kappa},
 \end{align}
 with $s_\ell = \mbox{sgn}(\ell)$ and $ \gamma_\perp \cdot \mathcal{D}^{(f)}_\perp \equiv \gamma^1 \mathcal{D}^{(f)}_x + \gamma^2 \mathcal{D}^{(f)}_y$. Solutions to the Eq. \eqref{eqsec3f11} which are confined within a boundary \textrm{R} can be represented as 
 \begin{align}\label{eqsec3f14}
     F^{(f) \,\pm}_{\lambda, \ell}(X) = \mathcal{C}^{(f) \,\pm}_{\lambda, \ell} \Psi^{(f) \,\pm}_{\lambda, \ell}(X),
 \end{align}
 where the two variables $\Psi^\pm_{\lambda, \ell}(X)$ and $\mathcal{C}^\pm_{\lambda, \ell}$ are
 \begin{align}\label{eqsec3f15}
    & \Psi^{(f)\, \pm}_{\lambda, \ell}(X) \equiv \mathfrak{a}_\pm  e^{-\frac{X}{2}} X^{\frac{|\ell_{\pm}|}{2}} {}_1 F_1(-\mathcal{N}^\pm_{\lambda, \ell, s_f};|\ell_\pm|+1;X),\nonumber\\
    & \mathfrak{a}_\pm \equiv \frac{1}{\ell_\pm!} \left( \frac{|eQ_f B|}{2\pi} \frac{(\mathcal{N}^{\pm}_{\lambda, \ell, s_f}+|\ell_\pm|)!}{\mathcal{N}^{\pm}_{\lambda, \ell, s_f}!}\right)^{\frac{1}{2}}, \no\\
    &\mathcal{C}^{(f) \,\pm}_{\lambda, \ell} = \left( \frac{|eQ_f B|}{2 \pi \int_0^{\alpha_B} dX (\Psi^\pm_{\lambda, \ell}(X))^2}\right)^{\frac{1}{2}},
 \end{align}
 with
 \begin{align}\label{eqsec3f16}
     \mathcal{N}^{\pm}_{\lambda, \ell, s_f} \equiv \lambda + \frac{s_f \ell_\mp - |\ell_\pm| -1}{2}.
 \end{align}
 This number is given for different charged quarks and spin in Tab. \ref{tablambda}. 
\begin{table}[!htb]
\centering
\begin{tabular}{|l|l|l|}
\hline
     $s_f>0$& $\quad \mathcal{N}^{+}_{\lambda, \,\ell\geq 0, \,+} = \lambda$& $\quad \mathcal{N}^{+}_{\lambda, \,\ell\leq -1, \,+} = \lambda + \ell$\\
     \hline
    $s_f>0$ & $\quad \mathcal{N}^{-}_{\lambda, \,\ell\geq 0, \,+} = \lambda-1$& $\quad \mathcal{N}^{-}_{\lambda, \,\ell\leq -1, \,+} = \lambda + \ell$\\
     \hline
     $s_f<0$& $\quad \mathcal{N}^{+}_{\lambda, \,\ell\geq 0, \,-} = \lambda-\ell-1$& $\quad  \mathcal{N}^{+}_{\lambda, \,\ell\leq -1, \,-} = \lambda -1$\\
     \hline
     $s_f<0$ & $\quad \mathcal{N}^{-}_{\lambda, \,\ell\geq 0, \,-} = \lambda-\ell-1$& $\quad \mathcal{N}^{-}_{\lambda, \,\ell\leq -1, \,-} = \lambda$\\
     \hline
\end{tabular}
\caption{The $\mathcal{N}^{\pm}_{\lambda, \ell, s_f}$ numbers for different charges, spins and orbital numbers.}
\label{tablambda}
\end{table}
For positive charges changing $\lambda \to \lambda -1$ and $\ell \to \ell +1$ turn positive spin to negative spin solutions, while for negative charges $\lambda \to \lambda -1$ and $\ell \to \ell -1$ turn negative spin to positive spin solutions. Likewise, the eigenmomenta are determined to be as 
 \begin{align}\label{eqsec3f17}
     \tilde{p}^\mu_{\lambda, \ell, \kappa} = \left( \epsilon^{(f)}_\lambda, 0, \kappa s_\ell \sqrt{2 \lambda |e Q_f B|}, p_z\right),
 \end{align}
 where $\epsilon^{(f)}_\lambda = \kappa \Omega j + E^{(f)}_{\lambda, \ell, \kappa}$. Due to the $\tilde{p}_{\lambda, \ell, \kappa}^2= \sigma^2$, we get $\epsilon^{(f)}_\lambda = \sqrt{p_z^2 + \sigma^2 + 2 \lambda |e Q_f B|}$. 
 
 As long as we don't impose boundary condition, the $\lambda$s are arbitrary numbers. We set a global boundary condition, namely the spectral boundary condition in which the net flux of each flavor going out of the boundary $r =R$ is zero
 \begin{align}\label{eqsec3f18}
     \int_{-\infty}^\infty dz \, \int_0^{2\pi} d\phi \bar{\psi}^{(f)} \gamma^r \psi^{(f)}\bigg|_{r=R} = 0,
 \end{align}
 with $\gamma^r \equiv \gamma^1 \cos\phi + \gamma^2 \sin\phi$. Substituting the Ritus ansatz given in the Eqs. \eqref{eqsec3f10} along with Eq. \eqref{eqsec3f14} and integrating over the angle, one get the following constraints to fulfill the boundary condition of the Eq. \eqref{eqsec3f18}
 \begin{align}\label{eqsec3f19}
 \begin{array}{ll}
    \mbox{for} \quad \ell\geq 0, &\quad {}_1 F_1(-\mathcal{N}^{+}_{\lambda, \ell, s_f};|\ell_+|+1;\alpha \, Q_f)=0,\\
    \mbox{for} \quad \ell\leq -1, &\quad {}_1 F_1(-\mathcal{N}^{-}_{\lambda, \ell, s_f};|\ell_-|+1;\alpha \, Q_f)=0.
    \end{array}
 \end{align}
 If we are given the $\ell$s for each spin direction, then the $\lambda$s will be derived numerically by solving these equations. Needless to say that changing the kind of boundary condition would lead to different values and therefore may alter the entire results. \footnote{We thank M. N. Chernodub for nice discussions on this topic.} We will show that implying the spectral boundary condition and setting the energies to be always positive, turn into a correlation between the strength of magnetic and rotation field.  In next section, we will present numerical results derived from the Dirac equation along with the solutions of the gap equation at zero and finite temperature.     
 \section{Numerical results}
 Having set up the fundamental relations for the effective action from a general view and giving some examples, and reviewing the Dirac equation, we proceed to investigate  chiral symmetry breaking/restoration in a magnetized and rotating QM in presence of non-vanishing $\mu$ and  $T$.  To explore the phase diagram, we need to some prior information about the properties of QM at $T=0$. These pieces of information contain knowledge about the values of coupling constant, roots of the Dirac equation (which are already discussed) as inputs to the gap equation and the relation between the magnetic and rotation fields. To this reason, we break this section into two parts. In the first subsection, we will study the  gap equation  at $T=0$. These data will serve as inputs in the phase transition studies. In the second subsection, we are going to solve numerically the gap equation at $T\neq 0$ and $\mu \neq 0$ and sketch the phase diagram as $T_c(\mu)$ and $T_c(R\Omega)$ plots for different cases. We are able to pinpoint exactly the position of CEP by the new technique which is  solving the gap-equation and its   derivative with respect to the order parameter "$\sigma$", simultaneously.

 \subsection{Results at T=0 and $\mu =$0}
 \begin{figure}
    \centering
        \includegraphics[width=8cm,height=5cm]{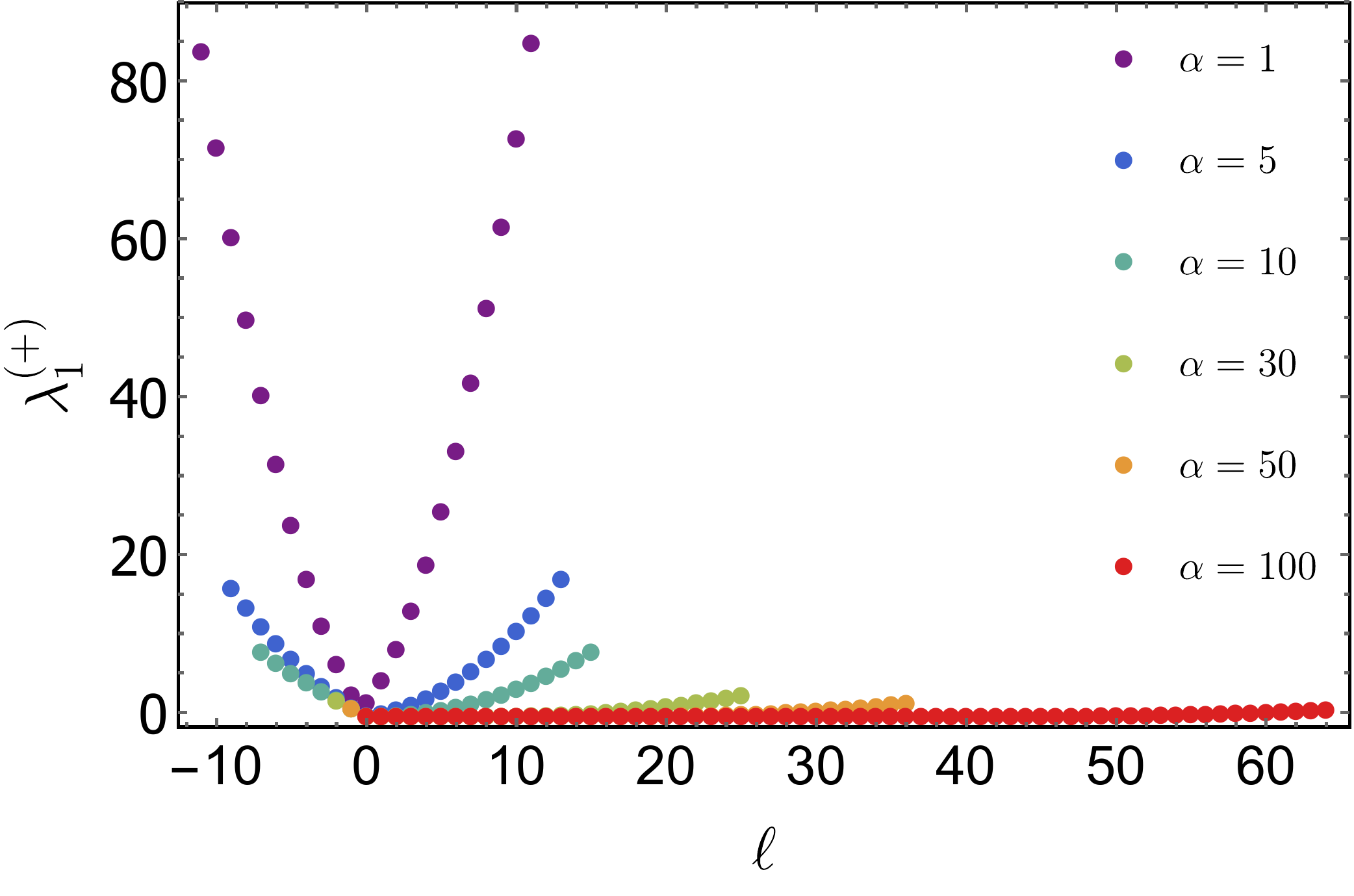}
        \hspace{0.25cm}
        \includegraphics[width=8cm,height=5cm]{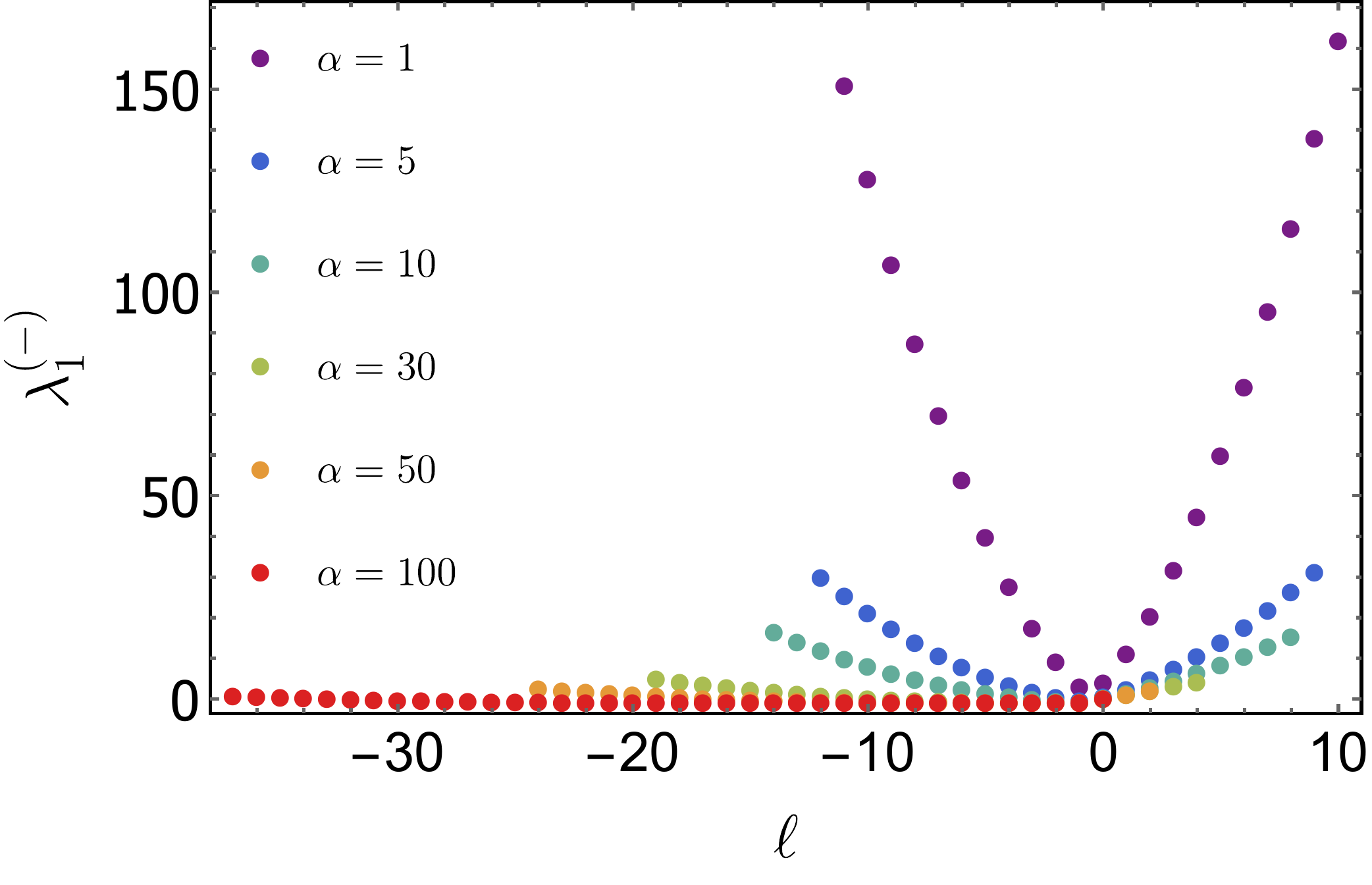}
    \caption{color online. Top panel: first root of up quark for different $\alpha$ values ($1, 5, 10, 30, 50, 100$). Bottom panel: the same one for down quark.}
    \label{fig:lambda-vs-ell}
\end{figure}
The $\lambda$ numbers are the key ingredients. They play the role of initial data to solve  the gap equation and represent the  physical points consistent with the boundary condition.  They are derived from the Eq. \eqref{eqsec3f19} for each flavor separately.  In Fig. \ref{fig:lambda-vs-ell} we show the first root of each $\ell$s for different $\alpha$ values ($1, 5, 10, 30, 50, 100$). The top plot indicates the up quark roots with $Q_f=2/3$ and the bottom one shows the down quark roots with $Q_f=-1/3$. 

\begin{figure}
        \includegraphics[width=8cm,height=5cm]{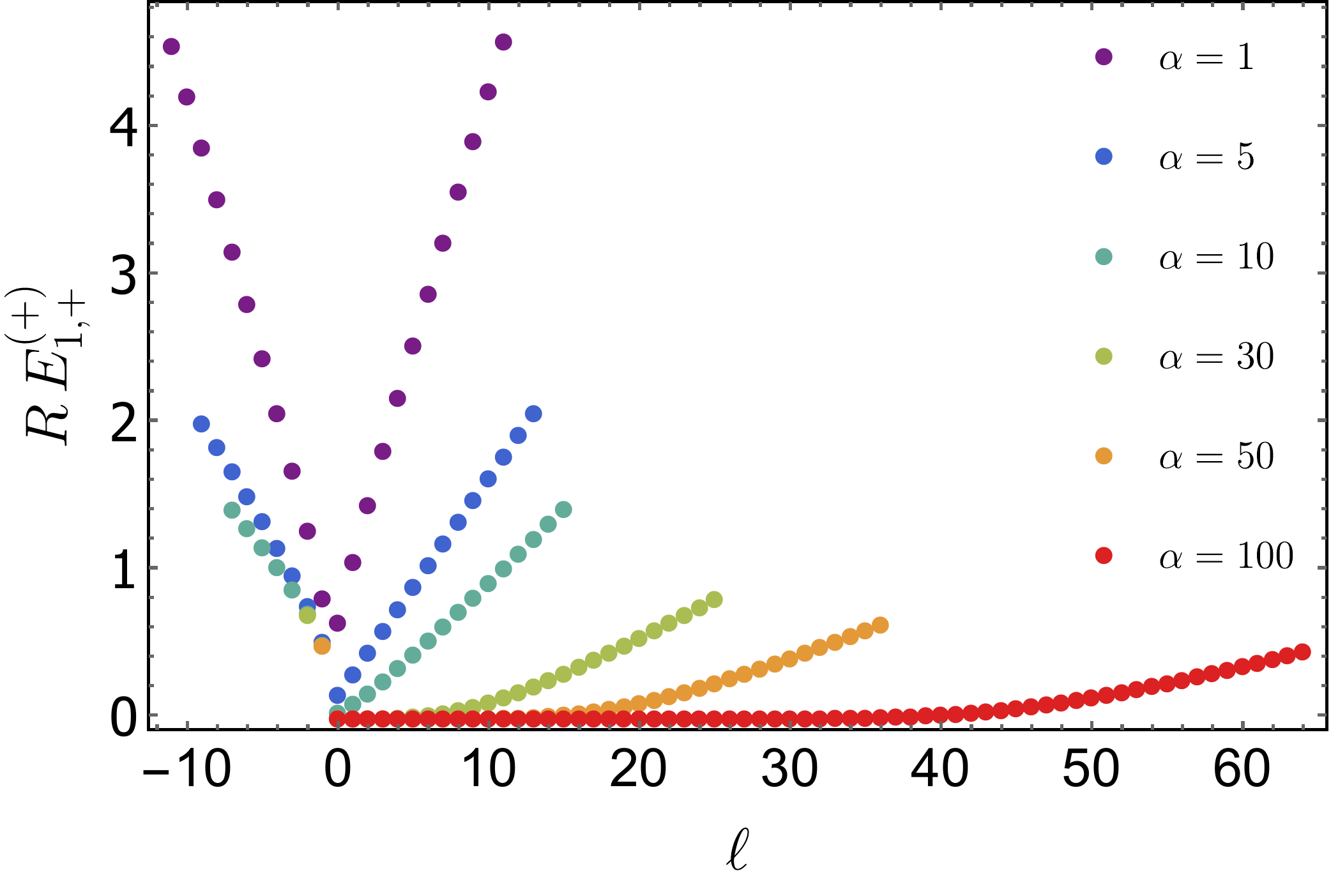}
        \hspace{0.25cm}
        \includegraphics[scale=0.355, valign=t]{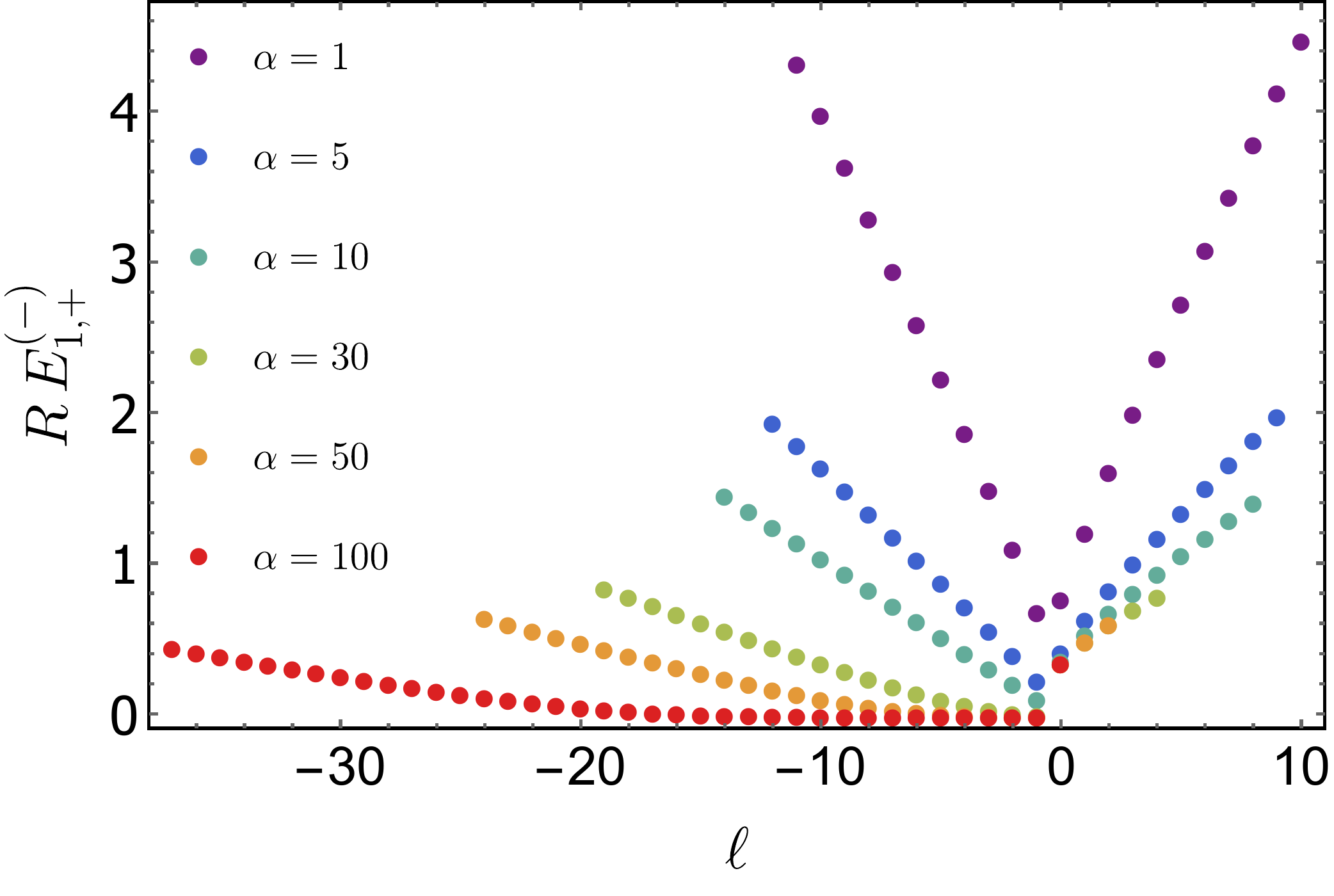}
    \caption{color online. Top panel: energy derived from first root of positive quark for different $\alpha$ at $R\Omega=0$. Bottom panel: the same plot for down quark.}
    \label{fig:E-vs-ell}
\end{figure}

\begin{figure}
        \includegraphics[width=8cm,height=5cm]{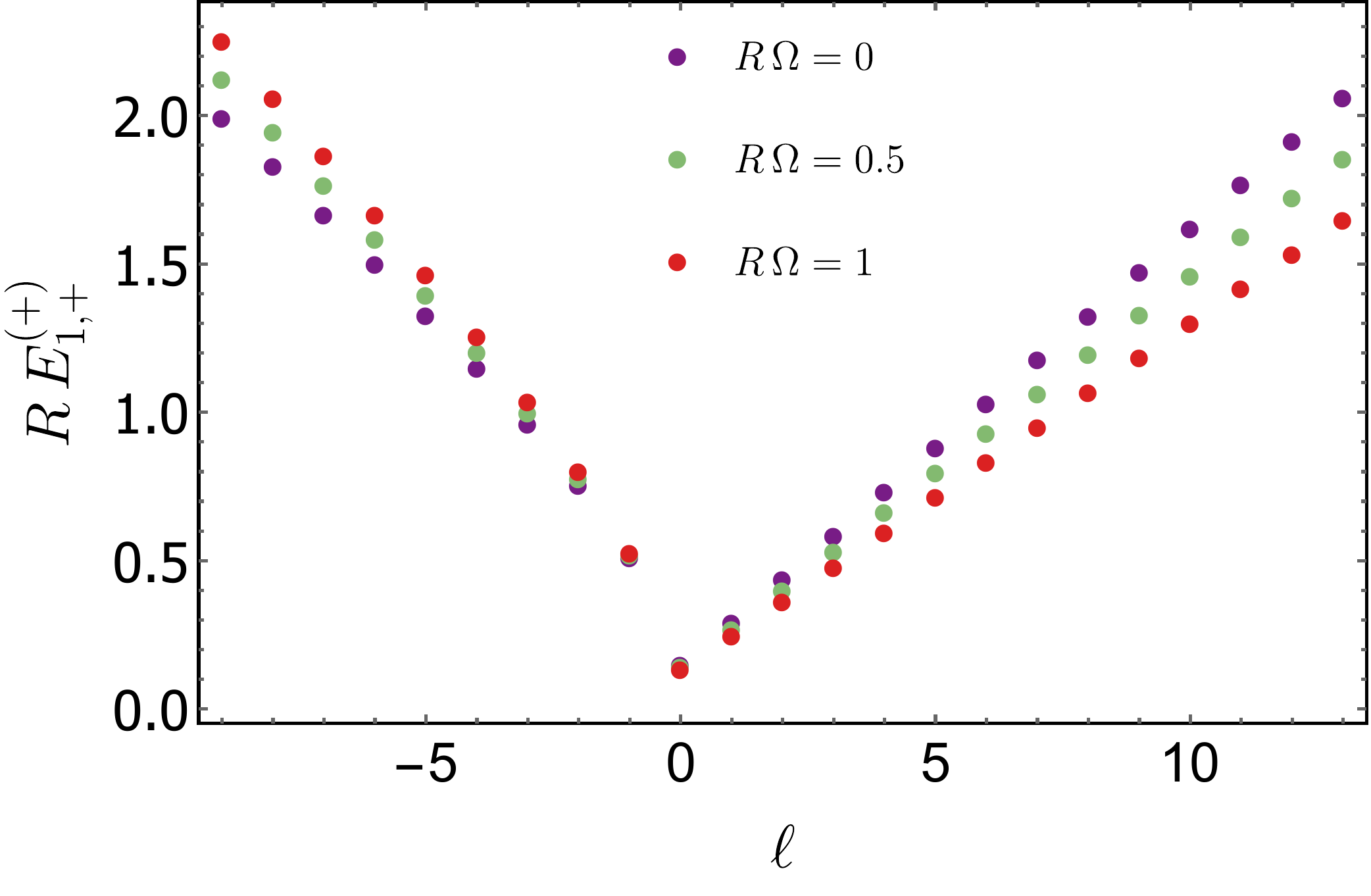}
        \hspace{0.3cm}
        \includegraphics[width=8cm,height=5cm]{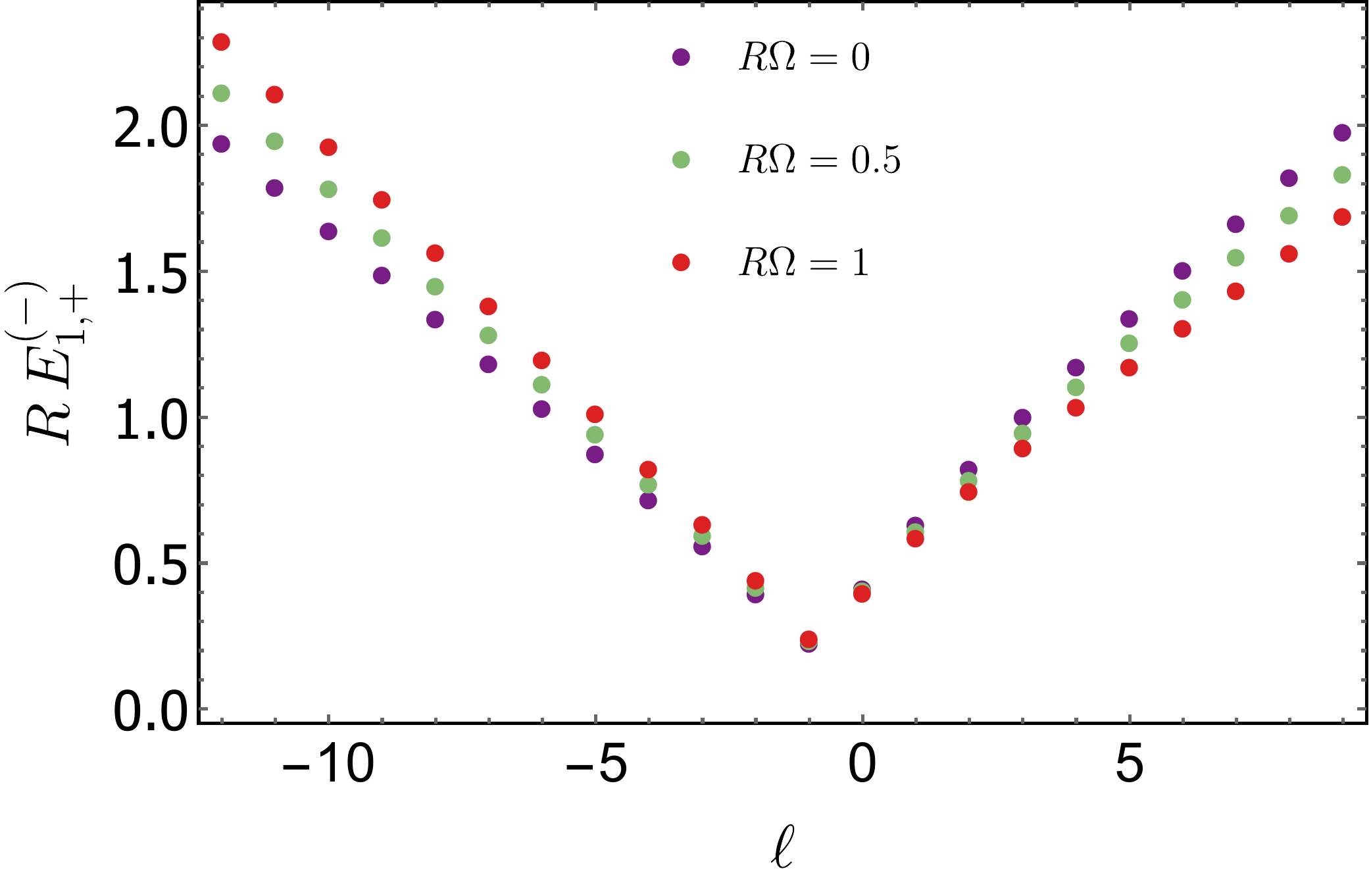}\vspace{.3cm}
        \includegraphics[width=8cm,height=5cm]{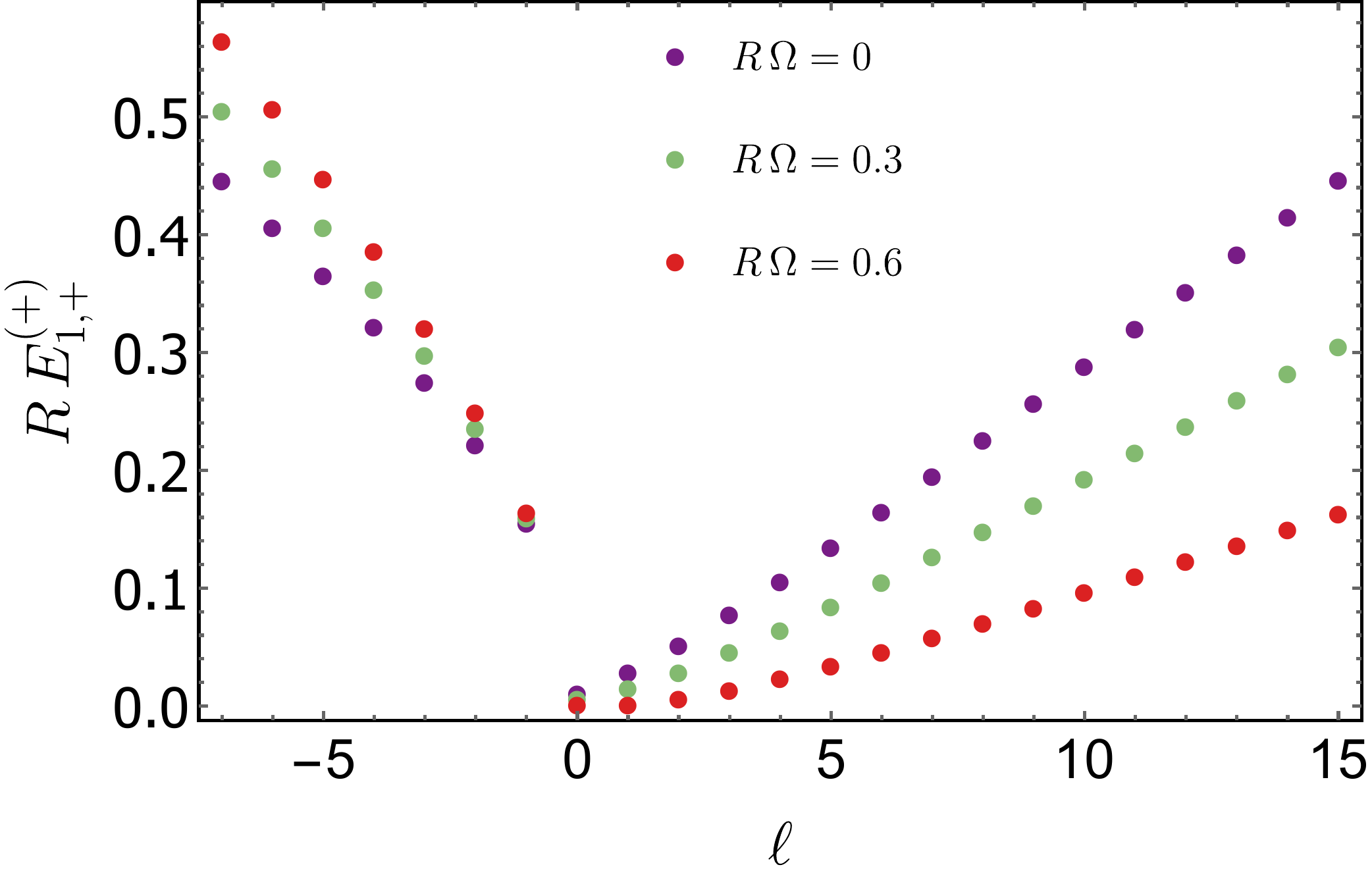}
        \hspace{0.3cm}
        \includegraphics[width=8cm,height=5cm]{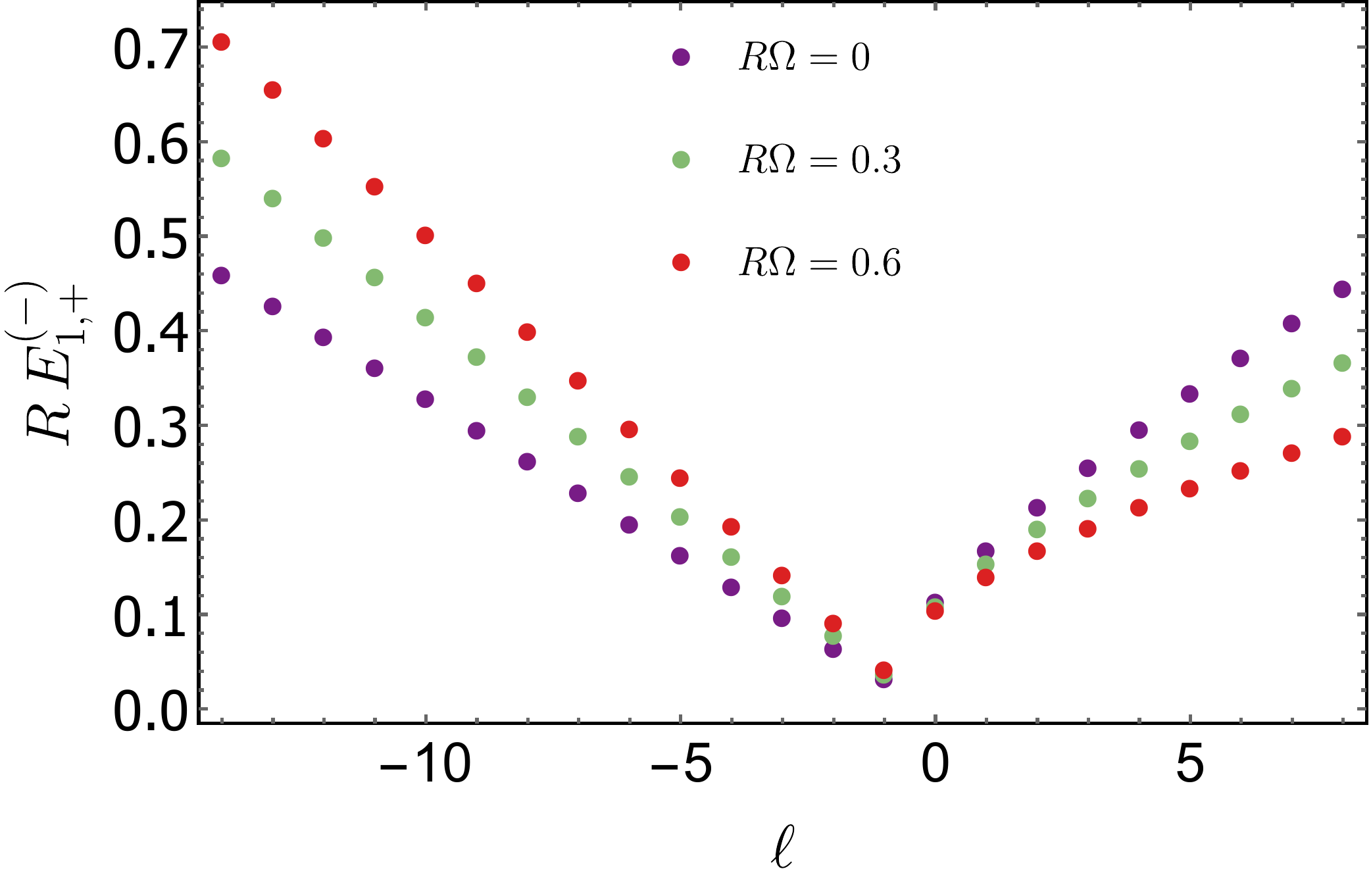}
    \caption{color online. Two top panels: energy levels derived from first root for  $\alpha=5$ with $(R \Omega=0, 0.5, 1)$. Upper part shows the up qaurk and lower part shows the down quark energy levels. Two bottom panel: the same ones for $\alpha=10$ with $(R \Omega=0, 0.3, 0.6)$.}
    \label{fig:E-vs-ell-alpha5}
\end{figure}

There are some points worth to comment on here. We run our numerical codes with 
\begin{align}\label{eqsec4f1}
    \Lambda = 500 \, \textrm{MeV}, \qquad R = 6 \, \textrm{fm},
\end{align}
where $\Lambda$ is the momentum cut-off and determines the validity region of NJL model. This cut-off puts a limit on the  energy of massless particles at rest
\begin{align}\label{eqsec4f2}
    \Lambda = \epsilon^{(f)}_\lambda(p_z=0, \sigma=0) =  \sqrt{\frac{4}{R^2} \lambda_\textrm{max} \, \alpha |Q_f|},
\end{align}
that in turn  gives the maximum value of $\lambda$ for each $\ell$.  Also, this leads  $"\ell"$ to be bounded from below and above $\ell_{\textrm{min}} \leq \ell \leq \ell_\textrm{max}$, and beyond these values there is no  solution to $\lambda$ compatible with the Eq. \eqref{eqsec4f2}. That is why the plots in Fig. \ref{fig:lambda-vs-ell} are shown in between $\ell_{\textrm{min}} \leq \ell \leq \ell_\textrm{max}$ for each $\alpha$.  Values of $(\ell_\textrm{min}, \ell_\textrm{max})$ depend on $\alpha$ and quark's charge in a way that for in small $\alpha$ they are almost symmetric $\ell_\textrm{min} \sim - \ell_\textrm{max}$, but increasing $\alpha$ leads to pushing the $\ell_\textrm{min} \to 0$ and $\ell_\textrm{max} \to \infty$ for positive quarks and $\ell_\textrm{max} \to 0$ and $\ell_\textrm{min} \to -\infty$ for negative quarks. This feature is due to the magnetic-orbit coupling, i.e. $e \vec{L}. \vec{B}$. Also, values of $\lambda$ decrease by increasing $\alpha$. These  are very obvious in the Fig. \ref{fig:lambda-vs-ell}. It is worthwhile to mention that the strength of magnetic field is given by the definition $eB = 2\alpha / R^2$ with $R$ set in the Eq. \eqref{eqsec4f1}
\begin{align}
    |eB| = 2 \frac{\alpha}{R^2} = 0.11 \alpha \,m_\pi^2.
\end{align}
In  Fig. \ref{fig:E-vs-ell}, we show dimensionless energy levels for massless positively and negatively quarks at zero momentum. These energy levels are
\begin{align}\label{eqsec4f3}
    R E^{(f)}_{\lambda, \ell, \kappa}(\sigma \to 0, p_z \to 0) = - \kappa j\, R\Omega + \sqrt{4 \lambda \, \alpha |Q_f|}.
\end{align}
The top panel of Fig. \ref{fig:E-vs-ell}  shows the energies for up quark derived from the first $\lambda$ of the top panel of Fig. \ref{fig:lambda-vs-ell} at $R \Omega =0$ and the bottom one shows the same quantity for down quark at $R \Omega =0$ derived from the first $\lambda$ of the bottom plot of the Fig. \ref{fig:lambda-vs-ell}.
As it is evident from the Fig. \ref{fig:E-vs-ell}, increasing the $\alpha$ leads to decreasing energies. In  Fig. \ref{fig:E-vs-ell-alpha5}, we show the $R\Omega$ dependence of energies for $\alpha =5$ and $\alpha=10$. The two top plots shows energies for $\alpha=5$ which is derived from the first $\lambda$ for $(R\Omega = 0, 0.5, 1)$. The upper(lower) part shows the energy levels for up(down) quarks. The two bottom plots show energy levels for $\alpha=10$ with $(R \Omega=0, 0.3, 0.6)$. We observe that for $\ell \geq 0$ increasing $R\Omega$ results in decreasing  energies, while for $\ell \leq -1$ the opposite case has happened. Also, increasing the $\alpha$ decreases the energies because $\lambda$ behaves in this kind. 
 
\begin{figure}[ht]
    \centering
        \includegraphics[width=8cm,height=5cm]{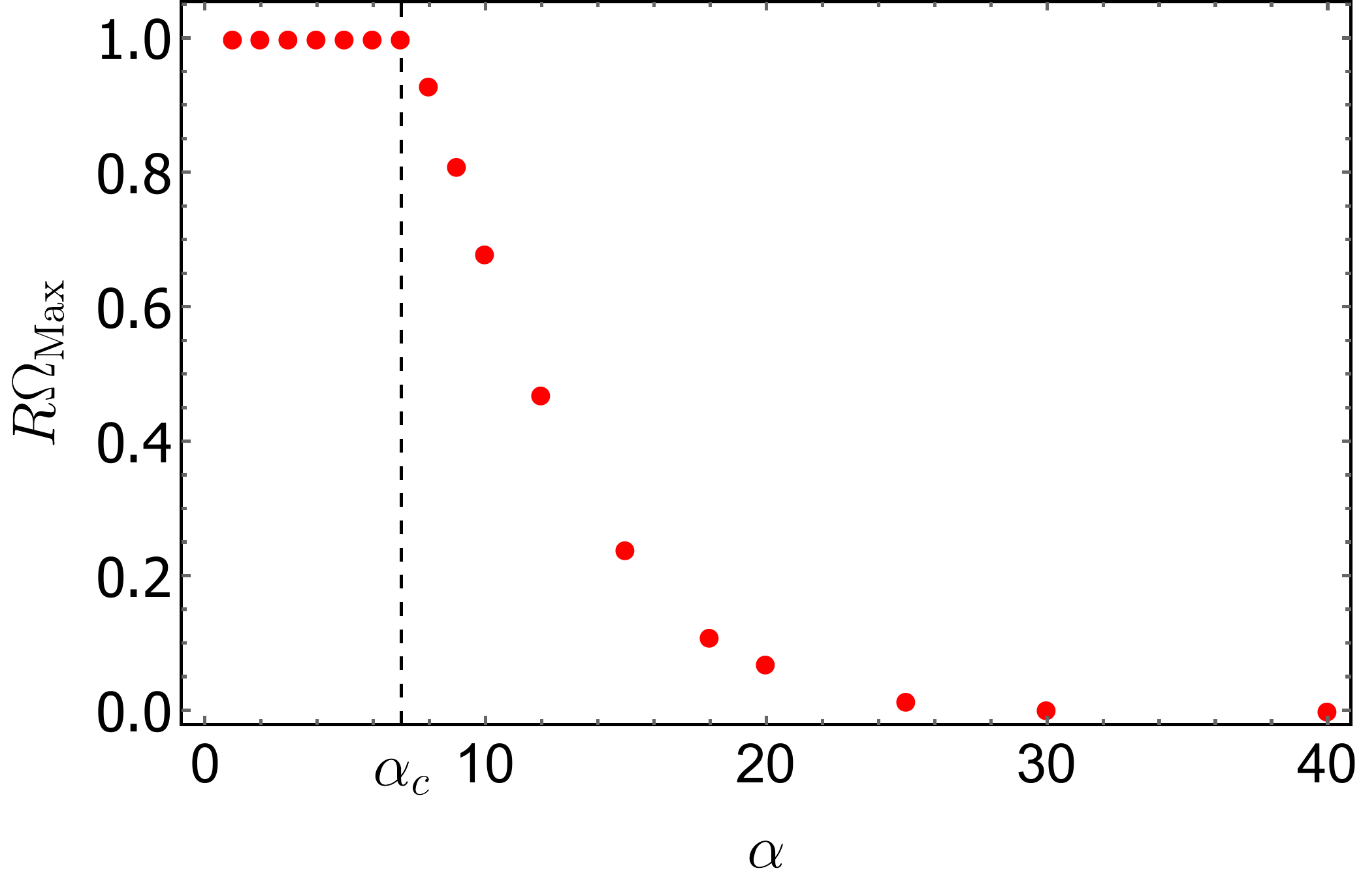}
    \caption{color online.  $R\Omega_\textrm{Max}$ v.s. $\alpha$. This states that strongly magnetized plasma can not rotate.}
    \label{fig:ro-vs-eB}
\end{figure}

 One of the main results of our paper is to find a non-trivial correlation between the magnetic and rotation field. To have a consistent quantization scheme, we take the  energies of the  Eq. \eqref{eqsec4f3}  to be always non-negative for each flavor and $\kappa = \pm 1$.   We find that  by keeping $\alpha$  fixed and varying the $R\Omega$ from 0 to 1, it can be found a reasonable region $0 \leq R \Omega \leq R\Omega_\textrm{Max}$ in which  $E^{(f)}_{\lambda, \ell, \kappa}\geq 0$. In the Fig. \ref{fig:ro-vs-eB} we sketch this region in terms of $\alpha$. At small $\alpha$ all the physical region $R\Omega_\textrm{Max} = 1$ is accessible, but at the point $\alpha_c=7$ this  $R\Omega_\textrm{Max}$ starts to shrink. At very strong magnetic field, this accessible part reaches to zero and therefore strongly magnetized plasma can not rotate anymore. This result is a clear consequence of the spectral boundary condition.

\begin{figure}
    \centering
        \includegraphics[width=8cm,height=5cm]{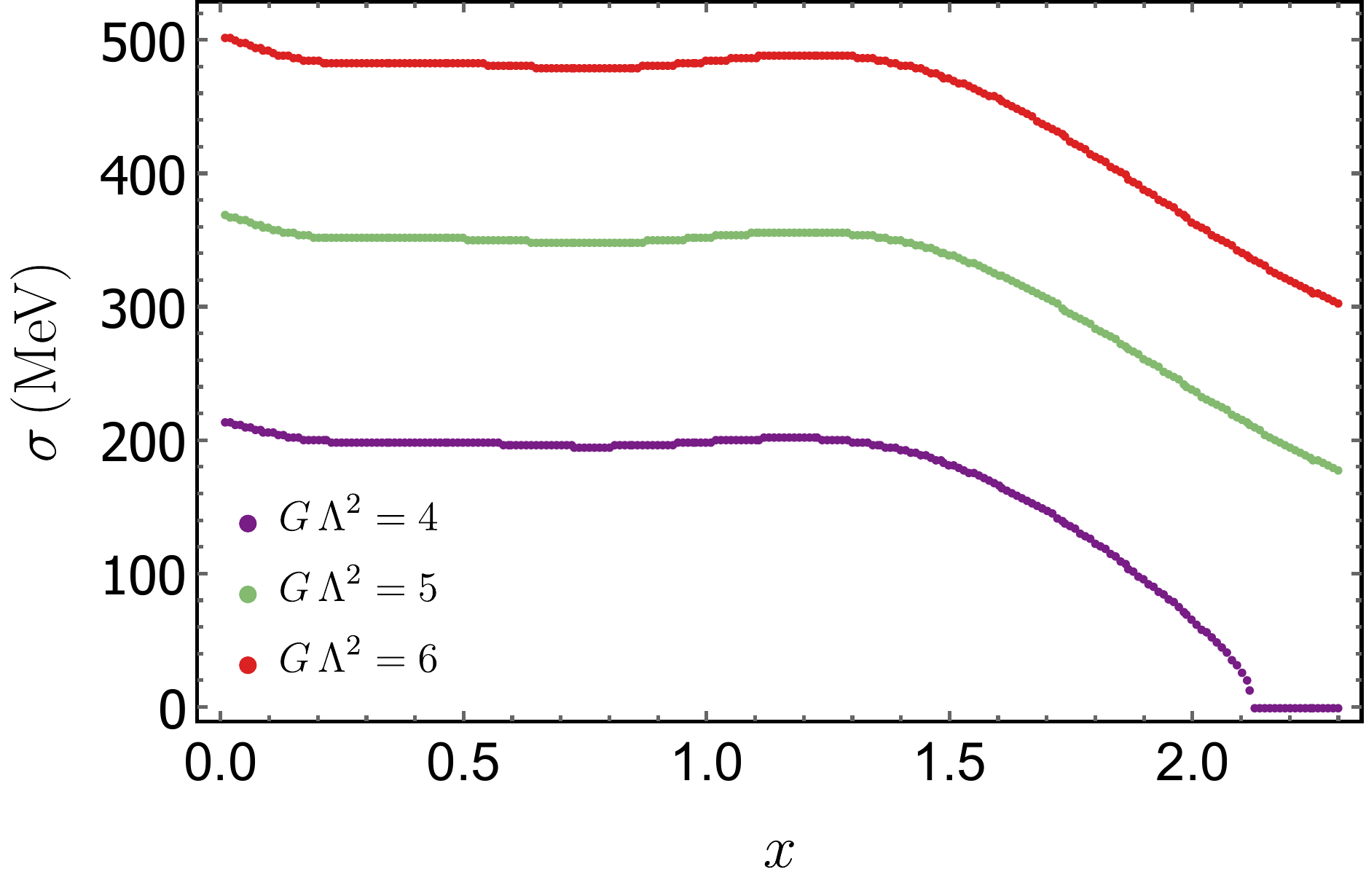}
    \caption{color online. Dynamical mass generated at $\alpha=7$ for $x \in [0, 2.3]$ with different couplings $(G \Lambda^2 = 4, 5, 6)$ from bottom to top.}
    \label{fig:m-vs-x-alpha6}
\end{figure}

\begin{figure}
    \centering
        \includegraphics[width=8cm,height=6cm]{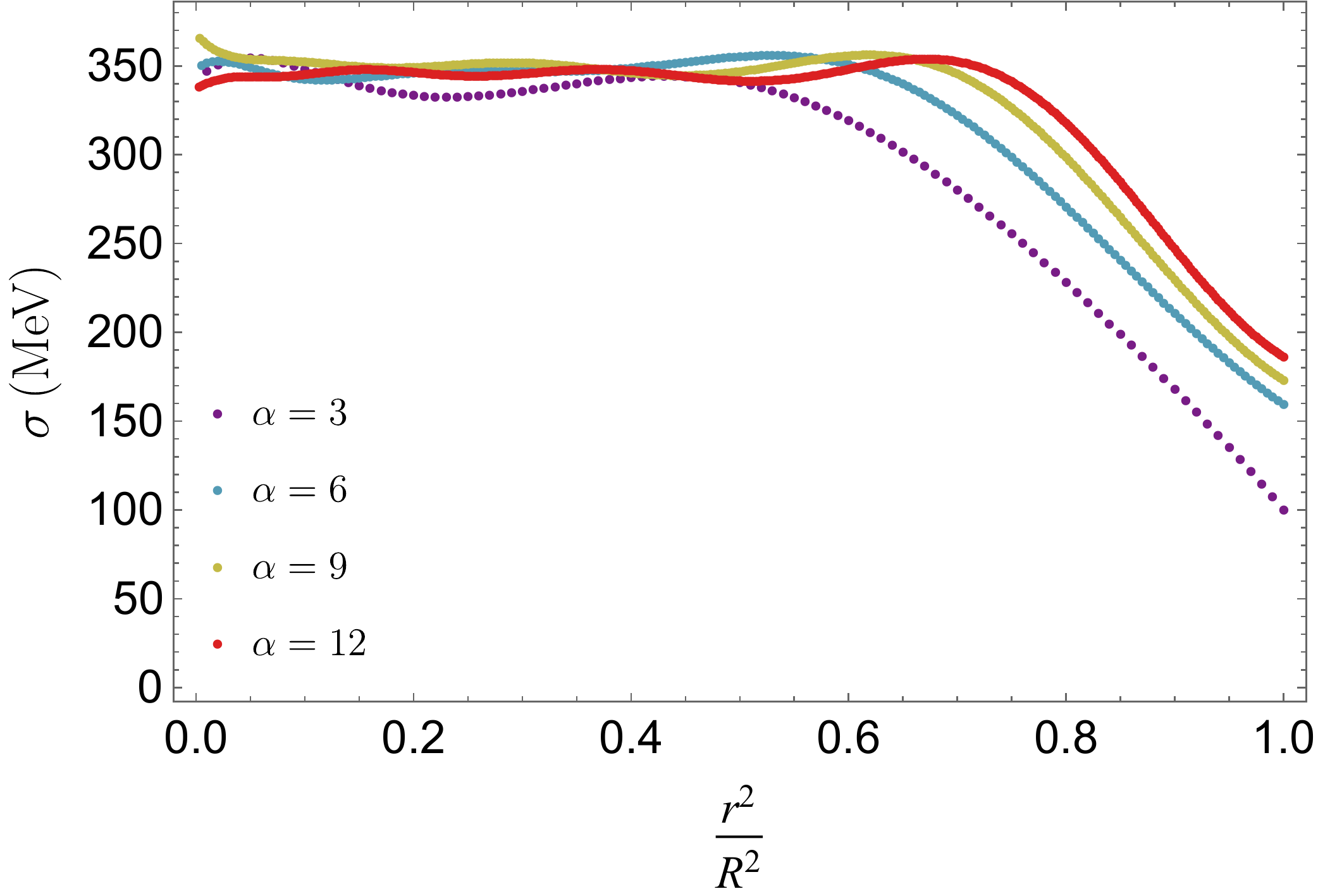}
    \caption{color online. Dynamical mass generated at coupling $G \Lambda^2=5$ for different $\alpha= (3, 6, 9, 12)$ in their corresponding interval $x$.}
    \label{fig:m-vs-x-G5}
\end{figure}

\begin{figure}
    \centering
        \includegraphics[width=8cm,height=5cm]{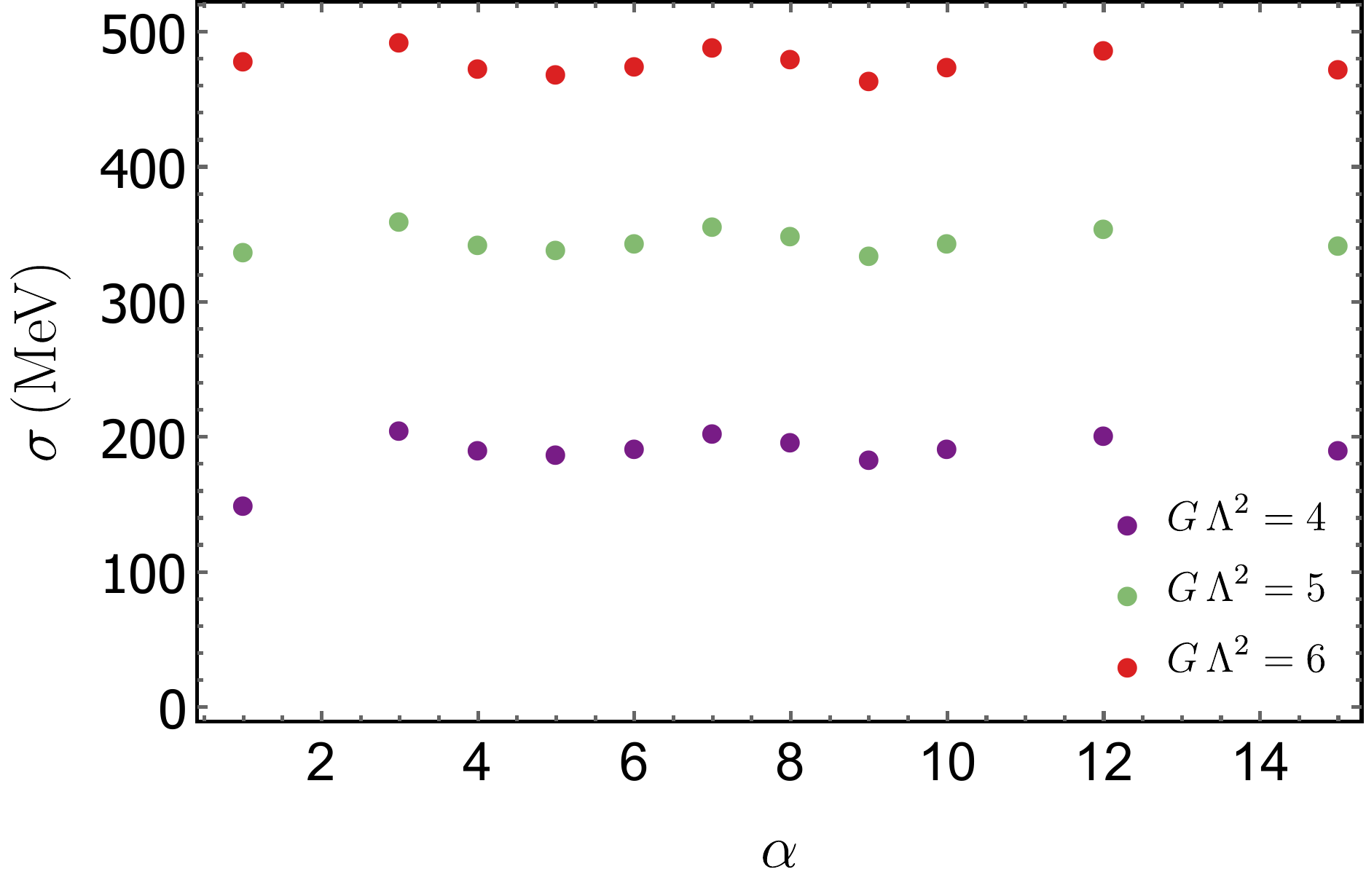}
    \caption{color online. Dynamical mass generated v.s $\alpha$ for couplings $G \Lambda^2= (4, 5, 6)$ from bottom to top, at $x= 0.1$.}
    \label{fig:m-vs-alpha-x01}
\end{figure}

Now, we are going to investigate the gap equation at $T=0$. The effective action in a cold rotating and magnetized plasma is 
\begin{align}\label{eqsec4f4}
    \mathcal{V}_\textrm{eff}^{B, \Omega} &= - \frac{\sigma^2}{2G} \no\\
    &+ N_c \sum_{f= u, d} \sum_{\ell = \ell_\textrm{min}}^{\ell_\textrm{max}} \int \frac{dp_z}{2\pi} \, f(p, \Lambda, \delta \Lambda)\, \mathcal{G}^{(f)}_{\lambda, \ell, s_f}(X) \, \epsilon_\lambda^{(f)},
\end{align}
with $\epsilon_\lambda^{(f)} = \sqrt{p_z^2 + \sigma^2 + \frac{4}{R^2} \lambda \alpha \,|Q_f|}$ and $\mathcal{G}_{\lambda, \ell, s_f}(X)$ is defined in Eq. \eqref{eqsec2f28}. Note that $\mathcal{C}_{\lambda, \ell} = \mathcal{C}^\pm_{\lambda, \ell}$ due to the boundary condition \eqref{eqsec3f19}. To smooth out the $p_z$ integration, we introduce the cut-off function as
\begin{align}\label{eqsec4f5}
    f(p, \Lambda, \delta \Lambda) = \frac{\sinh (\frac{\Lambda}{\delta \Lambda})}{ \cosh (\frac{p}{\delta \Lambda}) + \cosh(\frac{\Lambda}{\delta \Lambda})},
\end{align}
and $p \equiv \sqrt{p_z^2 + \frac{4}{R^2} \lambda \alpha \,|Q_f|}$. For the limit $\frac{\delta \Lambda}{\Lambda} \to 0$, the cut-off function tends to the step function
\begin{align}\label{eqsec4f6}
    \lim_{\frac{\delta \Lambda}{\Lambda} \to 0}  f(p, \Lambda, \delta \Lambda) = \Theta(\Lambda^2 - p^2).
\end{align}
Inserting the step function into the gap equation \eqref{eqsec4f4}, makes easier the $p_z$ integration and the gap equation $\partial \mathcal{V}_\textrm{eff}^{B, \Omega} / \partial \sigma = 0$ results in
\begin{widetext}
 \begin{align}\label{eqsec4f8}
   \sigma \bigg(1 - \frac{N_c G}{\pi} \sum_{f= u, d} \sum_{\ell = \ell_\textrm{min}}^{\ell_\textrm{max}} \mathcal{G}^{(f)}_{\lambda, \ell, s_f}(X)\, \tanh^{-1} \bigg(\sqrt{\frac{\Lambda^2 - \frac{4}{R^2} \lambda \alpha \, |Q_f|}{\Lambda^2 + \sigma^2}}\bigg) \Theta(\Lambda^2 - \frac{4}{R^2} \lambda \alpha \, |Q_f|)\bigg)=0.
\end{align}
\end{widetext}
The trivial solution to this equation is $\sigma=0$, but it does not lead to symmetry breaking. We find non-trivial solution and solve the latter equation numerically.
In the Fig. \ref{fig:m-vs-x-alpha6}, the non-trivial solution to the gap equation \eqref{eqsec4f8} is shown at $\alpha=7$ for different couplings. As we would expect, increasing the coupling causes to increasing the dynamical mass. 
In the  Fig. \ref{fig:m-vs-x-G5} it is seen the $\alpha$ dependence of dynamical mass  at fixed coupling, $G\Lambda^2=5$. Near the axis $x=0$, dynamical masses have no distinct $\alpha$ dependence and rotational magnetic inhibition is excluded \cite{Chen:2015hfc}. This is because the vorticity vanishes at $T=0$ gap equations \cite{Sadooghi:2021upd, Chernodub:2016kxh}. The condensate  falls off near the boundary region, since the condition $\partial_r \sigma \ll \sigma^2$ violates there. In this region chiral condensate grows with $\alpha$ due to the mode accumulation in which the magnetic field is enhanced for larger angular momenta \cite{Chen:2017xrj}.  \footnote{We appreciate Xu-Guang Huang because of helpful discussions.}
 In Fig. \ref{fig:m-vs-alpha-x01}, we show another realization of $\alpha$ dependence. The gap equation is solved at $x=0.1$ for different couplings and $\alpha$s. The profile $\sigma(\alpha)$ looks oscillatory at these small magnetic fields around the axis and no definitive conclusion can be made about the magneto or inverse-magneto catalysis.

\begin{figure}
    \centering
        \includegraphics[width=8cm,height=5cm]{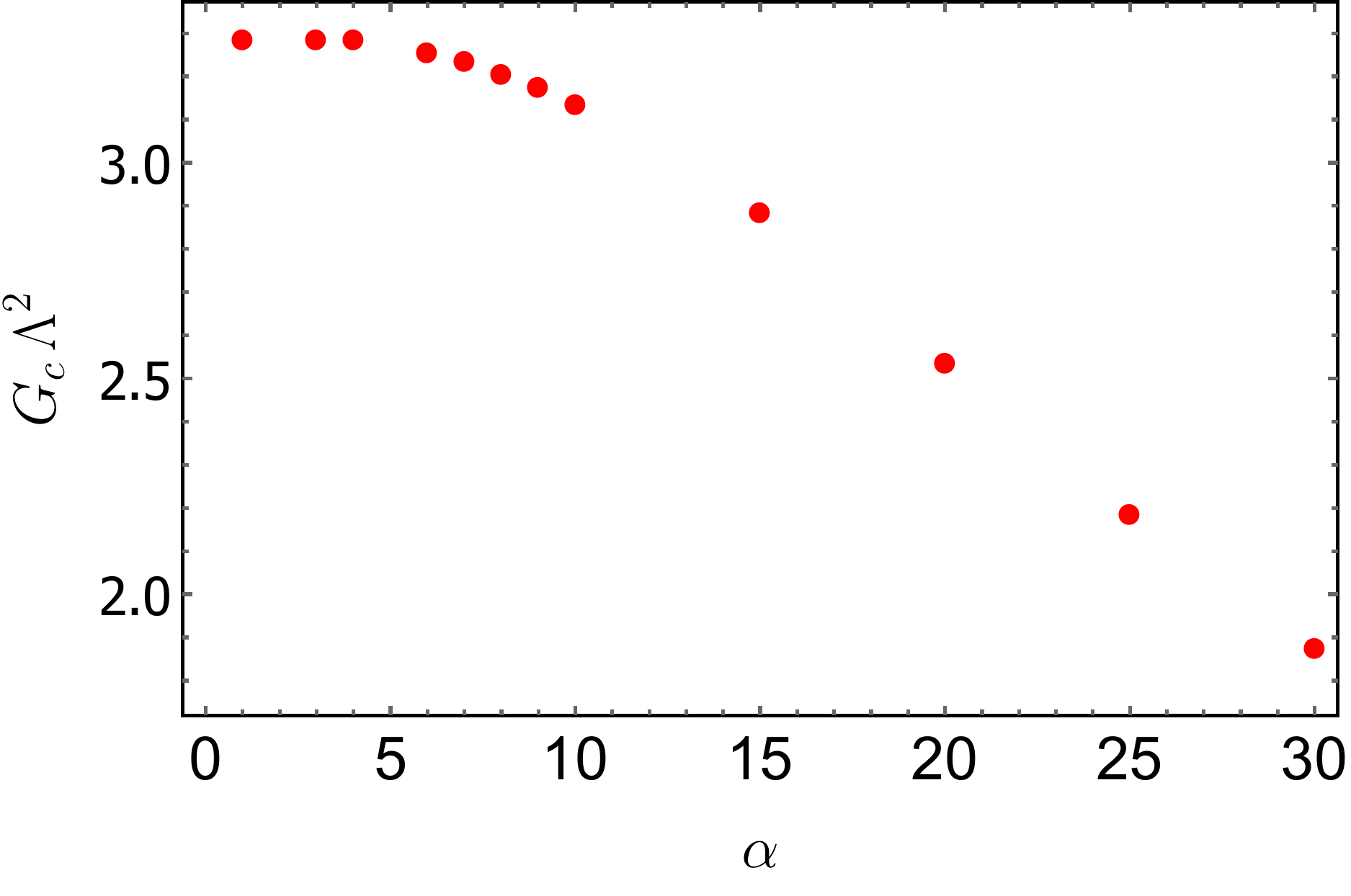}
    \caption{color online. Critical coupling v.s. $\alpha$. It is the minimum coupling  needed to develop a non-trivial solution at the very first $x$.}
    \label{fig:Gc-vs-alpha}
\end{figure}

Critical coupling is very important quantity in the gap equation. By definition, it is the minimum coupling needed to provide a non-trivial solution $\sigma \neq 0$, for the gap equation. In the Fig. \ref{fig:Gc-vs-alpha}, this parameter is plotted against $\alpha$. We derived it by searching the entire region $0 \leq x \leq \alpha$ for each $\alpha$ and we call the $G_c\Lambda^2$ as the minimum value of coupling which gives a solution $\sigma \neq 0$ at the very first $x$. Near the $\alpha \to 0$, it closes to $G_c\Lambda^2 \sim \frac{2\pi^2}{N_c N_f}$, but increasing the magnetic field turns this coupling to smaller values and at strong magnetized plasma $G_c \Lambda^2 \to 0$ which is a direct sign of the magneto-catalysis.    

\subsection{Results at T$\neq$ 0 and $\mu \neq$ 0}
In curved spaces, temperature has to be defined locally and thermal states are well-defined within the box that metric is flat. So, the notion of global temperature is ambiguous.  Red shift equation relates the local (Minkowski) temperature $T_0$ to a global temperature $T(r)$ as follows \cite{Santiago:2018kds, Tolman:1930ona}
\begin{align}\label{eqsec4f9}
    T(r) = \frac{T_0}{\sqrt{g_{00}(r)}},
\end{align}
where $T_0$ is the temperature seen by a local rest frame observer when $\sqrt{g_{00}(r)} \to 1$,  while $T(r)$ is the temperature seen by an accelerated observer. We infer from Eq. \eqref{eqsec4f9} that at very high gravity points, i.e. $g_{00}(r) \to 0$ accelerated observer sees a high temperature medium $T(r) \to \infty$, while in low gravity points when $g_{00}(r) \to \infty$ the medium seems to be  cold, $T(r) \to 0$. For rotating plasma according to the Eq. \eqref{eqsec3f1}, $T(r) = T_0/\sqrt{1 - r^2 \Omega^2}$ and therefore, if $r \Omega \to 0$ then $T(r) \to T_0$. To this reason, we solve the gap equation at points close to the axis which is $x=0.1$. 

\begin{figure}
    \centering
        \includegraphics[width=8cm,height=5cm]{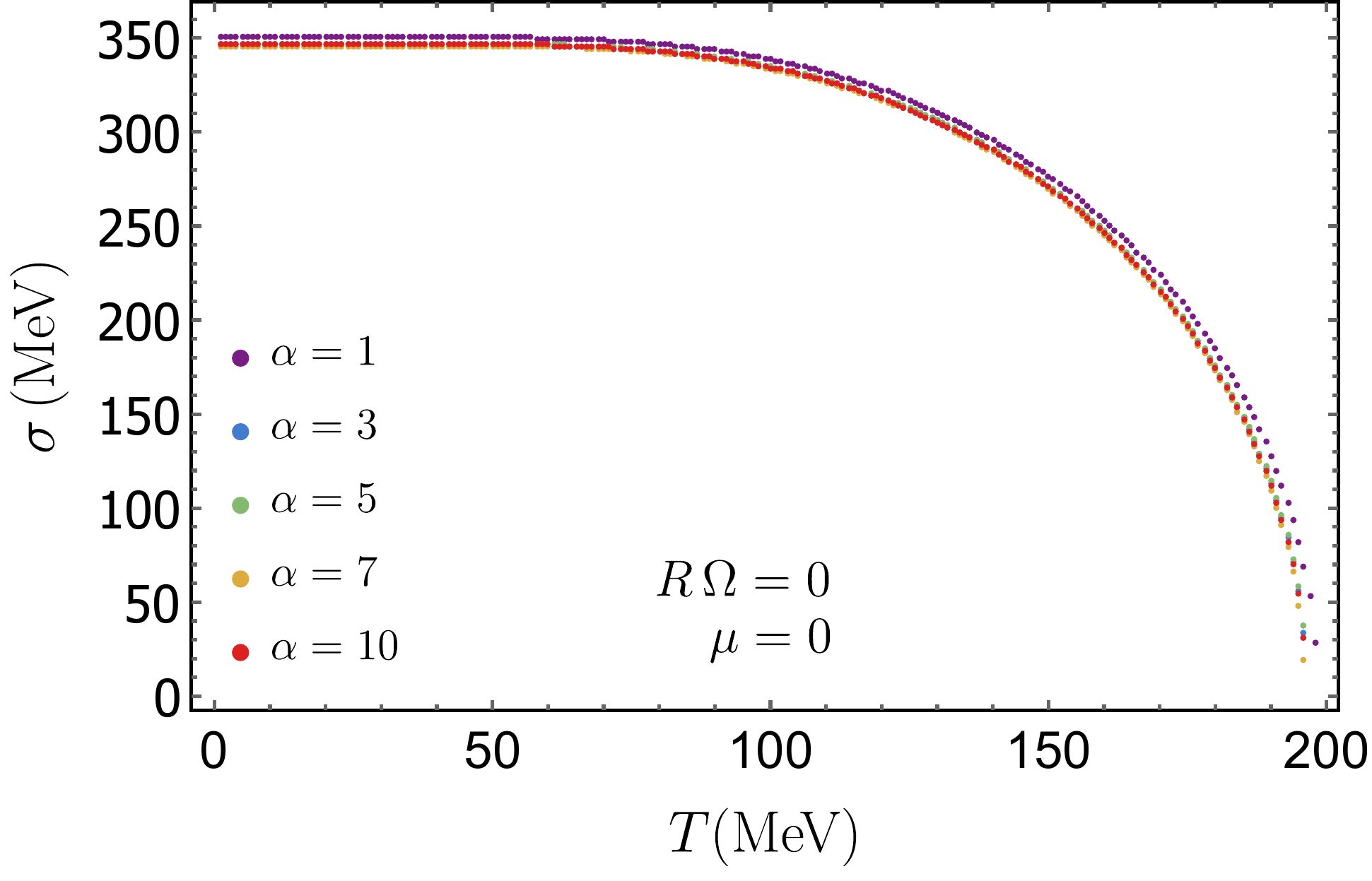}
        \hspace{0.3cm}
        \includegraphics[width=8cm,height=5cm]{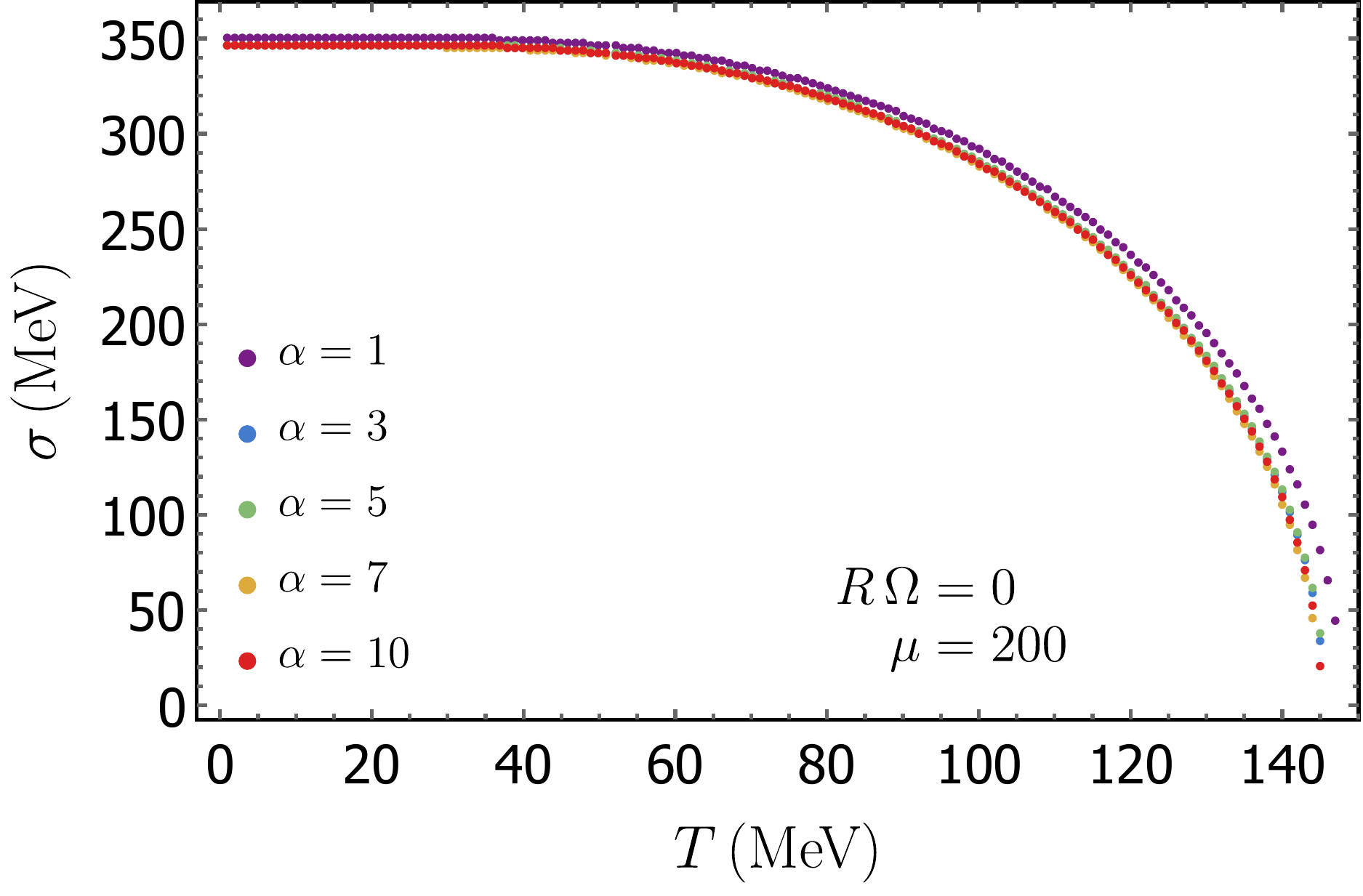}\\
        \includegraphics[width=8cm,height=5cm]{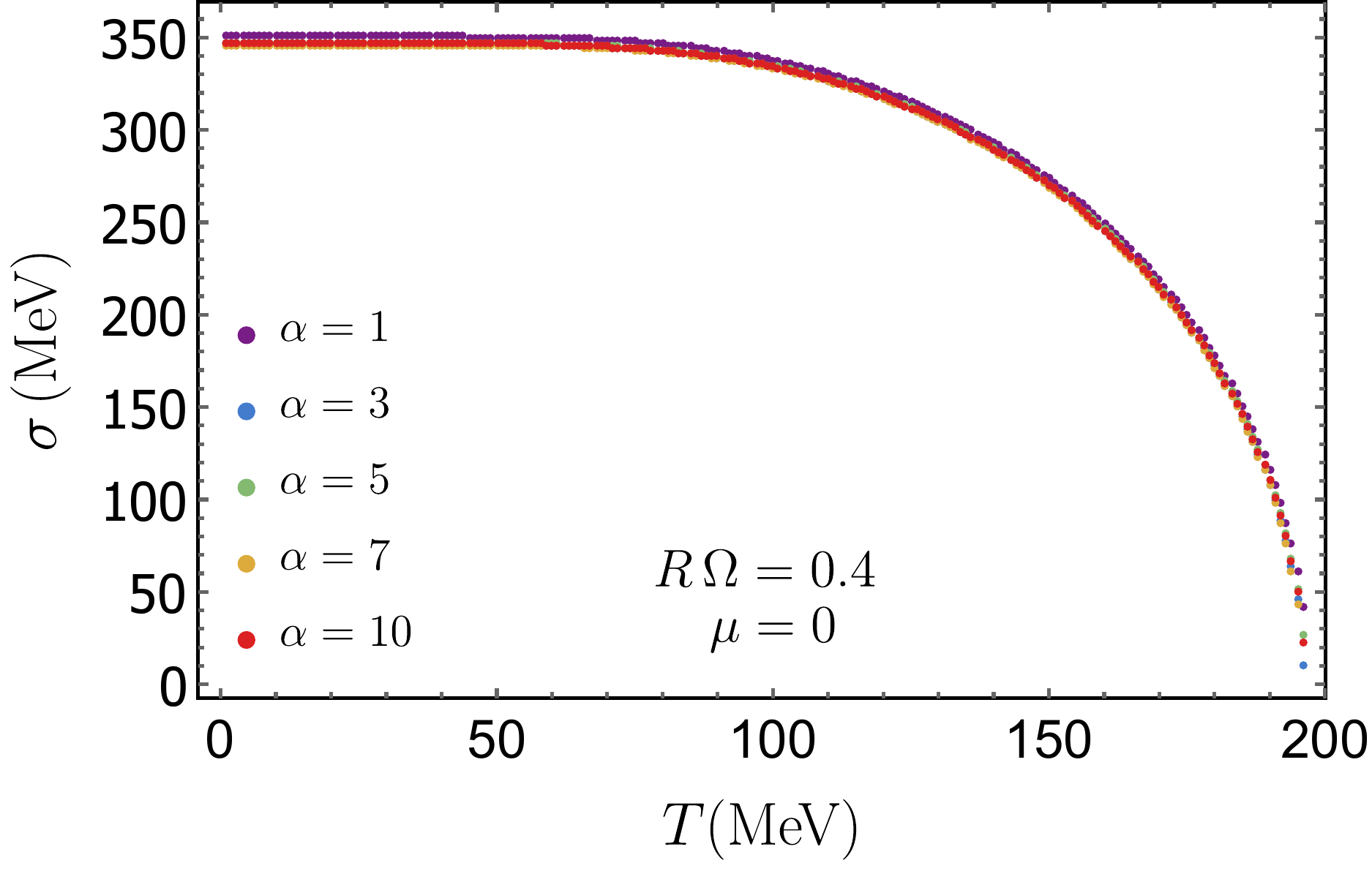}
        \hspace{0.3cm}
        \includegraphics[width=8cm,height=5cm]{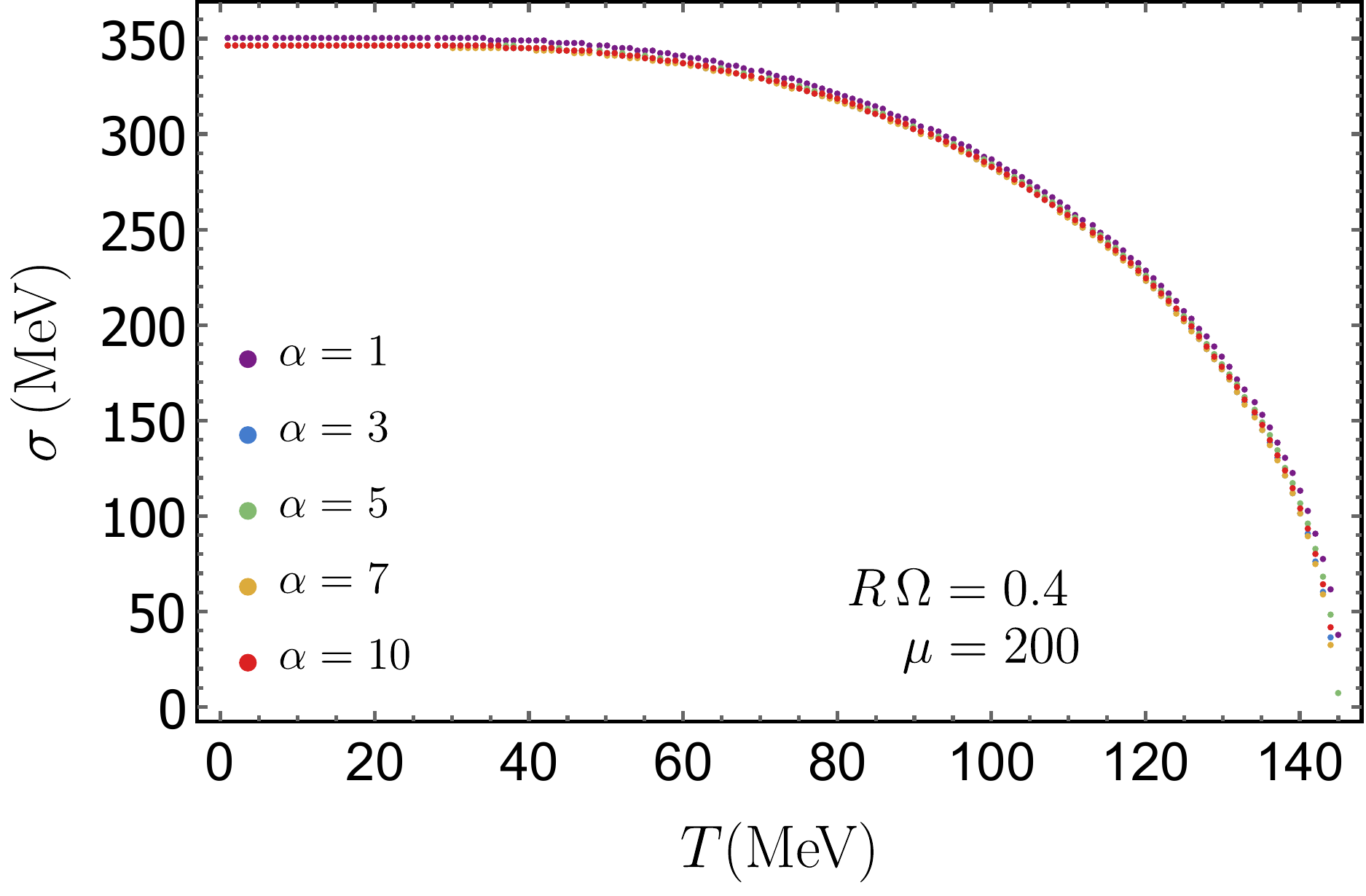}
    \caption{color online. Dynamical mass  $\sigma(T)$ for different $\alpha$ at different situations.  Two top plots indicate non-zero mass solution for $\mu=0$ and $\mu=200 \textrm{MeV}$ at $R\Omega= 0$. Right plots show the similar plot for $\mu=200$ at $(R\Omega= 0, 0.2, 0.4)$ from top to bottom. In each plot different colors stand for corresponding  $\alpha$. The interval $0\leq T \leq 50$ is magnified in the middle plots.}
    \label{fig:m-vs-T-1}
\end{figure}

Effective action at $T\neq 0$ for rotating and magnetized plasma with $N_f$ flavors and $N_c$ colors is given by the Eq. \eqref{eqsec2f28}. Hence, the non-trivial solution to the gap equation comes from the solutions of the following relation
\begin{widetext}
\begin{align}\label{eqsec4f11}
    1 + N_c G \sum_{f= u, d} \sum_{\ell = \ell_\textrm{min}}^{\ell_\textrm{max}} \sum_\lambda\,\int \frac{dp_z}{2\pi} \frac{1}{\epsilon^{(f)}_{\lambda}}\, \mathcal{S}^{(f)}_{\lambda, \ell}(x) \left( f_{FD}(E^{(f)}_{\lambda, \ell, -} + \mu) + f_{FD}(E^{(f)}_{\lambda, \ell, +} - \mu)-1\right) = 0,
\end{align}
\end{widetext}
where the Fermi-Dirac distribution is
\begin{align}
    f_{FD}(\varepsilon) = \frac{1}{1 + e^{\beta \, \varepsilon}}.
\end{align}
We solve the gap equation \eqref{eqsec4f11} for the chiral field $\sigma$ by keeping fixed $(\alpha, R\Omega, T, \mu, G, x)$ and  results are shown in  Fig. \ref{fig:m-vs-T-1}. All the plots of this subsection are sketched at $G \Lambda^2 =5$ and $x =0.1$. 
The two top plots display non-trivial solution for $\mu=0$ and $\mu=200 \textrm{MeV}$ at $R\Omega = 0$ from up to down and the two bottom plots show similar solution for $\mu=0$ and $\mu=200 \textrm{MeV}$ at $R\Omega = 0.4$. In each plot different colors correspond to different $(\alpha = 1, 3, 5, 7, 10)$. Increasing temperature leads to decreasing the mass and at a critical point $\sigma(T_c) \to 0$. Value of this critical point depends on $(\mu, \alpha, R\Omega)$ and we will discuss it later. 
A similar pattern of the $\sigma(\alpha)$ exist for the rest diagrams of the Fig. \ref{fig:m-vs-T-1}. It seems that $\sigma(\alpha)$ is a decreasing function for $1 \leq \alpha \leq 7$ and elsewhere it  increases.  

\begin{figure}
    \centering
        \includegraphics[width=8cm,height=5cm]{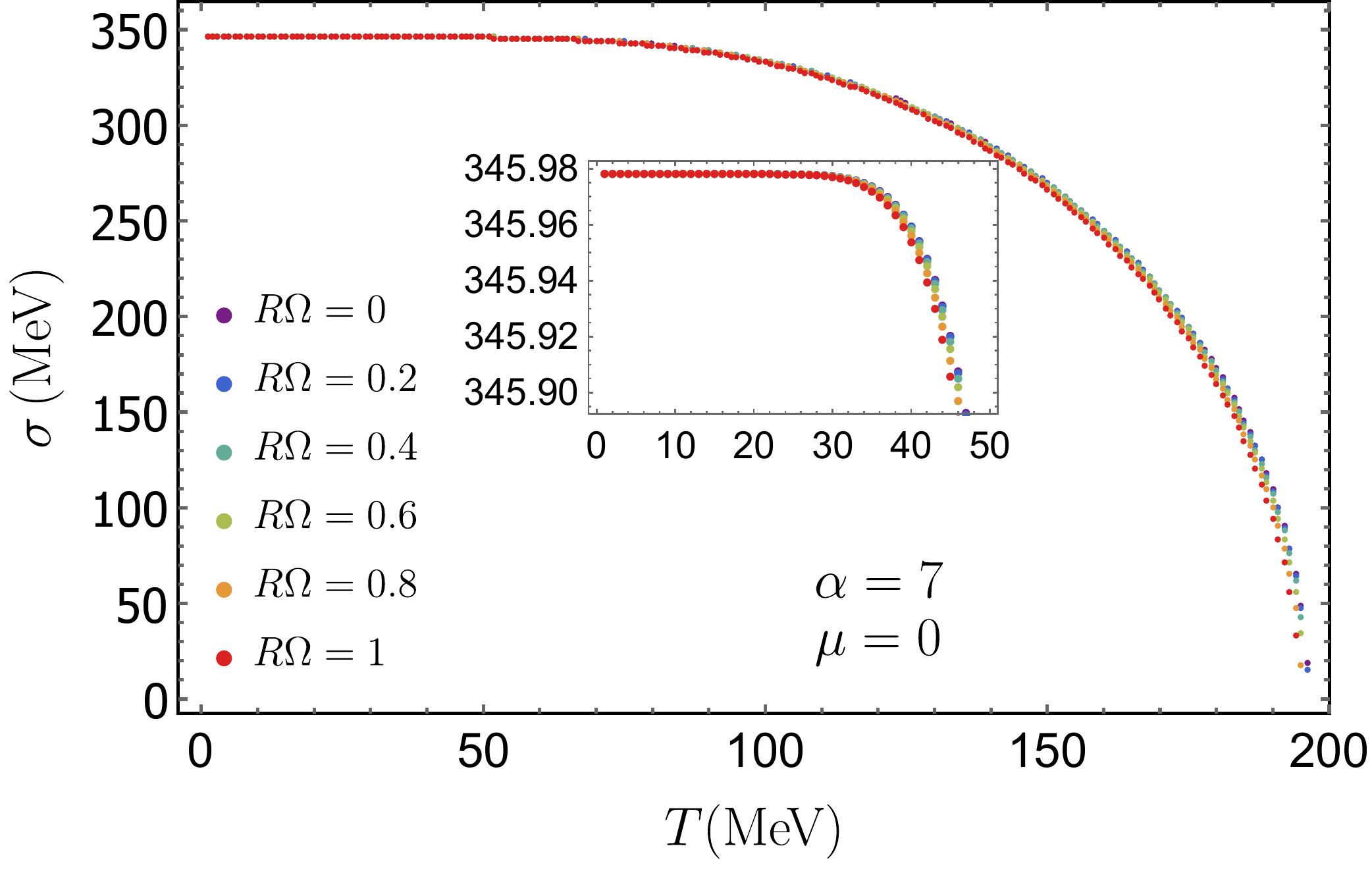}
        \includegraphics[width=8cm,height=5cm]{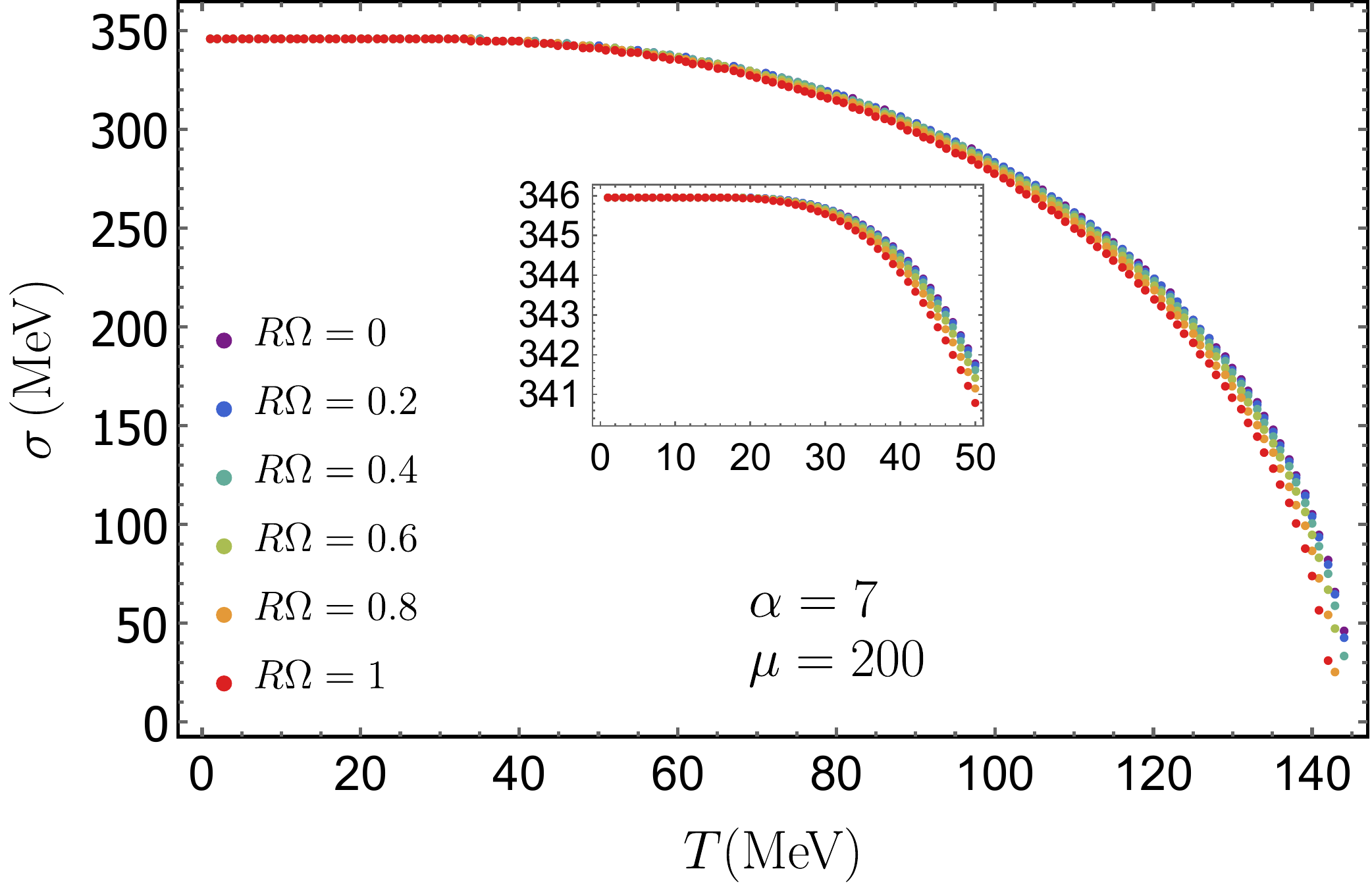}
    \caption{color online. Dynamical mass  $\sigma(T)$ T for different vorticities $(R \Omega = 0, 0.2, 0.4, 0.6, 0.8, 1)$ shown with their corresponding colors.  The inside plots show magnified the interval $0\leq T \leq 50$.}
    \label{fig:m-vs-T-2}
\end{figure}

\begin{figure}
    \centering
        \includegraphics[width=8cm,height=5cm]{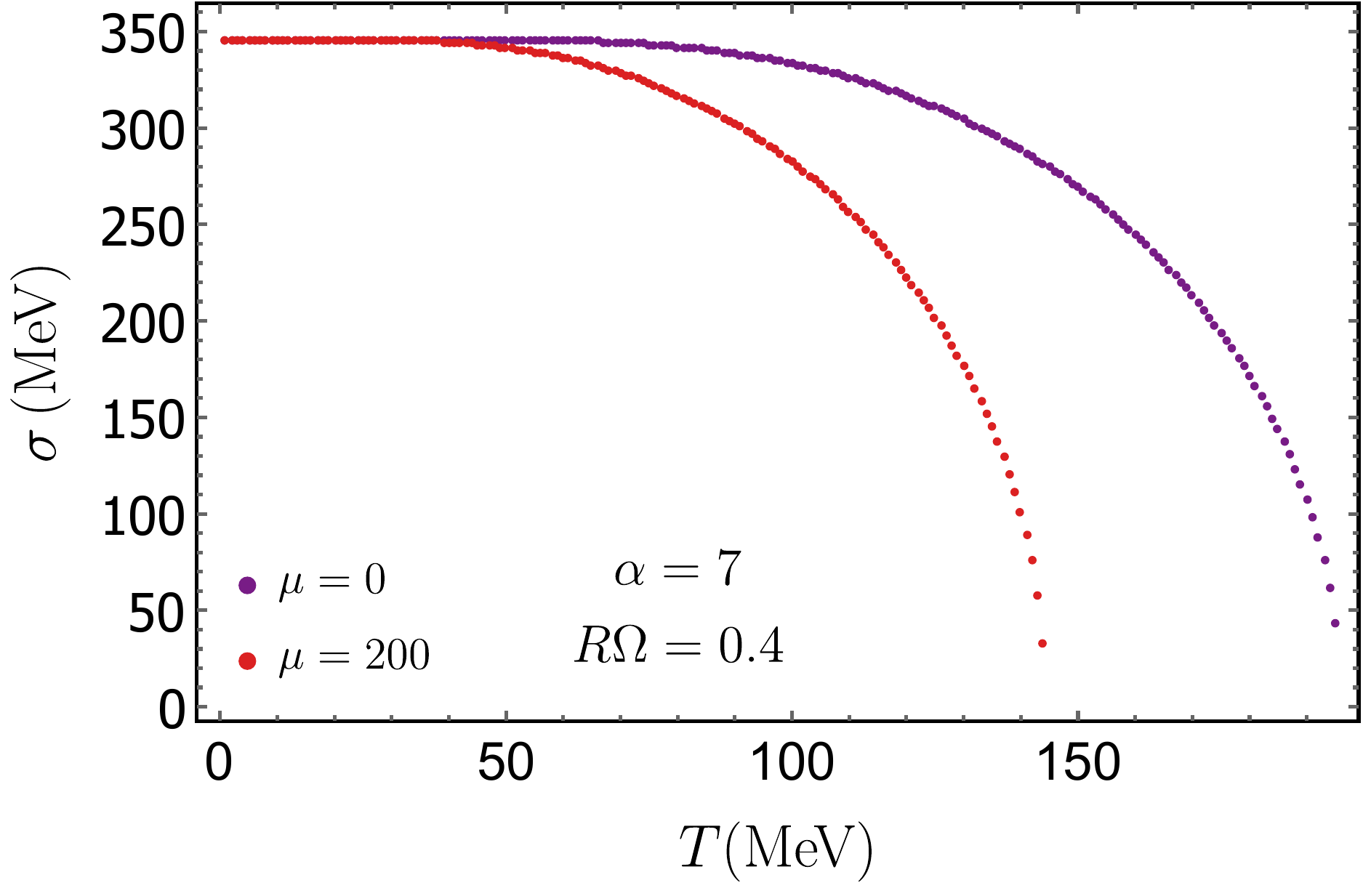}
        \includegraphics[width=8cm,height=5cm]{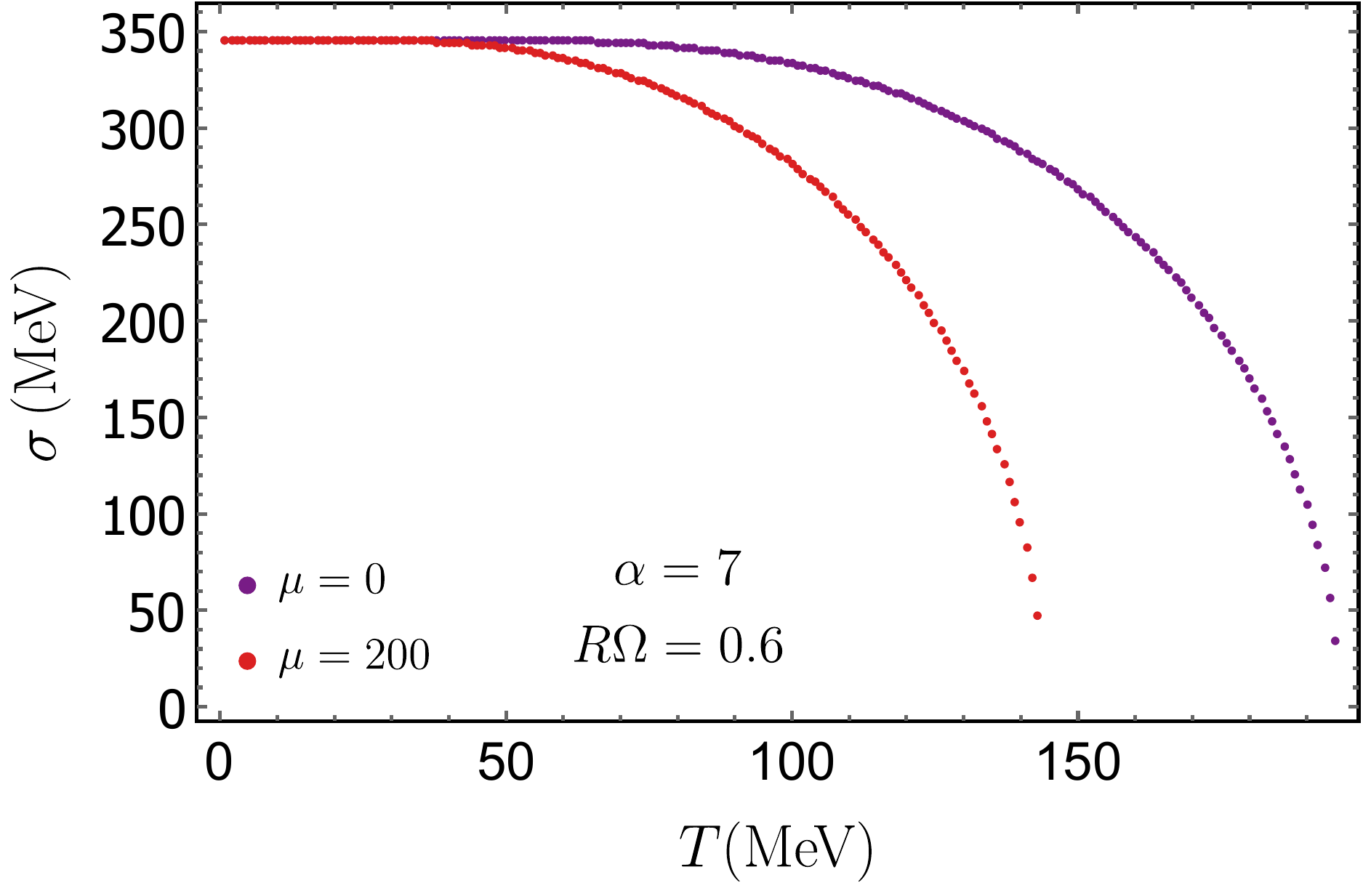}\\
        \includegraphics[width=8cm,height=5cm]{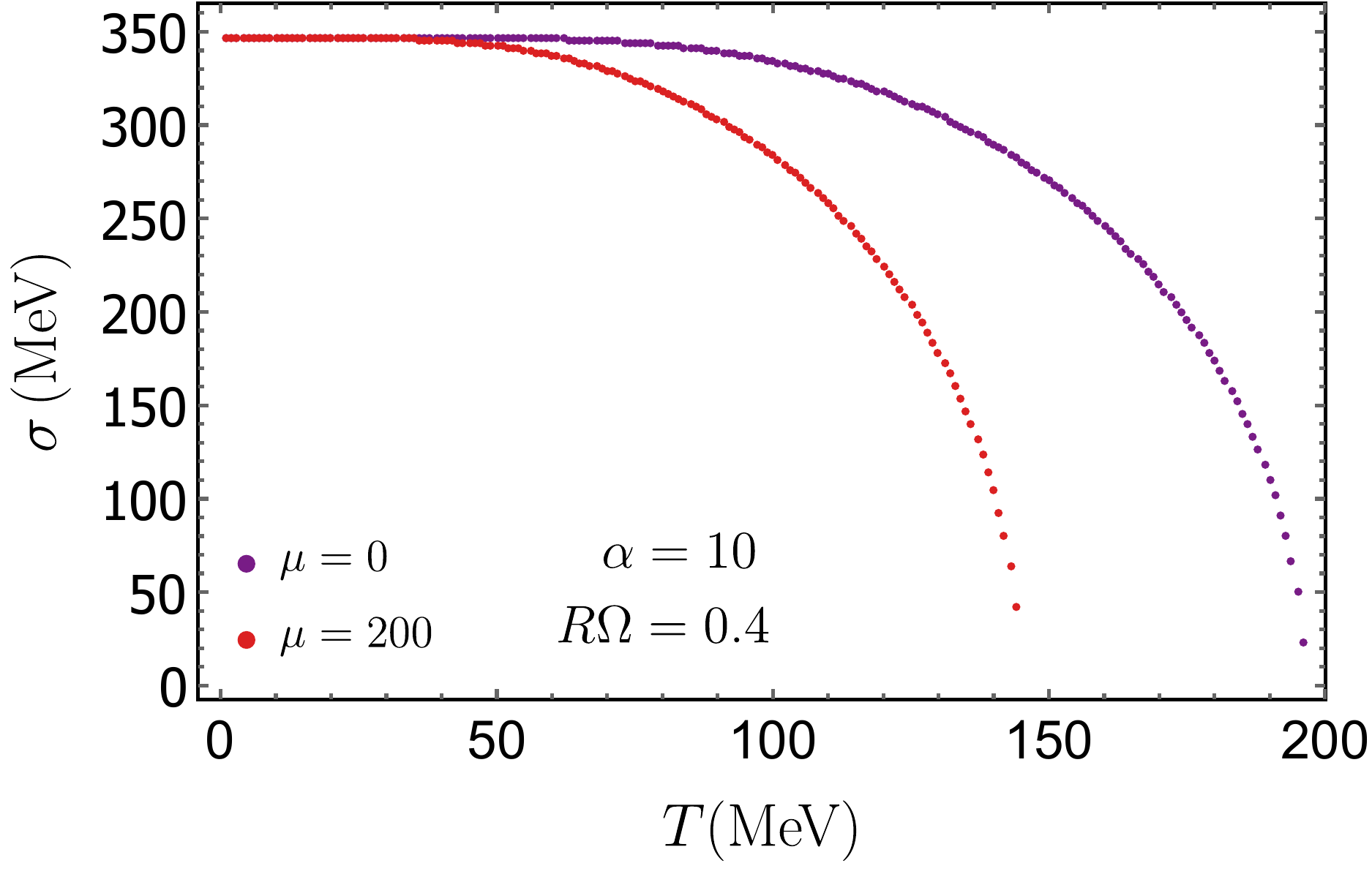}
        \includegraphics[width=8cm,height=5cm]{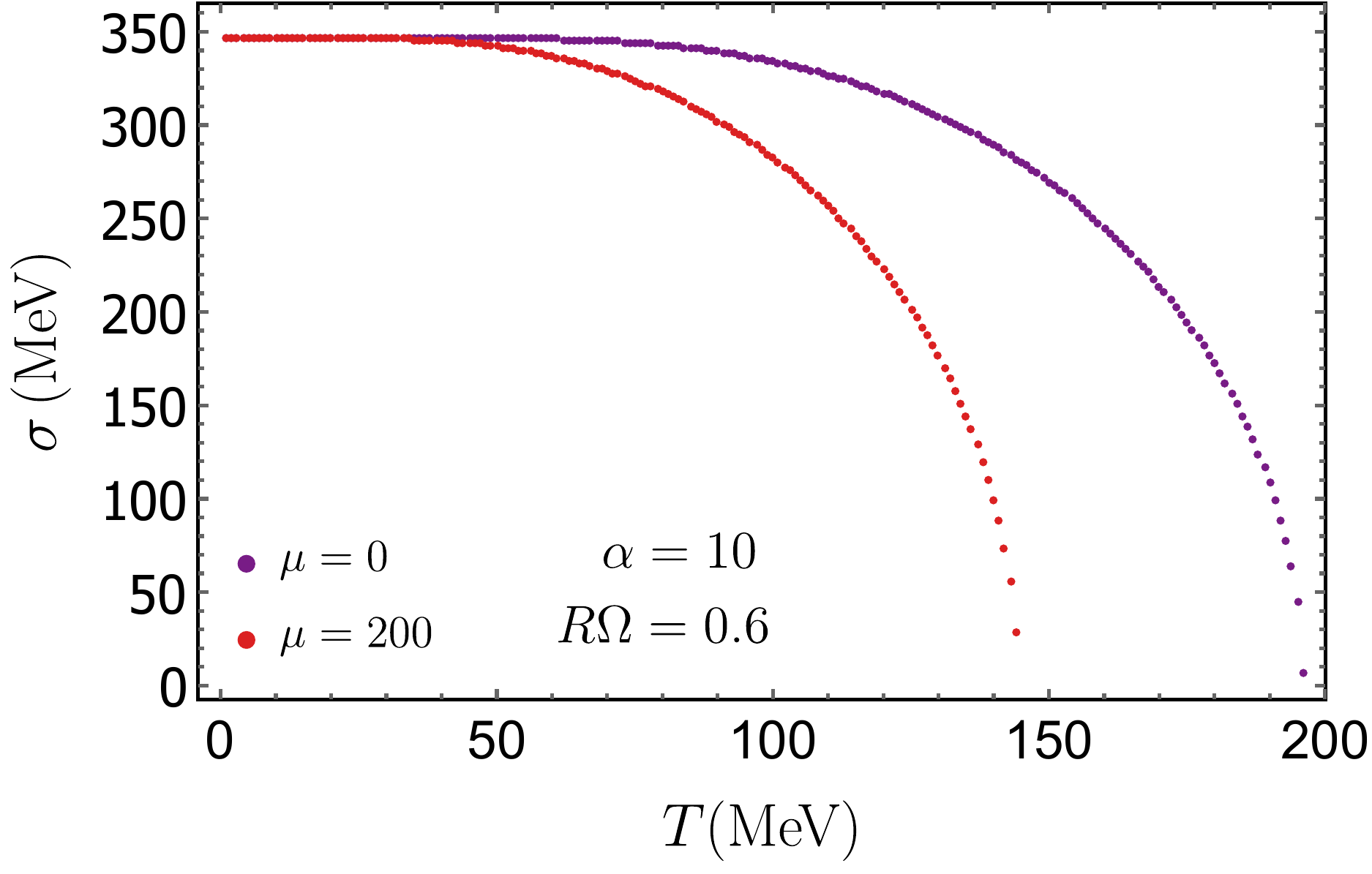}
    \caption{color online. The dynamical mass $\sigma(T)$  for different $R \Omega$ and $\mu$.  Two top plots indicate the $\alpha=7$ at $(R\Omega=0.4, 0.6)$ at $\mu=0$ and $\mu=200 \textrm{MeV}$. The two bottom ones show similar plots for $\alpha=10$ at $(R\Omega=0.4, 0.6)$ at  $\mu=0$ and $\mu=200 \textrm{MeV}$.}
    \label{fig:m-vs-T-3}
\end{figure}

In the Fig. \ref{fig:m-vs-T-2} we show $\sigma(T)$ for $\alpha =7$ at $\mu=0$  and $\mu=200 \textrm{MeV}$. 
Different colors stand for different vorticities $(R\Omega = 0, 0.2, 0.4, 0.6, 0.8, 1)$. The inside plots show magnified the interval $0\leq T\leq 50$. Looking very carefully, we observe  that increasing the rotations causes to decreasing the mass and it is a direct sign of inverse-rotational catalysis. We see this observation to be valid for  $1 \leq \alpha \leq 10$. In the Fig. \ref{fig:m-vs-T-3}, we plot $\sigma(T)$ for $\alpha =7$ and $\alpha=10$ which magenta(red) color shows the $\mu=0$($\mu=200$) solutions. Top plots indicate solution for $\alpha=7$ and bottom plots display the dynamical mass for $\alpha=10$. Likewise, the left part belongs to $R\Omega=0.4$ and the right part belongs to $R\Omega=0.6$. As we would expect, increasing the chemical potential decreases the dynamical mass that is a common feature in all NJL model calculations of phase diagrams.  

By definition, the critical point is a point of which $\sigma(T_c) \to 0$, i.e. the chiral symmetry is restored. However, the kind of phase transition is important. There are many ways to describe the order of phase transition. The regular way is to look for the derivatives of effective action with respect to the order parameter. If n$^\textrm{th}$ derivative of effective action would have singularity, it is the so called n$^\textrm{th}$-order phase transition. Another way is to explore the behavior of dynamical mass.  If phase transition is of second order, the trivial solution is the only one to the gap equation \eqref{eqsec4f11} at $T\geq T_c$ and  chiral symmetry is restored. If the kind of phase transition is of first order, at critical point there is at least two solutions to the gap equation  \eqref{eqsec4f11} besides the trivial one. In this kind, all the solutions have identical effective action and the system lives in a mixed state \cite{McLerran:2021zvt}.  Also, we can identify the first-order transition with a jump in the dynamical mass profile \cite{Asakawa:1989bq}. 

In the NJL and dense effective models at low densities we get second order phase transition, while at high densities we have first-order ones. The joining point of these two section is CEP and it defines the changing of phase transition pattern.  In the current paper, we are able to derive the location of CEP analytically and its derivation works as follows. As we already discussed, at critical point of first-order there are many solutions to the gap equation and they are equivalent in the effective action. You may assume that $\sigma_a = (\sigma_1, \sigma_2, \ldots, \sigma_N)$  is the set of ground state and its members are the solutions of the gap equation $\partial \mathcal{V}_\textrm{eff} / \partial \sigma_i=0$. Expansion of the effective action around each member is 
\begin{align}
    \mathcal{V}_\textrm{eff}(\sigma_i + \delta \sigma_i) =  \mathcal{V}_\textrm{eff}(\sigma_i)  + \frac{\delta \sigma_i^2}{2} \frac{\partial^2 \mathcal{V}_\textrm{eff}}{\partial \sigma^2}\bigg|_{\sigma = \sigma_i} + \ldots.\no
\end{align}
Since all the $\sigma_i$ have similar effective action,  solving $\partial^2 \mathcal{V}_\textrm{eff}/ \partial \sigma^2=0$ in addition to the $\partial \mathcal{V}_\textrm{eff} / \partial \sigma_i=0$ can locate the position of CEP. Basically, the CEP is the first point in phase diagram which meet this cut. Here, we give an example to show how this mechanism works. In the normal NJL model with the effective action given in the Eq.  \eqref{eqsec2f18} these two equations are given by
\begin{widetext}
\begin{align}
    &\frac{\partial \mathcal{V}_\textrm{eff}}{\partial \sigma} = \sigma + \frac{N_c N_f G}{\pi^2} \int_0^\Lambda p^2 dp \frac{\sigma}{\mathcal{E}_0} \left( f_{FD}(\mathcal{E}_0+\mu) + f_{FD}(\mathcal{E}_0-\mu)-1\right)=0,\nonumber\\
    & \frac{\partial^2 \mathcal{V}_\textrm{eff}}{\partial \sigma^2} = 1 + \frac{N_c N_f G}{\pi^2} \int_0^\Lambda p^2 dp \frac{\partial}{ \partial \sigma} \bigg( \frac{\sigma}{\mathcal{E}_0} \left( f_{FD}(\mathcal{E}_0+\mu) + f_{FD}(\mathcal{E}_0-\mu)-1\right)\bigg)=0.
\end{align}
\end{widetext}
The latter equations are identical to the previous methods to find the CEP \cite{Morones-Ibarra:2017avu}. We are able to reproduce the normal NJL phase diagram by solving this method.  Indeed, the second derivative give the susceptibility of the QM and its vanishing states the divergence of susceptibility around the first-order points.     

\begin{figure}
        \includegraphics[width=8cm,height=5cm]{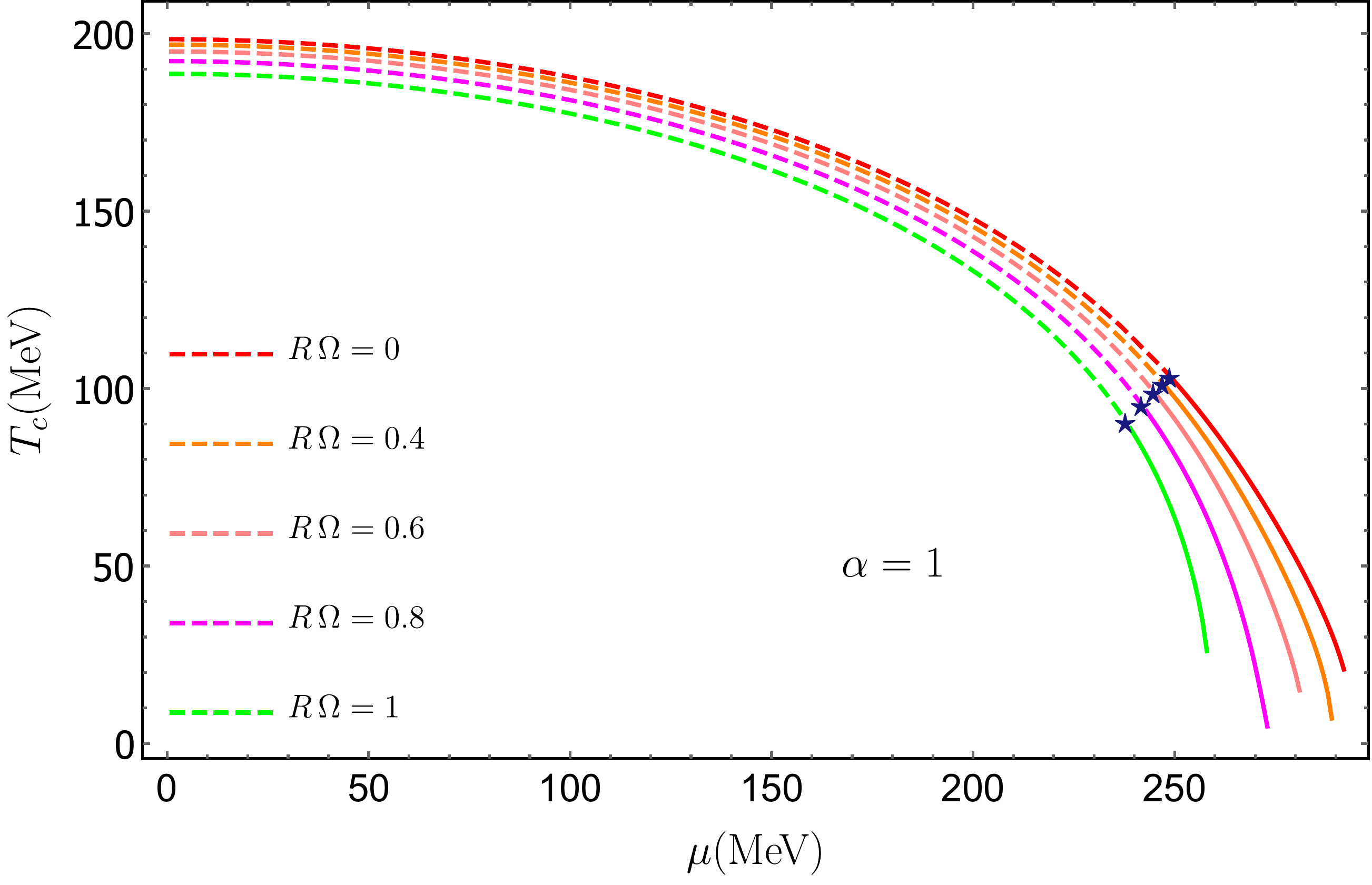}
        \includegraphics[width=8cm,height=5cm]{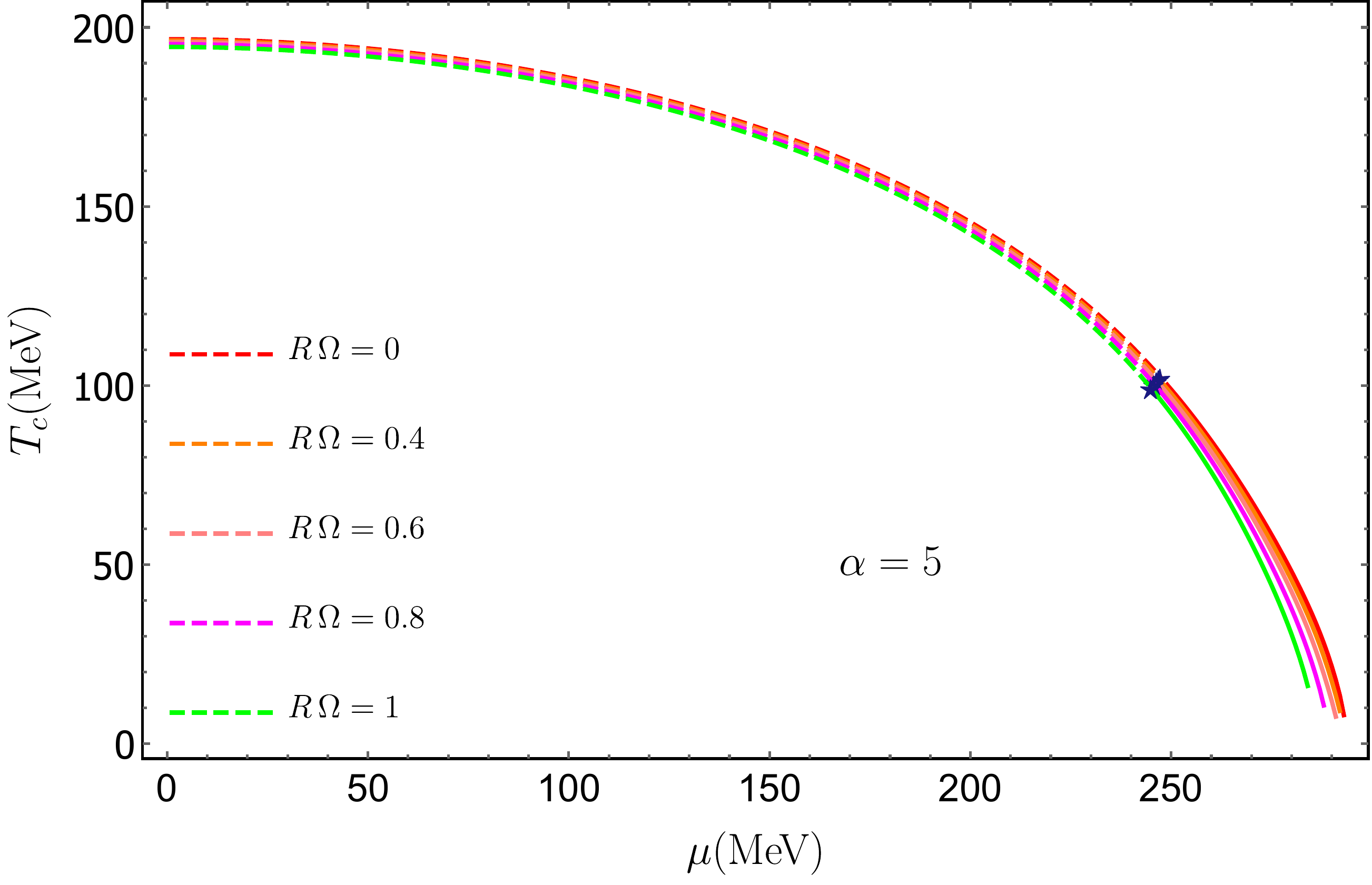}\\
        \includegraphics[width=8cm,height=5cm]{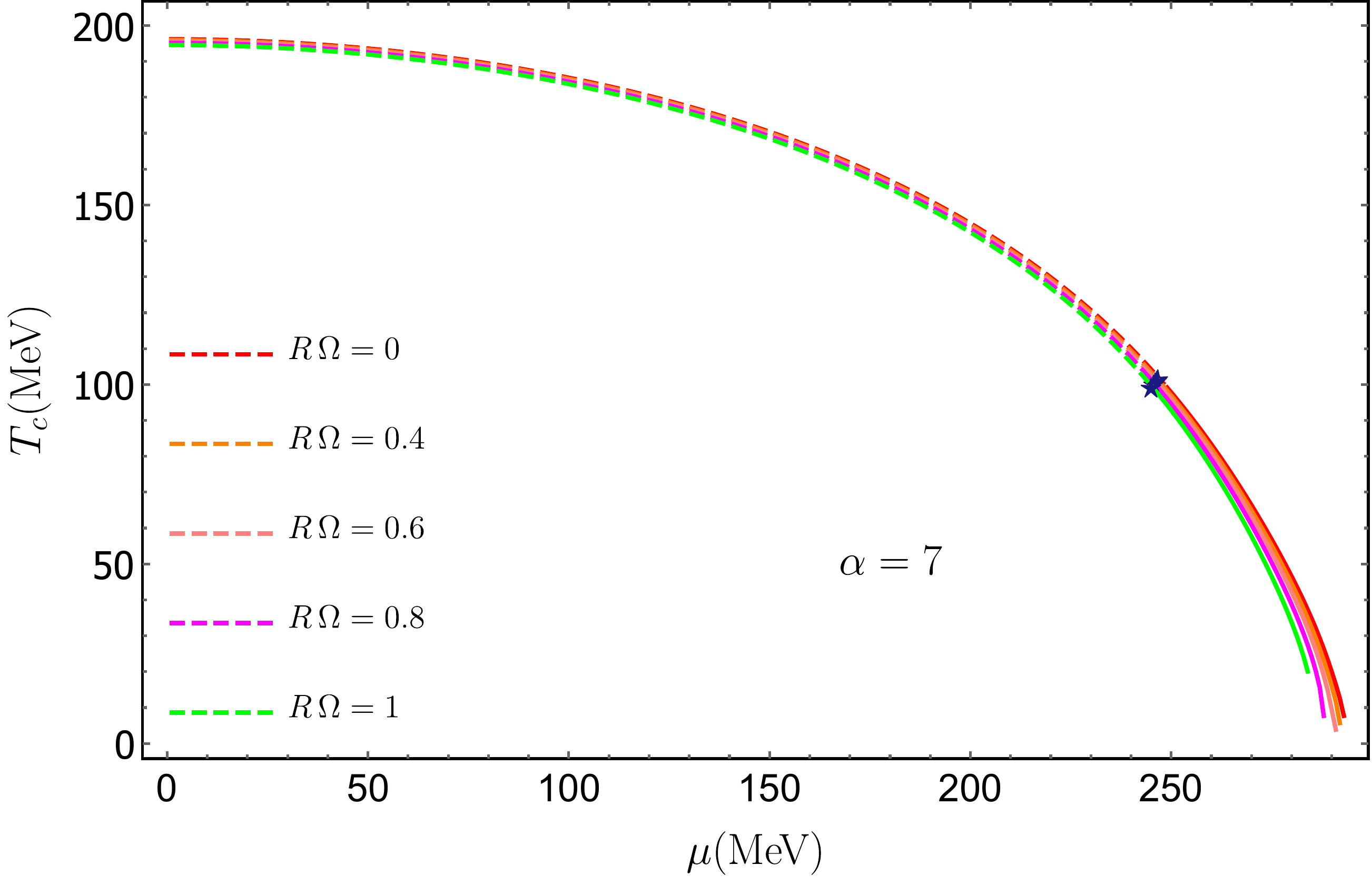}
        \includegraphics[width=8cm,height=5cm]{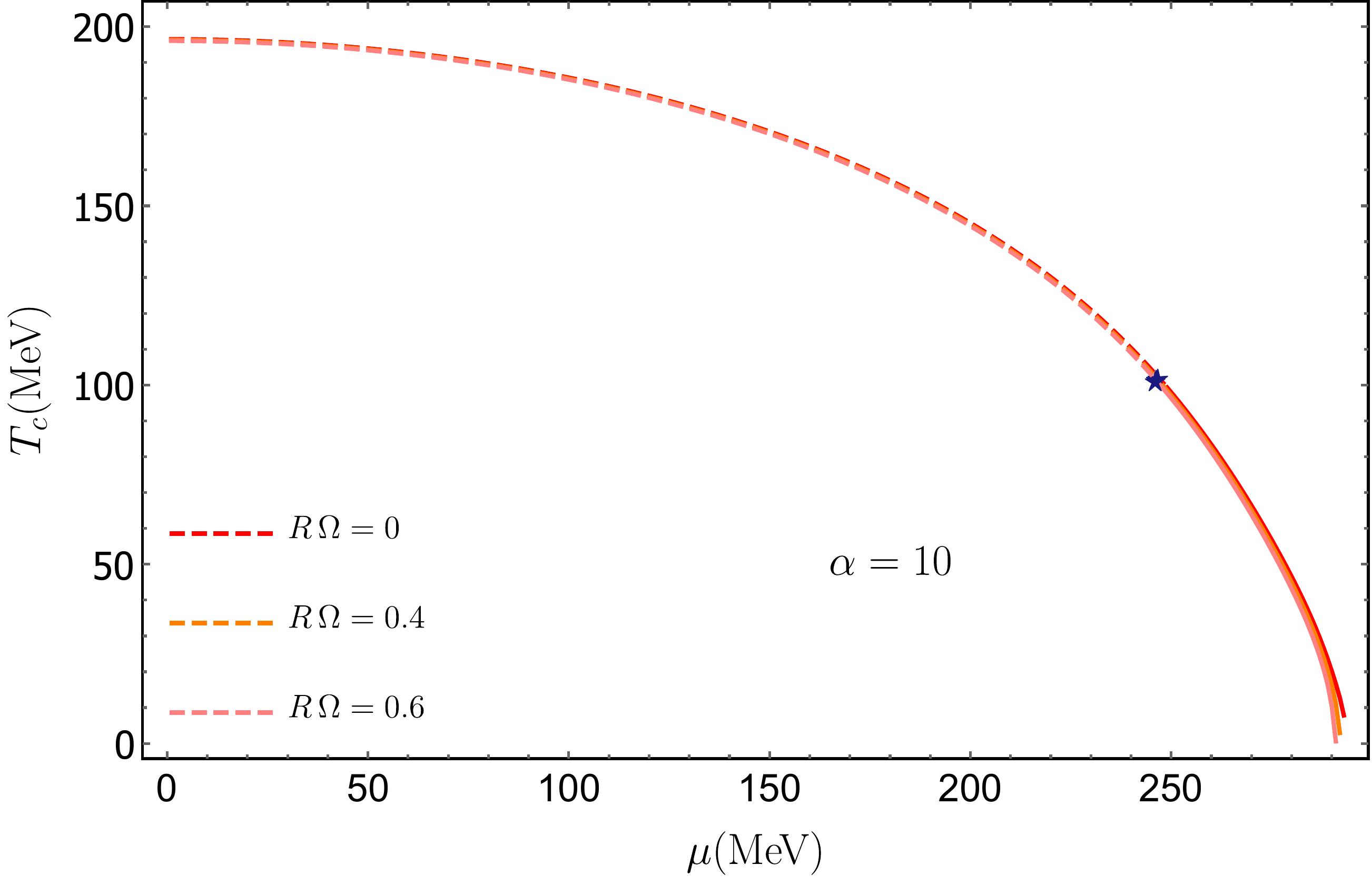}
    \caption{color online. Phase diagram $T_c(\mu)$ in different cases. Solid(dashed) lines represent the first(second) order phase transitions. The very top is for $\alpha=1$,  the next top is for $\alpha=5$, the third- top plot is for $\alpha=7$ and the very bottom is for $\alpha=10$. In each plot, the $R\Omega$ dependence is shown by different colors. For $\alpha=10$, plots are sketched until $R\Omega=0.6$ because $R\Omega_\textrm{Max}=0.68$.}
    \label{fig:Tc-vs-mu-1}
\end{figure}

 In Fig. \ref{fig:Tc-vs-mu-1} we sketch $T_c(\mu)$ for different $\alpha$ and $R\Omega$. Solid(dashed) lines define the first(second) order phase transitions. Their joining, which is the CEP, is marked with a star symbol. Top-left plot represents $\alpha=1$ at different $(R\Omega= 0, 0.4, 0.6, 0.8, 1)$, top-right plot shows $\alpha=5$ at different $(R\Omega= 0, 0.4, 0.6, 0.8, 1)$, bottom-left plot indicates $\alpha=7$ at different $(R\Omega= 0, 0.4, 0.6, 0.8, 1)$   and bottom-right plot shows $\alpha=10$ at different $(R\Omega= 0, 0.4, 0.6)$.  We should study the phase diagrams  by considering  the correlation between magnetic and rotation fields shown in the Fig. \ref{fig:ro-vs-eB}. That is why we show $\alpha=10$ up to $R\Omega=0.6$, because in this magnetic field strength $R\Omega_\textrm{Max} = 0.68$.  To illuminate our results, we fit precisely the transition points of  the Fig. \ref{fig:Tc-vs-mu-1} to the following function
 \begin{align}\label{eqsec4f16}
     T_c(\mu) = T_c^0 \left( 1 - \kappa_2 \left(\frac{\mu}{T_c^0}\right)^2 - \kappa_4 \left(\frac{\mu}{T_c^0}\right)^4\right).
 \end{align}
The parameters $(T_c^0, \kappa_2, \kappa_4)$ are given in the Appendix. \ref{AppA} for each $\alpha$ and $R\Omega$. Although, our model is very different from the Lattice models,  they are comparable and could hint us to see how the QM behaves under the rotation and magnetic field and what would we expect from other numerical methods \cite{Ding:2015ona}. Compared to the normal Lattice models (no rotation or magnetic fields), we find that the curvature of the phase diagram $\kappa_2$ is enhanced. The $\kappa_2$ decreases by increasing  the $R\Omega$ and $\alpha$.

\begin{figure}
    \centering
        \includegraphics[width=8cm,height=5cm]{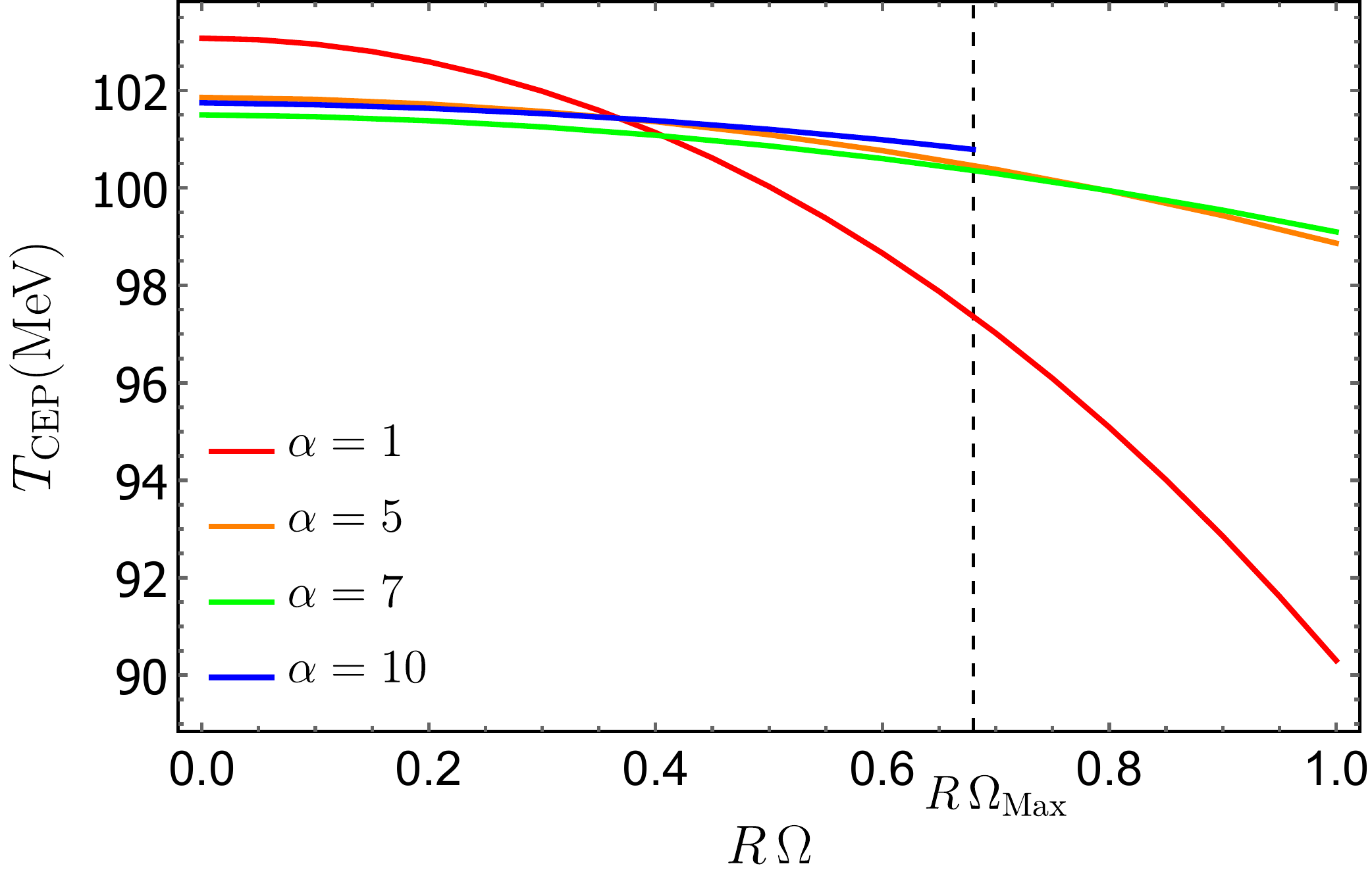}
    \caption{color online. Temperature of CEP  as a function of $R\Omega$ for different $\alpha$. Dashed line shows the $R\Omega_\textrm{Max}$ of $\alpha=10$.}
    \label{fig:Tc-vs-mu-2}
\end{figure}

\begin{figure}
    \centering
        \includegraphics[width=8cm,height=5cm]{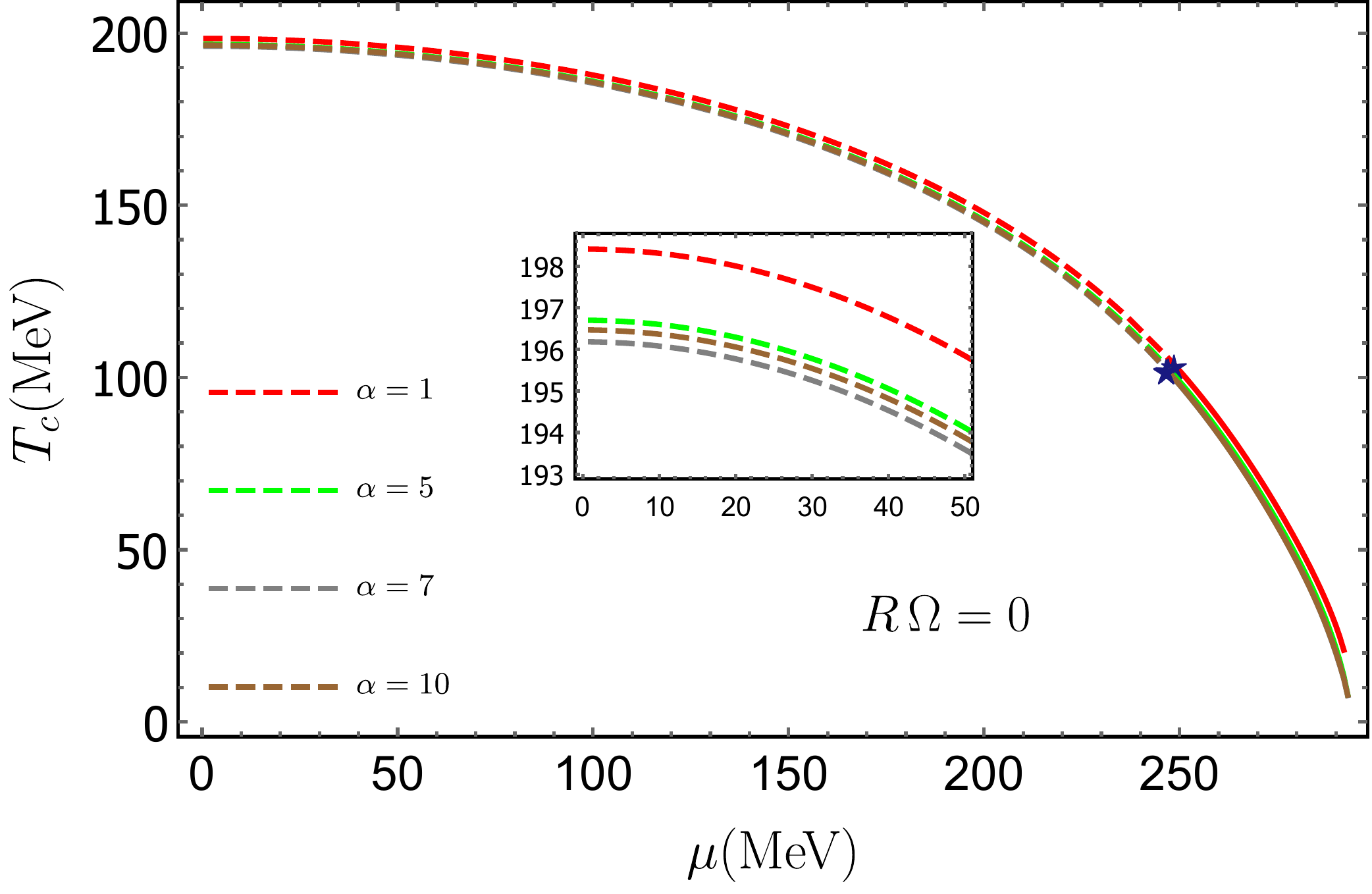}
        \includegraphics[width=8cm,height=5cm]{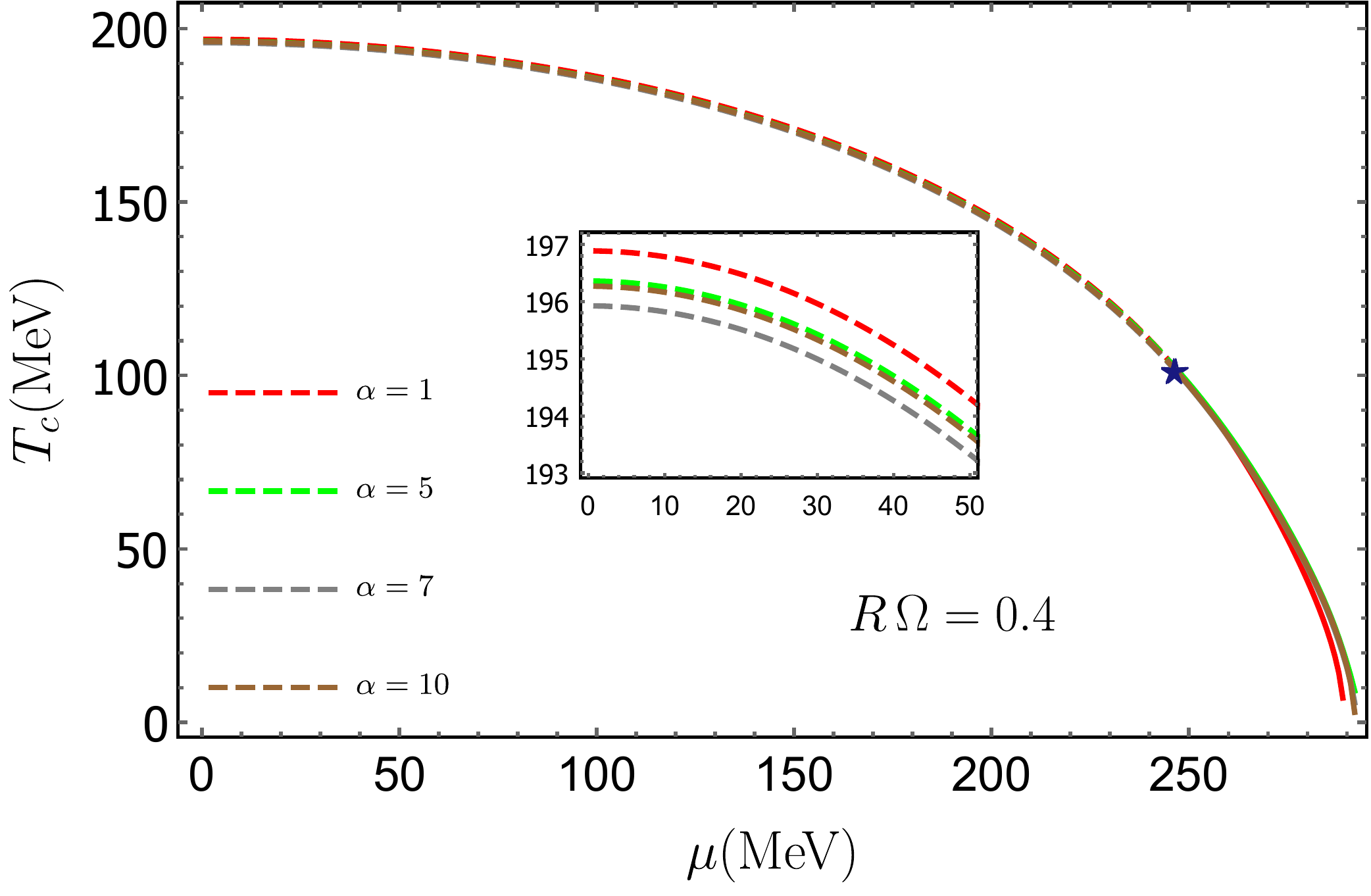}\\
        \includegraphics[width=8cm,height=5cm]{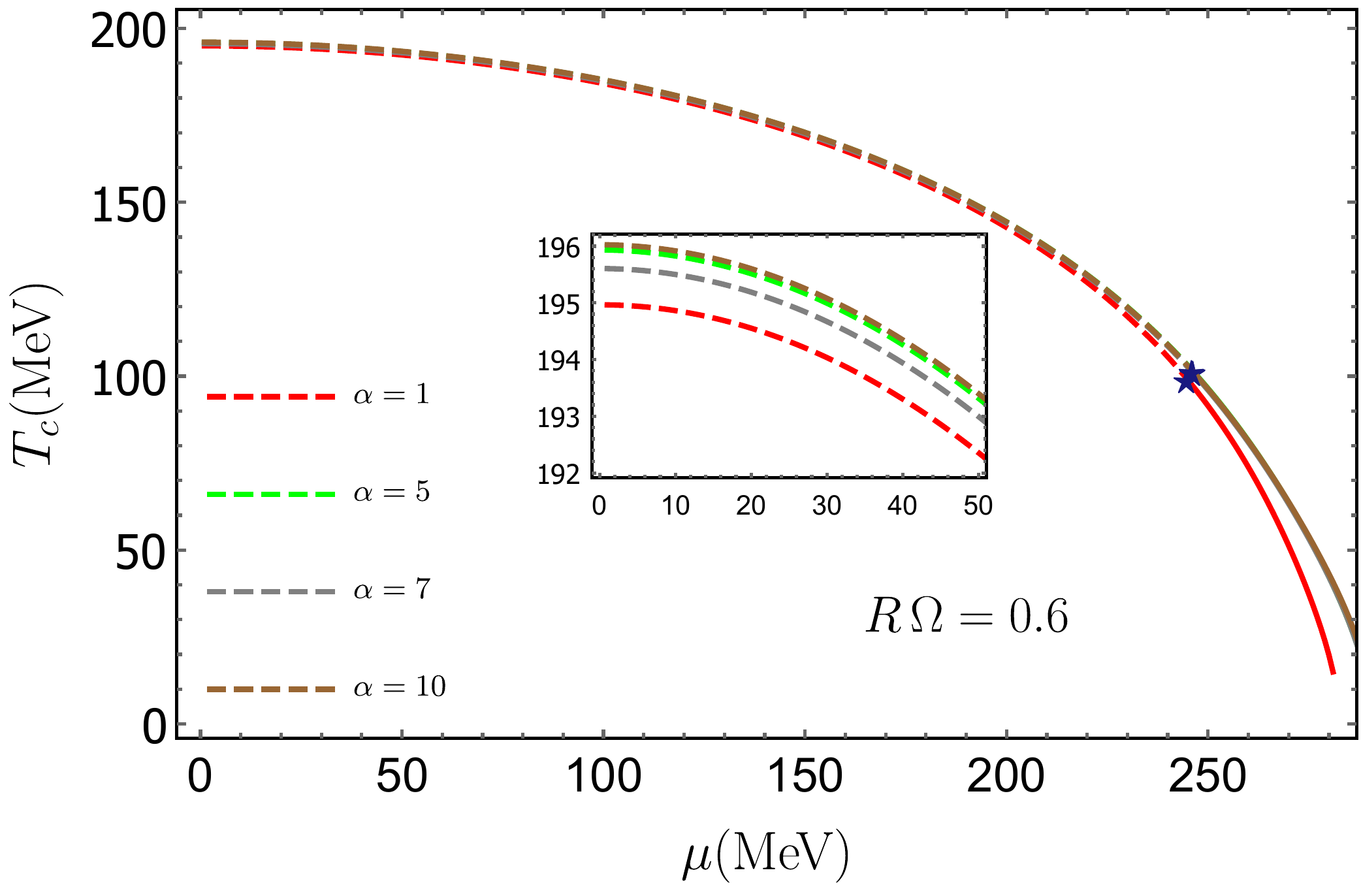}
    \caption{color online. Phase diagram $T_c(\mu)$ for different $R\Omega$. Top, middle and bottom plots show the phase diagram for $(R\Omega=0, 0.4, 0.6)$, respectively. In each plot different colors belong to $(\alpha= 1, 5, 7, 10)$ shown by different colors. The inside plots magnify the points  in interval $0\leq \mu \leq 50$. }
    \label{fig:Tc-vs-mu-4}
\end{figure}

We infer the inverse-rotational catalysis of these plots because increasing the $R\Omega$ causes decreasing $T_c(\mu)$.  
This property is very obvious in Fig. \ref{fig:Tc-vs-mu-2}. 
In this figure, we demonstrate $T_{\textrm{CEP}}(R\Omega)$ for different $\alpha$. As it is shown, the magnetic field has a complex impact on the phase diagram. We see inverse-magneto catalysis until a point $R\Omega \sim 0.39$,  which means that increasing $\alpha$ leads to decreasing $T_c$, while for $R\Omega > 0.39$ we see the magneto-catalysis. This pattern is valid up to $\alpha=7$ and for larger values of magnetic field it seems  magneto-catalysis to happen every where. Generally, magneto-catalysis is the dominant picture at large $R\Omega$.  We can also express these results from the numbers given in the table \ref{tab1} to table \ref{tab4} of the App. \ref{AppA}.  
\begin{figure}
    \centering
        \includegraphics[width=8cm,height=5cm]{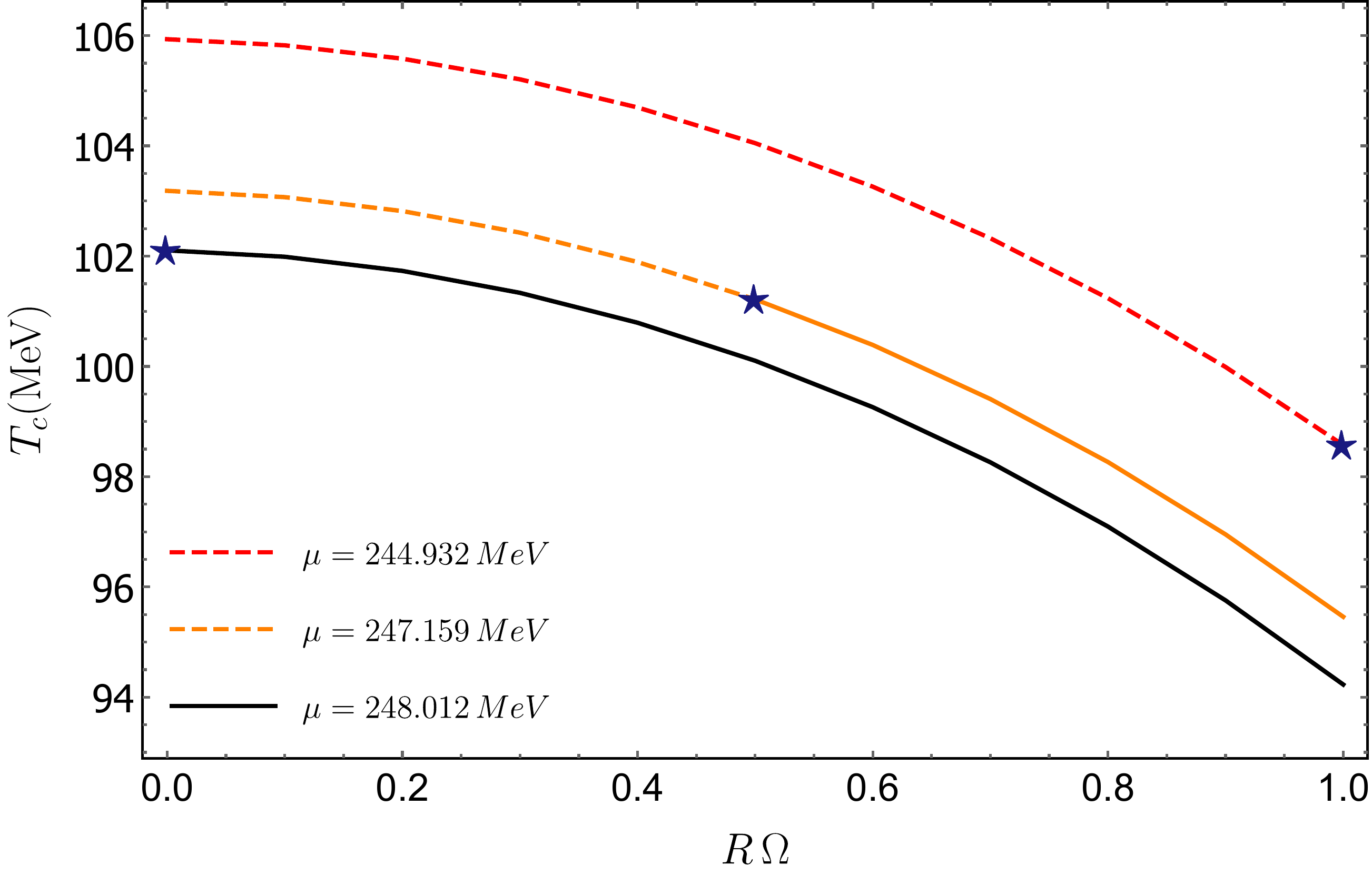}
    \caption{color online. Phase diagram $T_c(R\Omega)$ for $\alpha=7$ at different $\mu$. Solid(dashed) lines illustrate first(second) order phase transitions. Star marks indicate the location of CEP at each $\mu$.}
    \label{fig:Tc-vs-mu-3}
\end{figure}
In the Fig. \ref{fig:Tc-vs-mu-3}, we demonstrate $T_c(R\Omega)$ at $\alpha=7$ for different chemical potentials. Types of the lines and stars are similar to the Fig. \ref{fig:Tc-vs-mu-1}. For $\mu < \mu_1 = 244.932$ phase transition is second order every where in $0 \leq R\Omega \leq 1$, but at the point $\mu_1$ phase diagram starts to get  some first-order transition points. Increasing the chemical potential covers more points in the $T_c(R\Omega)$ plot to become first order and finally in the point $\mu_0=248.012$ all the points  in $0 \leq R\Omega \leq 1$ possess first-order transition. For $\mu \geq \mu_0$ the phase transition is always first order in the $R\Omega$ direction. This kind of interplay between the chemical potential and $R\Omega$ in the $T_c(R\Omega)$ plots, is observed everywhere in $0\leq \alpha \leq 10$. Likewise, the point that faster plasma changes sooner to become first-order, is definable within the inverse-rotational catalysis scenario.  The Fig. \ref{fig:Tc-vs-mu-4} shows $\alpha$ dependence in $T_c(R\Omega)$ plots.  Top-right, top-left and bottom plots indicate $(R\Omega=0, 0.4, 0.6)$, respectively and different colors in each plot belong to $(\alpha= 1, 5, 7, 10)$ as it is shown. The inside plots show magnified  points of plots in $0
\leq \mu \leq 50$. We observe that for $ 0\leq R\Omega \leq 0.39$ the dominant picture is inverse-magneto catalysis, while in faster plasma $R\Omega \geq 0.4$ the magneto-catalysis scenario is seen. This picture is valid up to $\alpha_c=7$. For larger magnetic field we see  the magneto-catalysis everywhere. Results of this figure are in complete agreement with the Fig.  \ref{fig:Tc-vs-mu-2}.
 \section{Conclusion}
Studying the phase diagrams of QM in presence of the rotation and magnetic field  is of great importance in today's high-energy research.  Either of the magnetic or rotation fields has its special non-trivial effects on the evolution of QGP and they can reveal many facts about the behavior of QGP under these extreme circumstances. In this work, we examine the chiral symmetry restoration/breaking of a QM with $N_f=2$ and $N_c=3$ within the effective NJL model accompanied by constant magnetic and rotation fields.  Utilizing the Ritus method has enabled us to present a master formula for the in-medium effective action of $"\sigma"$ field in curved spaces. In this regard, we gave some examples to clarify our framework and showed how it works in different situations. Likewise, we used the Ritus method to solve the Dirac equation in  the constantly rotating and magnetized plasma. We identified the energy eigenvalues along with the generalized Landau levels and to have a good quantization pattern, we take the positive energies. We imposed the spectral boundary condition and the profile of Landau levels as well as energy levels are sketched in terms of orbital number $\ell$ for different $\alpha$.  The very important result is the correlation between the magnetic and rotation fields and we concluded that strongly magnetized plasma is not able to rotate.  The critical coupling needed to provide a non-trivial solution to the gap equation decreases by increasing $\alpha$ and  it closes to the normal NJL value $G_c \Lambda^2 = \frac{2\pi^2}{N_c N_f}$ at small $\alpha$. 

Solutions of the gap equation at zero and finite temperature express many facts. At zero temperature, some features are in common with the normal NJL calculations. Since vorticity disappears at $T=0$ near the axis we didn't see rotational-magnetic inhibition. However, near the boundary the chiral condensate grows with $\alpha$ because of the mode accumulation and this confirms the surface-magnetic catalysis.   At finite temperature, the interplay between the chemical potential, magnetic and rotation fields makes the phase diagrams to be more complicated.  $T_c$ always decreases by increasing the $R\Omega$ which is nothing but the inverse-rotational catalysis. But $T_c(\alpha)$ is more complex in which for less rotating plasma ($0 \leq R\Omega \leq 0.38$) $T_c$ decreases by increasing $\alpha$ -the inverse-magneto catalysis-, while in fast rotating plasma ($0.38 < R\Omega \leq 1$) $T_c$ increases by increasing the $\alpha$ -the magneto catalysis-. Points of the phase diagrams are fitted to the polynomial function and we discover that the magnetic and rotation fields try to increase the curvature of the phase diagrams. This curvature decreases by increasing $\alpha$ and $R\Omega$. The position of CEP is found exactly by solving the gap equation and its derivatives, simultaneously. The $T_{\textrm{CEP}}$ always decrease by increasing the $R\Omega$ and at each fixed $\mu$, the $\frac{d T_{\textrm{CEP}}}{d R\Omega}$ is larger for smaller $\alpha$.
 
 Following the current paper, we can extend it to more cases. We study the chiral symmetry dynamics, but an interesting topic is confinement/deconfinement phase transition in presence of constant magnetic and rotation fields. Also, the results of this work can be more investigated within the other effective models such as quark-hadron models to see whether they  change. Including the strange quarks into the effective models is also very interesting.  Spectrum of mesons in the constant rotation and magnetic field deserves an independent study.
 \section*{Acknowledgement}
 We would like to thank M. Asadi for reading carefully the manuscript. We kindly appreciate M. N. Chernodub for editing the primary draft, giving valuable comments and having fruitful discussions. We also thank Xu-Guang Huang and Igor Shovkovy for their precious discussions and clarification of some issues.  
\begin{appendices}
\section{Parameters of fit}\label{AppA}
The parameters of the fit equation \eqref{eqsec4f16} are shown below for each $\alpha$ and $R\Omega$
\begin{table}[!htb]
\begin{minipage}{.7\linewidth}
    \centering
    \begin{tabular}{|c|c|c|c|}
    \hline
         $R\Omega$& $T_c^0$&$\kappa_2$&$\kappa_4$ \\
         \hline
         0&197.985&0.174&0.074\\
         \hline
         0.4&196.365&0.169&0.08\\
         \hline
         0.6&194.398&0.166&0.083\\
         \hline
         0.8&191.59&0.161&0.089\\
         \hline
         1&187.935&0.155&0.096\\
         \hline
    \end{tabular}
    \caption{Parameters of fit for $\alpha=1$.}
    \label{tab1}
    \end{minipage}%
    \begin{minipage}{.5\linewidth}
    \centering
    \begin{tabular}{|c|c|c|c|}
    \hline
         $R\Omega$& $T_c^0$&$\kappa_2$&$\kappa_4$ \\
         \hline
         0&194.445&0.106&0.122\\
         \hline
         0.4&194.055&0.105&0.124\\
         \hline
         0.6&193.524&0.101&0.127\\
         \hline
         0.8&192.929&0.1&0.128\\
         \hline
         1&192.194&0.1&0.13\\
         \hline
    \end{tabular}
    \caption{Parameters of fit for $\alpha=5$.}
    \label{tab2}
     \end{minipage}
\end{table}

\begin{table}[!htb]
\begin{minipage}{.7\linewidth}
    \centering
    \begin{tabular}{|c|c|c|c|}
    \hline
         $R\Omega$& $T_c^0$&$\kappa_2$&$\kappa_4$ \\
         \hline
         0&193.917&0.106&0.122\\
         \hline
         0.4&193.598&0.104&0.124\\
         \hline
         0.6&193.189&0.1&0.127\\
         \hline
         0.8&192.817&0.102&0.126\\
         \hline
         1&192.396&0.107&0.125\\
         \hline
    \end{tabular}
    \caption{Parameters of fit for $\alpha=7$.}
    \label{tab3}
\end{minipage}%
    \begin{minipage}{.4\linewidth}
    \centering
    \begin{tabular}{|c|c|c|c|}
    \hline
         $R\Omega$& $T_c^0$&$\kappa_2$&$\kappa_4$ \\
         \hline
         0&194.199&0.106&0.123\\
         \hline
         0.4&193.931&0.104&0.125\\
         \hline
         0.6&193.632&0.102&0.127\\
         \hline
    \end{tabular}
    \caption{Parameters of fit for $\alpha=10$.}
    \label{tab4}
    \end{minipage}
\end{table}
\end{appendices}
\bibliographystyle{fullsort}
\bibliography{myreferences}

\providecommand{\href}[2]{#2}\begingroup\raggedright\begin{thebibliography}{10}

\bibitem{Shuryak:2004cy}
E.~V. Shuryak, ``{What RHIC experiments and theory tell us about properties of
  quark-gluon plasma?},'' {\em Nucl. Phys. A} {\bf 750} (2005) 64--83,
  \href{http://www.arXiv.org/abs/hep-ph/0405066}{{\tt hep-ph/0405066}}.

\bibitem{ALICE:2012eyl}
{\bf ALICE} Collaboration, B.~Abelev {\em et al.}, ``{Long-range angular
  correlations on the near and away side in $p$-Pb collisions at
  $\sqrt{s_{NN}}=5.02$ TeV},'' {\em Phys. Lett. B} {\bf 719} (2013) 29--41,
  \href{http://www.arXiv.org/abs/1212.2001}{{\tt 1212.2001}}.

\bibitem{STAR:2005gfr}
{\bf STAR} Collaboration, J.~Adams {\em et al.}, ``{Experimental and
  theoretical challenges in the search for the quark gluon plasma: The STAR
  Collaboration's critical assessment of the evidence from RHIC collisions},''
  {\em Nucl. Phys. A} {\bf 757} (2005) 102--183,
  \href{http://www.arXiv.org/abs/nucl-ex/0501009}{{\tt nucl-ex/0501009}}.

\bibitem{Alford:1997zt}
M.~G. Alford, K.~Rajagopal, and F.~Wilczek, ``{QCD at finite baryon density:
  Nucleon droplets and color superconductivity},'' {\em Phys. Lett. B} {\bf
  422} (1998) 247--256, \href{http://www.arXiv.org/abs/hep-ph/9711395}{{\tt
  hep-ph/9711395}}.

\bibitem{Fukushima:2010bq}
K.~Fukushima and T.~Hatsuda, ``{The phase diagram of dense QCD},'' {\em Rept.
  Prog. Phys.} {\bf 74} (2011) 014001,
  \href{http://www.arXiv.org/abs/1005.4814}{{\tt 1005.4814}}.

\bibitem{Skokov:2009qp}
V.~Skokov, A.~Y. Illarionov, and V.~Toneev, ``{Estimate of the magnetic field
  strength in heavy-ion collisions},'' {\em Int. J. Mod. Phys. A} {\bf 24}
  (2009) 5925--5932, \href{http://www.arXiv.org/abs/0907.1396}{{\tt
  0907.1396}}.

\bibitem{STAR:2017ckg}
{\bf STAR} Collaboration, L.~Adamczyk {\em et al.}, ``{Global $\Lambda$ hyperon
  polarization in nuclear collisions: evidence for the most vortical fluid},''
  {\em Nature} {\bf 548} (2017) 62--65,
  \href{http://www.arXiv.org/abs/1701.06657}{{\tt 1701.06657}}.

\bibitem{Miransky:2015ava}
V.~A. Miransky and I.~A. Shovkovy, ``{Quantum field theory in a magnetic field:
  From quantum chromodynamics to graphene and Dirac semimetals},'' {\em Phys.
  Rept.} {\bf 576} (2015) 1--209,
  \href{http://www.arXiv.org/abs/1503.00732}{{\tt 1503.00732}}.

\bibitem{Mizher:2010zb}
A.~J. Mizher, M.~N. Chernodub, and E.~S. Fraga, ``{Phase diagram of hot QCD in
  an external magnetic field: possible splitting of deconfinement and chiral
  transitions},'' {\em Phys. Rev. D} {\bf 82} (2010) 105016,
  \href{http://www.arXiv.org/abs/1004.2712}{{\tt 1004.2712}}.

\bibitem{Fukushima:2012xw}
K.~Fukushima and J.~M. Pawlowski, ``{Magnetic catalysis in hot and dense quark
  matter and quantum fluctuations},'' {\em Phys. Rev. D} {\bf 86} (2012)
  076013, \href{http://www.arXiv.org/abs/1203.4330}{{\tt 1203.4330}}.

\bibitem{Johnson:2008vna}
C.~V. Johnson and A.~Kundu, ``{External Fields and Chiral Symmetry Breaking in
  the Sakai-Sugimoto Model},'' {\em JHEP} {\bf 12} (2008) 053,
  \href{http://www.arXiv.org/abs/0803.0038}{{\tt 0803.0038}}.

\bibitem{Evans:2010xs}
N.~Evans, T.~Kalaydzhyan, K.-y. Kim, and I.~Kirsch, ``{Non-equilibrium physics
  at a holographic chiral phase transition},'' {\em JHEP} {\bf 01} (2011) 050,
  \href{http://www.arXiv.org/abs/1011.2519}{{\tt 1011.2519}}.

\bibitem{Fraga:2008um}
E.~S. Fraga and A.~J. Mizher, ``{Can a strong magnetic background modify the
  nature of the chiral transition in QCD?},'' {\em Nucl. Phys. A} {\bf 820}
  (2009) 103C--106C, \href{http://www.arXiv.org/abs/0810.3693}{{\tt
  0810.3693}}.

\bibitem{Bali:2011qj}
G.~S. Bali, F.~Bruckmann, G.~Endrodi, Z.~Fodor, S.~D. Katz, S.~Krieg,
  A.~Schafer, and K.~K. Szabo, ``{The QCD phase diagram for external magnetic
  fields},'' {\em JHEP} {\bf 02} (2012) 044,
  \href{http://www.arXiv.org/abs/1111.4956}{{\tt 1111.4956}}.

\bibitem{Endrodi:2015oba}
G.~Endrodi, ``{Critical point in the QCD phase diagram for extremely strong
  background magnetic fields},'' {\em JHEP} {\bf 07} (2015) 173,
  \href{http://www.arXiv.org/abs/1504.08280}{{\tt 1504.08280}}.

\bibitem{Bruckmann:2013oba}
F.~Bruckmann, G.~Endrodi, and T.~G. Kovacs, ``{Inverse magnetic catalysis and
  the Polyakov loop},'' {\em JHEP} {\bf 04} (2013) 112,
  \href{http://www.arXiv.org/abs/1303.3972}{{\tt 1303.3972}}.

\bibitem{Preis:2010cq}
F.~Preis, A.~Rebhan, and A.~Schmitt, ``{Inverse magnetic catalysis in dense
  holographic matter},'' {\em JHEP} {\bf 03} (2011) 033,
  \href{http://www.arXiv.org/abs/1012.4785}{{\tt 1012.4785}}.

\bibitem{Agasian:2008tb}
N.~O. Agasian and S.~M. Fedorov, ``{Quark-hadron phase transition in a magnetic
  field},'' {\em Phys. Lett. B} {\bf 663} (2008) 445--449,
  \href{http://www.arXiv.org/abs/0803.3156}{{\tt 0803.3156}}.

\bibitem{Fukushima:2008xe}
K.~Fukushima, D.~E. Kharzeev, and H.~J. Warringa, ``{The Chiral Magnetic
  Effect},'' {\em Phys. Rev. D} {\bf 78} (2008) 074033,
  \href{http://www.arXiv.org/abs/0808.3382}{{\tt 0808.3382}}.

\bibitem{Kharzeev:2007jp}
D.~E. Kharzeev, L.~D. McLerran, and H.~J. Warringa, ``{The Effects of
  topological charge change in heavy ion collisions: 'Event by event P and CP
  violation'},'' {\em Nucl. Phys. A} {\bf 803} (2008) 227--253,
  \href{http://www.arXiv.org/abs/0711.0950}{{\tt 0711.0950}}.

\bibitem{Chernodub:2016kxh}
M.~N. Chernodub and S.~Gongyo, ``{Interacting fermions in rotation: chiral
  symmetry restoration, moment of inertia and thermodynamics},'' {\em JHEP}
  {\bf 01} (2017) 136, \href{http://www.arXiv.org/abs/1611.02598}{{\tt
  1611.02598}}.

\bibitem{Chernodub:2017ref}
M.~N. Chernodub and S.~Gongyo, ``{Effects of rotation and boundaries on chiral
  symmetry breaking of relativistic fermions},'' {\em Phys. Rev. D} {\bf 95}
  (2017), no.~9, 096006, \href{http://www.arXiv.org/abs/1702.08266}{{\tt
  1702.08266}}.

\bibitem{Wang:2018sur}
X.~Wang, M.~Wei, Z.~Li, and M.~Huang, ``{Quark matter under rotation in the NJL
  model with vector interaction},'' {\em Phys. Rev. D} {\bf 99} (2019), no.~1,
  016018, \href{http://www.arXiv.org/abs/1808.01931}{{\tt 1808.01931}}.

\bibitem{Zhang:2020jux}
Z.~Zhang, C.~Shi, X.~Luo, and H.-S. Zong, ``{Chiral phase transition in a
  rotating sphere},'' {\em Phys. Rev. D} {\bf 101} (2020), no.~7, 074036,
  \href{http://www.arXiv.org/abs/2003.03765}{{\tt 2003.03765}}.

\bibitem{Braguta:2021jgn}
V.~V. Braguta, A.~Y. Kotov, D.~D. Kuznedelev, and A.~A. Roenko, ``{Influence of
  relativistic rotation on the confinement-deconfinement transition in
  gluodynamics},'' {\em Phys. Rev. D} {\bf 103} (2021), no.~9, 094515,
  \href{http://www.arXiv.org/abs/2102.05084}{{\tt 2102.05084}}.

\bibitem{Chernodub:2020yaf}
M.~N. Chernodub and V.~E. Ambrus, ``{Phase diagram of helically imbalanced QCD
  matter},'' {\em Phys. Rev. D} {\bf 103} (2021), no.~9, 094015,
  \href{http://www.arXiv.org/abs/2005.03575}{{\tt 2005.03575}}.

\bibitem{Liu:2017spl}
Y.~Liu and I.~Zahed, ``{Pion Condensation by Rotation in a Magnetic field},''
  {\em Phys. Rev. Lett.} {\bf 120} (2018), no.~3, 032001,
  \href{http://www.arXiv.org/abs/1711.08354}{{\tt 1711.08354}}.

\bibitem{Liu:2017zhl}
Y.~Liu and I.~Zahed, ``{Rotating Dirac fermions in a magnetic field in 1+2 and
  1+3 dimensions},'' {\em Phys. Rev. D} {\bf 98} (2018), no.~1, 014017,
  \href{http://www.arXiv.org/abs/1710.02895}{{\tt 1710.02895}}.

\bibitem{Chen:2019tcp}
H.-L. Chen, X.-G. Huang, and K.~Mameda, ``{Do charged pions condense in a
  magnetic field with rotation?},''
  \href{http://www.arXiv.org/abs/1910.02700}{{\tt 1910.02700}}.

\bibitem{Chen:2021aiq}
H.-L. Chen, X.-G. Huang, and J.~Liao, ``{QCD phase structure under rotation},''
  {\em Lect. Notes Phys.} {\bf 987} (2021) 349--379,
  \href{http://www.arXiv.org/abs/2108.00586}{{\tt 2108.00586}}.

\bibitem{Sadooghi:2021upd}
N.~Sadooghi, S.~M.~A. Tabatabaee~Mehr, and F.~Taghinavaz, ``{Inverse
  magnetorotational catalysis and the phase diagram of a rotating hot and
  magnetized quark matter},'' {\em Phys. Rev. D} {\bf 104} (2021), no.~11,
  116022, \href{http://www.arXiv.org/abs/2108.12760}{{\tt 2108.12760}}.

\bibitem{Ritus:1972ky}
V.~I. Ritus, ``{Radiative corrections in quantum electrodynamics with intense
  field and their analytical properties},'' {\em Annals Phys.} {\bf 69} (1972)
  555--582.

\bibitem{Chen:2017xrj}
H.-L. Chen, K.~Fukushima, X.-G. Huang, and K.~Mameda, ``{Surface Magnetic
  Catalysis},'' {\em Phys. Rev. D} {\bf 96} (2017), no.~5, 054032,
  \href{http://www.arXiv.org/abs/1707.09130}{{\tt 1707.09130}}.

\bibitem{Chen:2015hfc}
H.-L. Chen, K.~Fukushima, X.-G. Huang, and K.~Mameda, ``{Analogy between
  rotation and density for Dirac fermions in a magnetic field},'' {\em Phys.
  Rev. D} {\bf 93} (2016), no.~10, 104052,
  \href{http://www.arXiv.org/abs/1512.08974}{{\tt 1512.08974}}.

\bibitem{Ding:2015ona}
H.-T. Ding, F.~Karsch, and S.~Mukherjee, ``{Thermodynamics of
  strong-interaction matter from Lattice QCD},'' {\em Int. J. Mod. Phys. E}
  {\bf 24} (2015), no.~10, 1530007,
  \href{http://www.arXiv.org/abs/1504.05274}{{\tt 1504.05274}}.

\bibitem{Tabatabaee:2020efb}
S.~M.~A. Tabatabaee and N.~Sadooghi, ``{Wigner function formalism and the
  evolution of thermodynamic quantities in an expanding magnetized plasma},''
  {\em Phys. Rev. D} {\bf 101} (2020), no.~7, 076022,
  \href{http://www.arXiv.org/abs/2003.01686}{{\tt 2003.01686}}.

\bibitem{Sadooghi:2016jyf}
N.~Sadooghi and F.~Taghinavaz, ``{Dilepton production rate in a hot and
  magnetized quark-gluon plasma},'' {\em Annals Phys.} {\bf 376} (2017)
  218--253, \href{http://www.arXiv.org/abs/1601.04887}{{\tt 1601.04887}}.

\bibitem{Kapusta:2006pm}
J.~I. Kapusta and C.~Gale, {\em {Finite-temperature field theory: Principles
  and applications}}.
\newblock Cambridge Monographs on Mathematical Physics. Cambridge University
  Press, 2011.

\bibitem{Santiago:2018kds}
J.~Santiago and M.~Visser, ``{Tolman temperature gradients in a gravitational
  field},'' {\em Eur. J. Phys.} {\bf 40} (2019), no.~2, 025604,
  \href{http://www.arXiv.org/abs/1803.04106}{{\tt 1803.04106}}.

\bibitem{Tolman:1930ona}
R.~Tolman and P.~Ehrenfest, ``{Temperature Equilibrium in a Static
  Gravitational Field},'' {\em Phys. Rev.} {\bf 36} (1930), no.~12, 1791--1798.

\bibitem{McLerran:2021zvt}
L.~D. McLerran, ``{Lecture on Quarkyonic Effective Field Theory},'' {\em Acta
  Phys. Polon. B} {\bf 52} (2021), no.~3, 229--241.

\bibitem{Asakawa:1989bq}
M.~Asakawa and K.~Yazaki, ``{Chiral Restoration at Finite Density and
  Temperature},'' {\em Nucl. Phys. A} {\bf 504} (1989) 668--684.

\bibitem{Morones-Ibarra:2017avu}
J.~R. Morones-Ibarra, A.~Enriquez-Perez-Gavilan, A.~I.~H. Rodriguez, F.~V.
  Flores-Baez, N.~B. Mata-Carrizalez, and E.~V. Ordo\~nez, ``{Chiral symmetry
  restoration and the critical end point in QCD},'' {\em Open Phys.} {\bf 15}
  (2017), no.~1, 1039--1044.

\end{thebibliography}\endgroup
\end{document}